\documentclass[fleqn,usenatbib]{mnras}
\pdfoutput=1

\usepackage[T1]{fontenc}

\DeclareRobustCommand{\VAN}[3]{#2}
\let\VANthebibliography\thebibliography
\def\thebibliography{\DeclareRobustCommand{\VAN}[3]{##3}\VANthebibliography}

\usepackage{graphicx}	
\usepackage{amsmath}	
\usepackage{amssymb}	
\usepackage{newtxtext,newtxmath}

\newcommand{\reduceme}{\mbox{R\raisebox{-0.35ex}{E}D%
\hspace{-0.05em}\raisebox{0.85ex}{uc}\hspace{-0.90em}%
\raisebox{-.35ex}{{m}}\hspace{0.05em}E}}

\title[Synthetic RGB photometry of bright stars]{Synthetic RGB
photometry of bright stars: definition of the standard
photometric system and UCM library of spectrophotometric spectra}

\author[N. Cardiel et al.]{%
Nicol\'{a}s Cardiel,$^{1,2}$\thanks{E-mail: cardiel@ucm.es (NC)}
Jaime Zamorano,$^{1,2}$
Salvador Bar\'{a},$^{3}$
Alejandro S\'{a}nchez de Miguel$^{1,4,5}$
\newauthor
Cristina Cabello,$^{1,2}$
Jes\'{u}s Gallego,$^{1,2}$
Luc\'{i}a Garc\'{\i}a,$^{1}$
Rafael Gonz\'{a}lez,$^{1}$
\newauthor
Jaime Izquierdo,$^{1}$
Sergio Pascual,$^{1,2}$
Jos\'{e} Robles,$^{1}$
Ainhoa S\'{a}nchez,$^{1}$
and Carlos Tapia$^{1}$
\\
$^{1}$Departamento de F\'{\i}sica de la Tierra y Astrof\'{\i}sica,
Fac.~CC.~F\'{\i}sicas, Universidad Complutense de Madrid, Plaza de las
Ciencias~1, E-28040, Spain\\
$^{2}$Instituto de F\'{\i}sica de Part\'{\i}culas y del Cosmos, IPARCOS,
Fac.~CC.~F\'{\i}sicas, Universidad Complutense de Madrid, Plaza de las
Ciencias~1, E-28040 Madrid, Spain\\
$^{3}$Departamento de F\'{\i}sica Aplicada, Universidade de Santiago de
Compostela, E-15782 Santiago de Compostela, Galicia, Spain\\
$^{4}$Environment and Sustainability Institute, University of Exeter, Penryn,
Cornwall TR10 9FE, UK\\
$^{5}$Instituto de Astrof\'{\i}sica de Andaluc\'{\i}a, Glorieta de la
Astronom\'{\i}a, s/n,C.P.18008 Granada, Spain
}

\date{Accepted XXX. Received YYY; in original form ZZZ}

\pubyear{2021}

\begin{document}
\label{firstpage}
\pagerange{\pageref{firstpage}--\pageref{lastpage}}
\maketitle

\begin{abstract}
Although the use of RGB photometry has exploded in the last decades due to the
advent of high-quality and inexpensive digital cameras equipped with Bayer-like
color filter systems, there is surprisingly no catalogue of bright stars that
can be used for calibration purposes. Since due to their excessive brightness,
accurate enough spectrophotometric measurements of bright stars typically
cannot be performed with modern large telescopes, we have employed historical
13-color medium-narrow-band photometric data, gathered with quite reliable
photomultipliers, to fit the spectrum of 1346~bright stars using stellar
atmosphere models. This not only constitutes a useful compilation of bright
spectrophotometric standards well spread in the celestial sphere, the UCM
library of spectrophotometric spectra, but allows the generation of a catalogue
of reference RGB magnitudes, with typical random uncertainties $\sim 0.01$~mag.
For that purpose, we have defined a new set of spectral sensitivity curves,
computed as the median of 28~sets of empirical sensitivity curves from the literature,
that can be used to establish a standard RGB photometric system. Conversions
between RGB magnitudes computed with any of these sets of empirical RGB curves
and those determined with the new standard photometric system are provided.
Even though particular RGB measurements from single cameras are not expected to
provide extremely accurate photometric data, the repeatability and multiplicity
of observations will allow access to a large amount of exploitable data in many
astronomical fields, such as the detailed monitoring of light pollution and its
impact on the night sky brightness, or the study of meteors, solar system
bodies, variable stars, and transient objects. In addition, the RGB magnitudes
presented here make the sky an accessible
and free laboratory for the calibration of the cameras themselves.
\end{abstract}





\section{Introduction}

Scientific and commercial-grade RGB cameras provide affordable means for the
acquisition of quantitative radiance data with large fields-of-view, using
a large number of pixels, and moderate multispectral content. Manufactured with
different technologies, and widely different in terms of their absolute
sensitivities, noise levels, pixel sizes and processing capabilities, all of
them share the key common feature of detecting light in three spectral channels
broadly comparable across devices, centered in the red, green, and blue regions
of the visible spectrum, providing that way a relatively
homogenous basis for data sharing and processing. 

The sustained improvement in the performance of the CCD and CMOS RGB sensor
arrays enabled the development of an increasing number of scientific
applications in astronomy, by both the professional and amateur communities,
which is expected to grow in the coming years. Among other
examples, commercial-grade RGB cameras have proven to be valuable science tools
for night sky brightness measurement \citep{2018hanel_etal, 2018jechow_etal,
2019bertolo_etal, 2019d_jechow_etal, 2019a_jechow_etal, 2019b_jechow,
2019c_jechow_etal, 2020kollath_etal}, radiometry of artificial light polluting
sources from Earth-orbit platforms \citep{2014kyba_etal, 2017stefanov_etal,
2018zheng_etal, 2019sanchezdemiguel_etal, 2020sanchezdemiguel_etal}, as well as
from airborne \citep{2012kuechly_etal, 2020bouroussis_topalis}, and ground
based stations \citep{2015dobler_etal, 2018meier, 2019bara_et_al}, meteor and
fireball detection \citep{2009JIMO...37...28G}, planetary astronomy
\citep{2014Mousis_etal}, and variable stars \citep{2016aavso_dslr_manual}.

Besides, the exponential growth of the consumer optoelectronics segment opens
unprecedented opportunities for large-scale citizen science projects \citep[see
e.g.][]{2019kiba_loss_of_the_night, 2020azotea_project}: recent market studies
\citep{2017LDV_capital} estimate that by 2022 as much as~45 billion cameras
(defined as the combination of an objective lens plus a focal plane
spatially resolved sensor) will be operative worldwide in different supports,
from hand-held smartphones and classical photographic units to home appliances
and artificial vision systems. 

Despite these facts, a consistent astronomical magnitude system has not yet
been defined for RGB photometry. The possibility of reporting fluxes
(irradiances) in astronomical magnitudes, and surface brightnesses (radiances)
in magnitudes per square arcsecond in the native photometric bands of these
widely used devices is appealing. The relative similarity of the $R$, $G$, and
$B$ channels across camera models, and their expected stability in the
foreseeable future (as far as cameras continue to be developed for human
vision-driven applications) are two key features enabling large-scale
broadly-homogeneous data gathering and monitoring across extended periods of
time.

We develop in this work a complete RGB photometric system, characterized by a
set of basic filters, the use of photon-based quantities, and with zero points
defined in the absolute (AB) scale. We also provide a catalogue of RGB star
magnitudes corresponding to a good subset of the brightest stars in the
celestial sphere, that can be used as a reference for calibration purposes. The
focus on the bright stars comes from the need to provide observers using
wide-field digital cameras with the possibility of properly calibrating their
images, even with short exposure times. The goal is to reach the same level of
calibration accuracy attained in radiometric measurements performed in
laboratory within the optical range \citep[see][Table
10-4]{1998radiometrybook}: 1--2.5\% with tungsten lamps, or even 0.1\% with
self-calibrating detectors.

The absolute spectrophotometric flux calibration of bright star spectra is not
an easy task with modern large telescopes and spectrographs,
mainly due to the difficulty to avoid light losses while
preserving the desired spectral resolutions using narrow slits, but also
because of other observational problems like avoiding detector saturation or
lack of linearity of the detectors themselves when used at very high count
rates, the proper correction of atmospheric extinction or differential
refraction (when observing at non-negligible airmasses, with the slit position
angle different from the parallactic angle, while covering relatively wide
wavelength ranges), or the inhomogeneous illumination of the spectrograph focal
plane in very short exposure times due to the limited speed of camera shutters,
to mention a few.  Although the use of neutral density filters can help to
alleviate some of these problems, this approach not only requires a good
spectrophotometric calibration of the density filters themselves, but also
translates into using modern telescopes to observe bright stars with large
exposure times, something that typically is not easy to get approved by
telescope allocation committees.

For all those reasons we decided to base our work on historical, but
quite reliable and homogeneous, medium-narrow-band photometric data, that can
be used to fit star model spectra able reproduce the observed spectral energy
distribution of bright stars. The fitted models have then been used to compute
synthetic RGB magnitudes, fulfilling our initial goal. 

A brief description of the practical computation of synthetic magnitudes
followed in this work is summarised in Section~\ref{sec:synthetic_magnitudes}.
Section~\ref{sec:data_sample} describes the photometric data and initial sample
selection, while the model fitting procedure and final sample definition, based
on comparisons between synthetic magnitudes and additional photometric
measurements, are presented in Section~\ref{sec:spectrum_fitting}.  To avoid
the problem of choosing the RGB sensitivity curves of a particular camera, we
decided to define a set of reference RGB spectral sensitivity curves, using
median values from existing sensitivity curves of well-known cameras, that can
be used to establish an RGB standard photometric system. The definition of
these curves is presented in Section~\ref{sec:synthetic_rgb_magnitudes},
together with the RGB magnitudes for the 1346 stars constituting the UCM
library of spectrophotometric standards, and a discussion concerning the
conversion between magnitudes measured under the standard RGB system
defined here and those derived employing the RGB sensitivity
curves of individual cameras. The conclusions of this work are summarized in
Section~\ref{sec:conclusions}, while Appendices~A and~B
include the graphical comparison of the
results of applying the adopted fitting technique in 39 stars
with available spectrophotometric data from the literature, and a table with
polynomial coefficients that allow the computation of the expected differences
between the standard RGB system and 28~particular digital cameras,
respectively.


\section{Computation of synthetic magnitudes}
\label{sec:synthetic_magnitudes}

Synthetic magnitudes in this work have been determined using the Python package
synphot \citep{2018ascl.soft11001S}\footnote{\url{https://synphot.readthedocs.io/en/latest/}}, which
facilitates the computation of photometric properties from user-defined
bandpasses and spectra. This package follows the photon-counting formalism,
expected for modern CCD detectors \citep[see e.g.][]{2014MNRAS.444..392C},
where the number of photons, instead of the arriving energy, is the relevant
property to be considered. In this way, for a particular bandpass defined in
the wavelength interval ranging from $\lambda_i$ to $\lambda_f$, magnitudes are
computed following
\begin{align}
\label{eq:mag_photons}
m = -2.5\,\log_{10} 
\frac{\int_{\lambda_i}^{\lambda_f} 
n_\gamma(\lambda)\; T(\lambda)\; {\rm d}\lambda}%
{\int_{\lambda_i}^{\lambda_f} 
n_{\gamma,r}(\lambda) \; T(\lambda)\; {\rm d}\lambda} =
-2.5\,\log_{10}
\frac{N_\gamma}{N_{\gamma,r}},
\end{align}
where $n_\gamma(\lambda)$ is the number of photons per unit time and per unit
spectral bandwidth at the wavelength $\lambda$ through a unit area
(\mbox{photons~s$^{-1}$~cm$^{-2}$~\AA$^{-1}$}; flux units
known as \texttt{PHOTLAM} in synphot) for the desired target. Similarly,
$n_{\gamma,r}(\lambda)$ has the same physical meaning for the spectrum to be
used to set the $m\!=\!0$ reference point. In addition, $T(\lambda)$ is the
system spectral sensitivity response.  Sometimes, it is also useful to compute
$N_\gamma$, the integrated number of photons
(photons~s$^{-1}$~cm$^{-2}$) within the bandpass, modulated by
the spectral sensitivity response, although logically the absolute number
depends on the particular normalization of $T(\lambda)$. In
this sense, it is important to note that if the averaged number of photons
within the bandpass is the sought parameter, a proper normalization must be
performed using $T(\lambda)$ as the weighting factor
\begin{align}
\label{eq:averaged_photons}
\langle n_{\gamma} \rangle = \frac{\int_{\lambda_i}^{\lambda_f} 
n_\gamma(\lambda)\; T(\lambda)\; {\rm d}\lambda}%
{\int_{\lambda_i}^{\lambda_f} T(\lambda)\; {\rm d}\lambda} =
\frac{N_\gamma}{\int_{\lambda_i}^{\lambda_f} T(\lambda)\; {\rm d}\lambda},
\end{align}
with a similar expression for the reference spectrum $\langle n_{\gamma,r}
\rangle$, interchanging $n_\gamma(\lambda)$ by $n_{\gamma,r}(\lambda)$, and
$N_\gamma$ by $N_{\gamma,r}$, in the
above equation. The denominator of the last equation, that works as
normalization factor and is computed in synphot as the bandpass equivalent
width, is actually also present in both the numerator and
denominator of Eq.~\ref{eq:mag_photons}, but cancels out and does not appear
explicitly.

Since the number of photons is directly related to the incoming
flux densities, magnitudes can also be computed as
\begin{align}
\label{eq:mag_flux_densities}
m = -2.5\,\log_{10} 
\frac{\int_{\lambda_i}^{\lambda_f} 
f(\lambda)\; \frac{\lambda}{h\, c}\;T(\lambda)\; {\rm d}\lambda}%
{\int_{\lambda_i}^{\lambda_f} 
f_r(\lambda) \; \frac{\lambda}{h\, c}\;T(\lambda)\; {\rm d}\lambda},
\end{align}
being $f(\lambda)$ the flux density of the target
(\mbox{erg\,s$^{-1}$\,cm$^{-2}$\,\AA$^{-1}$}; flux units known as \texttt{FLAM}
in synphot), and $f_r(\lambda)$ the flux density of the reference spectrum. In
this case, we have not simplified the $h\,c$ factor in the last equation in
order to keep the traceability of the units involved, being both the numerator
and denominator in that fraction given in \mbox{photons~s$^{-1}$\,cm$^{-2}$}.

We have checked that the computed synthetic magnitudes with synphot agree with
the measurements performed with
pyphot\footnote{\url{https://mfouesneau.github.io/docs/pyphot/}}, another
Python package providing tools to compute synthetic photometry. In no case the
differences found were larger than~0.0002~mag. In addition, we also performed
our own integrations by directly programming the equations shown in this
section, using trapezoidal integrations with the Numpy function
trapz\footnote{\url{https://numpy.org/doc/stable/reference/generated/numpy.trapz.html}},
being the largest differences also smaller than 0.0002~mag. However, it is
worth noticing that when applying a different integration strategy, namely the
Simpson's rule with the scipy function
simpson\footnote{\url{https://docs.scipy.org/doc/scipy/reference/generated/scipy.integrate.simpson.html}},
the differences are slightly larger, reaching in some cases 0.01~mag. Since the
Simpson's rule approximates the original function using piecewise quadratic
functions, it is clear that this fact has a non-negligible effect on the
computations. For that reason, we advocate the use of synphot or pyphot, when
delegating in third-party software packages when computing synthetic
magnitudes, or the implementation of the simple trapezoidal integration, when
employing user-defined code.


\section{Data sample}
\label{sec:data_sample}

\subsection{Historical 13-color photometric data}

\begin{figure}
\includegraphics[width=\columnwidth]{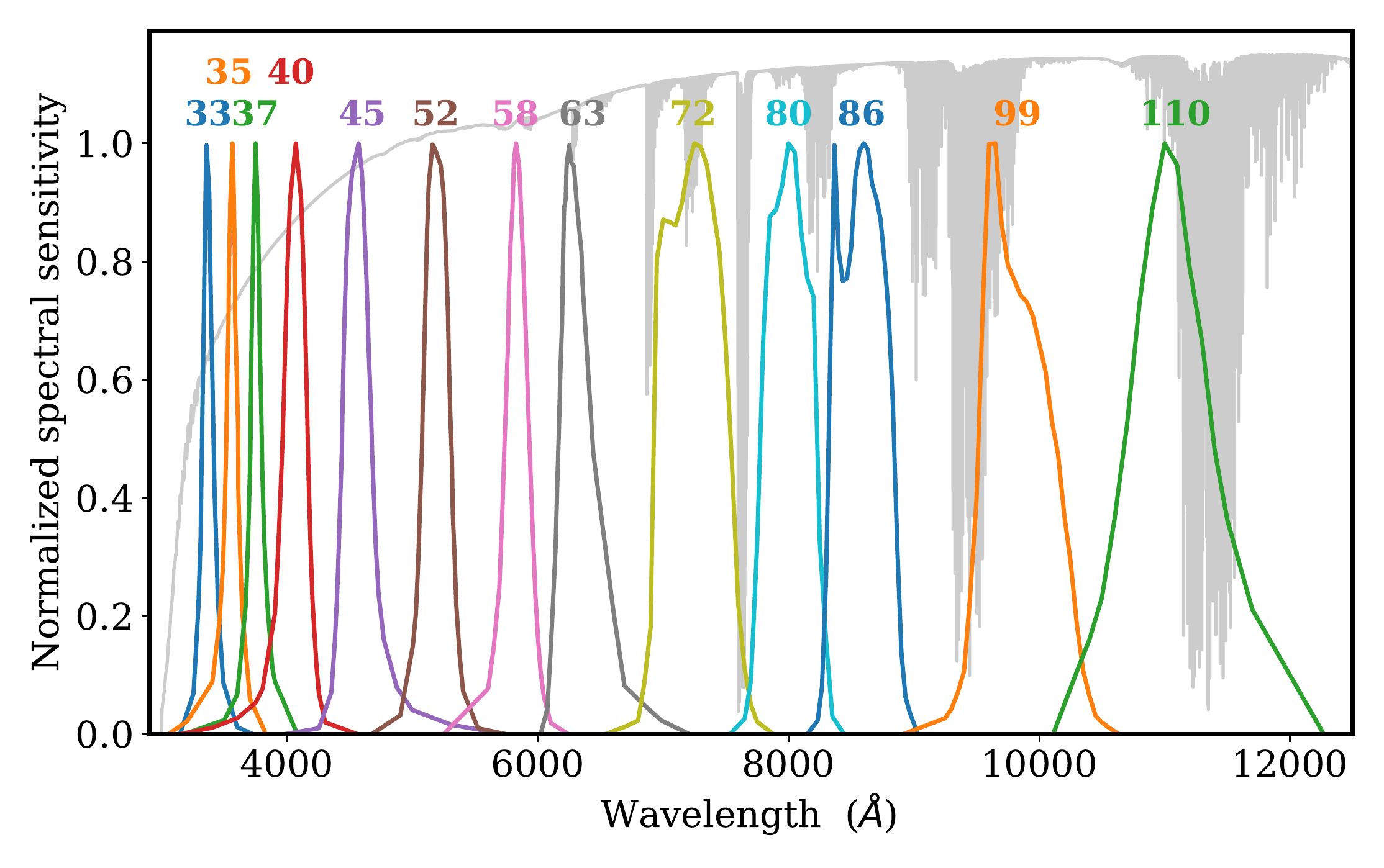}
\caption{Spectral sensitivity curves for the 13 filters, as given by \citet[see
their Table~1]{1975RMxAA...1..299J}. The name of each filter indicates its
approximate effective wavelength. Just for illustration, we are also displaying
the atmospheric telluric absorption (upper gray line) computed with the help of
the ESO SKYCALC tool \citep{2014A&A...568A...9M}.}
\label{fig:response_curves}
\end{figure}

\begin{table}
\centering
\caption{Basic properties of the filters composing the 13C photometric system:
(1) filter identification, (2) bandpass average wavelength, (3) equivalent
width, and (4) full width at half maximum.}
\label{tab:filter_properties}
\begin{tabular}{crcc} 
\hline
Filter & $\lambda_0$~(\AA) & Equivalent width (\AA) &  FWHM (\AA)\\
\hline
33 & 3374 & 116 & 143 \\
35 & 3540 & 132 & 224 \\
37 & 3749 & 133 & 213 \\
40 & 4034 & 240 & 349 \\
45 & 4603 & 285 & 383 \\
52 & 5187 & 259 & 250 \\
58 & 5822 & 230 & 247 \\
63 & 6360 & 326 & 392 \\
72 & 7239 & 589 & 463 \\
80 & 8002 & 435 & 348 \\
86 & 8585 & 483 & 380 \\
99 & 9828 & 583 & 575 \\
110 & 11088 & 830 & 872 \\
\hline
\end{tabular}
\end{table}

The classical 13-color (hereafter 13C) medium-narrow-band photometric system
\citep{1967CoLPL...6...85J, 1969CoLPL...8....1M, 1975RMxAA...1..299J,
1982RMxAA...5..149S} was created with the aim to obtain homogeneous and
calibrated photometric measurements of bright stars, with a high level of
accuracy.  Each filter is identified by a number that indicates its approximate
effective wavelength. The spectral sensitivity curves for all the filters are
displayed in Fig.~\ref{fig:response_curves}, whereas some basic filter
properties are provided in Table~\ref{tab:filter_properties}. The seven bluest
filters, with effective wavelengths ranging from 3370~\AA\ to~5820~\AA,
provided quantitative information concerning the behaviour
of the continuum of
early-type stars, including the Balmer jump. The six reddest filters, covering
the 6360~\AA\ to~11090~\AA\ region, were selected to avoid conspicuous telluric
atmospheric features.  

Apart from the initial references, the usefulness of the
13C photometric system was clearly demonstrated through numerous stellar
studies derived from its use
\citep[see e.g.][and references therein]{1976RMxAA...1..327S,
1978BAAS...10..687A, 1979RMxAA...4..233S, 1979RMxAA...4..301S, 
1979RMxAA...4..307S,
1979AN....300..295M, 1981RMxAA...6..159C,
1982RMxAA...5..137S, 1982RMxAA...5..149S, 1982RMxAA...5..173A,
1983PASP...95...35S,
1984RMxAA...9...53S, 1984RMxAA...9..141S,
1985IAUS..111..553C, 1985RMxAA..11....7S,
1989A_AS...78..511P, 1993RMxAA..26..100B,
1997PASP..109..958B}.

\subsection{The star sample}

\begin{figure}
\includegraphics[width=\columnwidth]{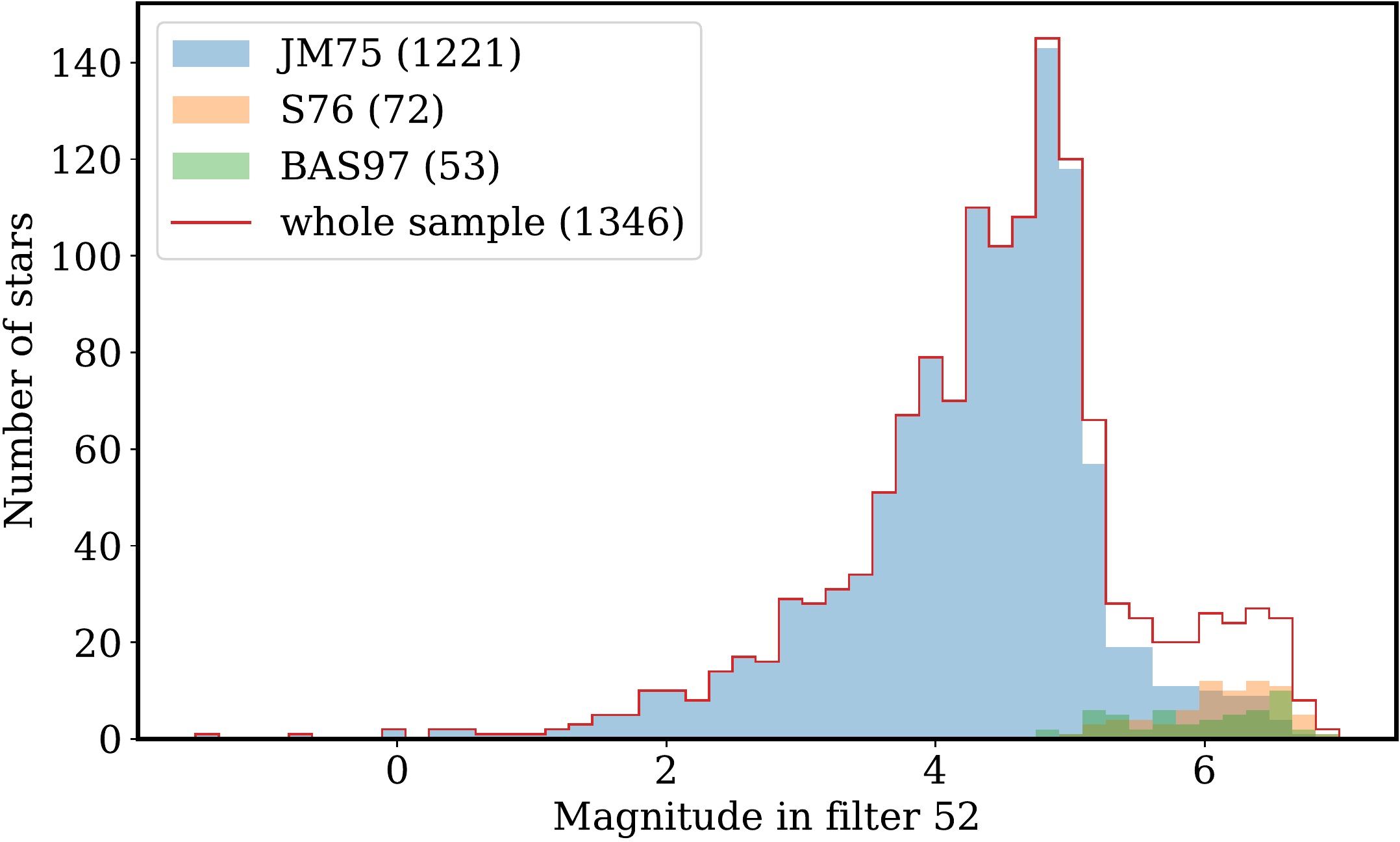}
\caption{Distribution of the final star selection according to the stellar
magnitude in the 52 filter. The number of stars belonging to each source is
displayed between parenthesis in the figure key. The red line delineates the
coadded histogram.  It is clear that most of the stars come from JM75, with a
small contribution of S76 and BAS97 in the faint regime of the sample.}
\label{fig:hist_mag52}
\end{figure}

\begin{figure*}
\includegraphics[width=\textwidth]{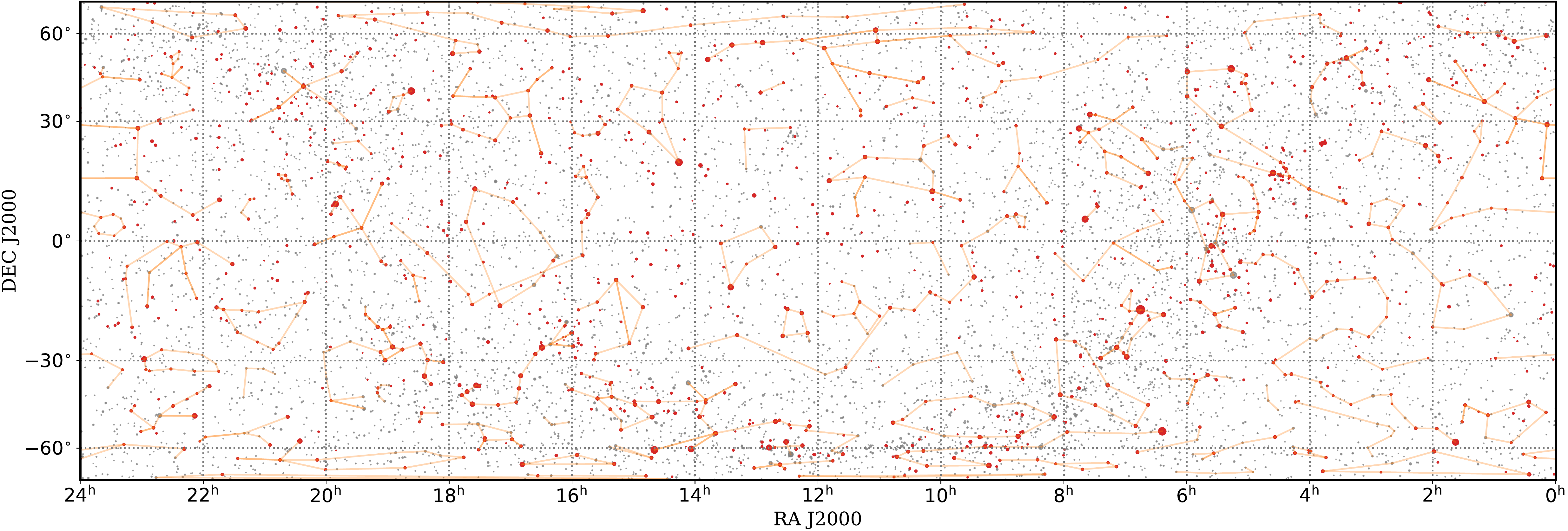}
\caption{Lambert cylindrical equal-area projection in J2000 equatorial
coordinates showing the distribution of the stars belonging to the Bright Star
Catalog \citep{1964cbs..book.....H}. The 1346 stars composing the UCM library
of spectrophotometric standards are highlighted as red filled cicles.
Simple constellations shapes \citep{2005Constellations}, displayed with
light orange lines, are also shown to facilitate the visual identification of
the stars. The size of each star symbol is inversely proportional to its
Johnson $V$~magnitude.}
\label{fig:lambert_radec}
\end{figure*}

We have assembled 13C photometric data of bright stars, belonging to the Bright
Star Catalog \citep{1964cbs..book.....H}, coming from three different sources:
1380~stars from \citet[hereafter JM75]{1975RMxAA...1..299J}, 81~stars from
\citet[hereafter S76]{1976RMxAA...1..327S}, and 71~stars from \citet[hereafter
BAS97]{1997PASP..109..958B}\footnote{By comparing repeated stars between JM75
and BAS97, we noted that there is an erratum in the header description of
Table~1 in BAS97: the column labelled as 58$-$63 is actually 52$-$63.}. The
largest contribution comes from the JM75 sample, that basically covered all the
stars brighter than the fifth visual magnitude north of declination
$-20^{\degr}$, and most of the stars brighter than the fourth visual magnitude
below that declination. It is important to note that 163 stars from the JM75
sample ($\sim 12$\% of the objects) do not have photometric data for the
5~reddest filters (72, 80, 86, 99 and 110).  On the other hand, the S76 and
BAS97 samples correspond to solar-type and A0-K0 supergiant stars,
respectively.  Taking into account that there are one and nine stars in common
between JM75 and S76, and between JM75 and BAS97, respectively, the total
initial number of different stars is 1522. As explained later
(Sect.~\ref{sec:cleaning_the_sample}), the initial list
was cleaned by removing stars with poor results in the fitting process or by
discrepancies with the available Johnson $B$ and~$V$ photometric data retrieved
from the Simbad database.
The final sample, listed in Table~\ref{tab:bigtable}, comprises a total of
1346~stars, with a magnitude distribution as shown in
Fig.~\ref{fig:hist_mag52}, being the stars well spread over
the whole celestial sphere, as displayed in Fig.~\ref{fig:lambert_radec}.

\begin{figure}
\includegraphics[width=\columnwidth]{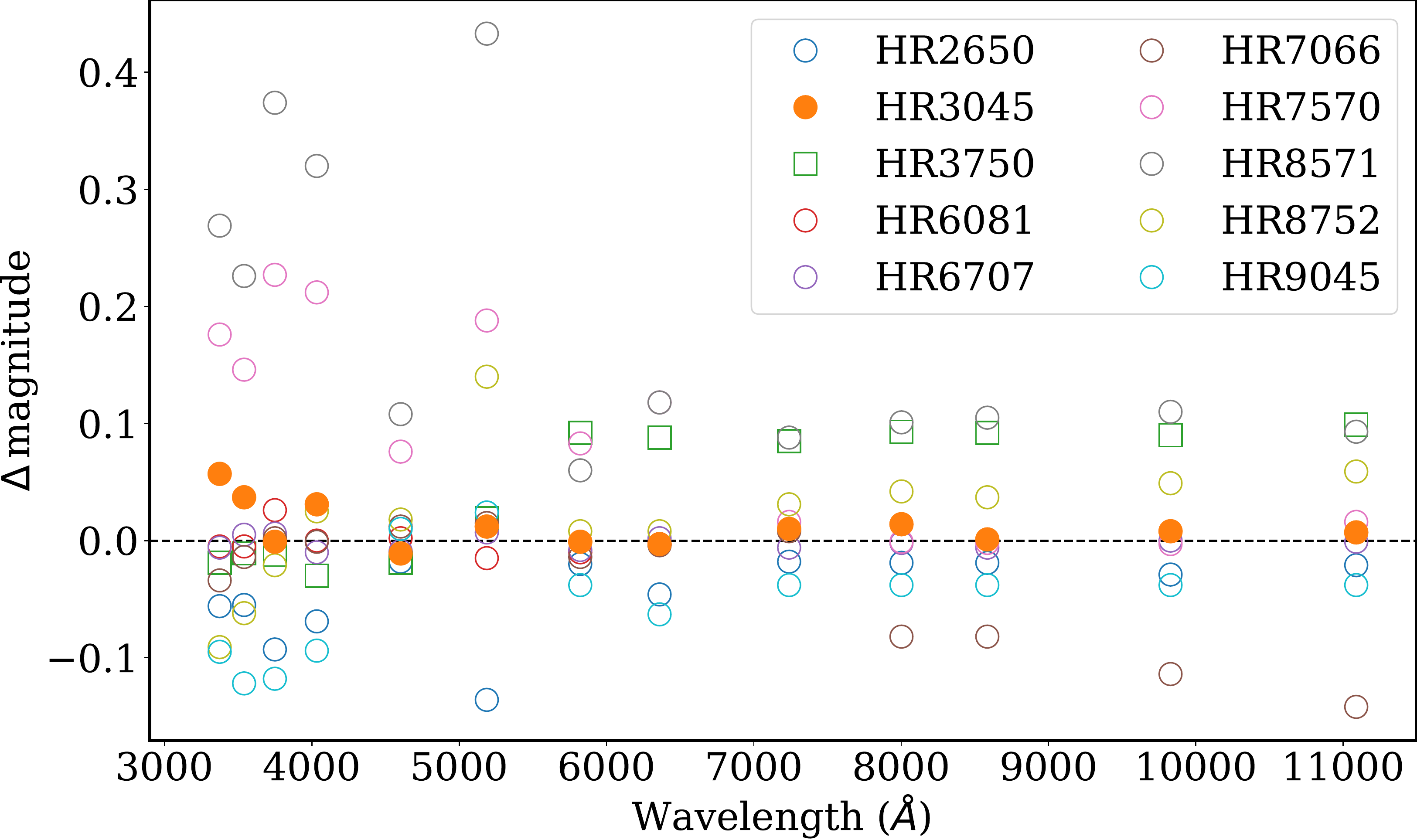}
\caption{Differences in the photometric measurements of 1~star from S76
(HR3750, open square) and 9~stars from BAS97 (the remaining objects) in common
with the JM75 sample. Open circles indicate variable stars. HR3750 is a binary
system. The r.m.s.\ for HR3045, the only non-variable and non-binary star, is
$0.018$~mag.}
\label{fig:repeated_stars}
\end{figure}

The comparison of the common stars in the three original sources is shown in
Fig.~\ref{fig:repeated_stars}. Even though most of these repeated stars are
variable, the scatter of the photometric measurements (except for the bluest
filters) is compatible with a $~\sim 0.1$~mag dispersion.  Focusing on HR3045,
the only star that is not variable and that does not belong to a binary system,
the r.m.s.\ for the 13 photometric measurements is~0.018~mag, which is
perfectly compatible with the probable errors of single observations in JM75
and S76, reported by \citet[see his Table 2]{1982RMxAA...5..149S} to be $\sim
0.02$~mag.  The photometric measurements for these 10 repeated stars have been
averaged for the subsequent work.


\section{Spectrum fitting}
\label{sec:spectrum_fitting}

\begin{table*}
\centering
\caption{First 10 rows of the table with the final sample, constituted by 1346
stars, comprising the UCM library of spectrophotometric standards.  The full
table is electronically available at~\url{http://guaix.ucm.es/rgbphot}.  Column
description: (1)~HR number; (2)--(3)~star coordinates (J2000), as provided by
Simbad; (4)~identificacion of the companion star for double
star systems (see Table~\ref{tab:double_stars}); (5)~additional name when the star is a known variable according to
Simbad; (6)~variability type from \citet{2017ARep...61...80S}, 
as provided by Simbad; 
(7)--(19)~absolute flux densities in each of the C13 
photometric bandpasses (in units of \mbox{erg~s$^{-1}$~cm$^{-2}$~\AA$^{-1}$});
(20)--(22)~effective temperature, surface gravity and 
metallicity derived from the best CK04 fit to the available photometric data, 
with the corresponding uncertainties derived from the bootstrapped spectra
(note that the quoted
uncertainties are simply lower limits to the expected random errors in the
stellar atmospheric parameters; see discussion in
Sect.~\ref{sec:uncertainties_fitting}); 
(23)~scaling factor to convert the
fitted CK04 model fluxes into absolute flux densities using
Eq.~(\ref{eq:flux_normalization}); 
(24)--(25)~Johnson~$B$ and~$V$ magnitudes
(VEGA system), extracted from Simbad; 
(26)--(27)~synthetic Johnson~$B$ and $V$
magnitudes (VEGA system), measured in the best CK04 fit, with their
corresponding uncertainties estimated from the bootstrapped spectra;
(28)--(30)~synthetic RGB magnitudes (AB system), measured in 
the best CK04 fit, using the bandpass definitions given in 
Table~\ref{tab:median_RGB_table}, with uncertainties estimated from 
bootstrapping.}
\label{tab:bigtable}
\begin{tabular}{crrccccccc} \hline
\multicolumn{1}{c}{(1)} & \multicolumn{1}{c}{(2)} & \multicolumn{1}{c}{(3)} & \multicolumn{1}{c}{(4)} & \multicolumn{1}{c}{(5)} & \multicolumn{1}{c}{(6)} & \multicolumn{1}{c}{(7)} & \multicolumn{1}{c}{(8)} & \multicolumn{1}{c}{(9)} & \multicolumn{1}{c}{(10)} \\ 
\multicolumn{1}{c}{HR} & \multicolumn{1}{c}{RA (\textdegree)} & \multicolumn{1}{c}{DEC (\textdegree)} & \multicolumn{1}{c}{Double} & \multicolumn{1}{c}{Variable} & \multicolumn{1}{c}{Var.Type} & \multicolumn{1}{c}{$f_{33}^{\star}$ ($\times 10^{-9}$)} & \multicolumn{1}{c}{$f_{35}^{\star}$ ($\times 10^{-9}$)} & \multicolumn{1}{c}{$f_{37}^{\star}$ ($\times 10^{-9}$)} & \multicolumn{1}{c}{$f_{40}^{\star}$ ($\times 10^{-9}$)} \\ \hline
0003 &    1.3339247 &   $-$5.7076189 & --- & BC Psc & RS & 0.008065 & 0.009427 & 0.011144 & 0.024608 \\ 
0005 &    1.5658922 &   58.4367280 & --- & V640 Cas & CST: & 0.006548 & 0.006900 & 0.007224 & 0.012313 \\ 
0015 &    2.0969161 &   29.0904311 & --- & alf And & ACV: & 1.034568 & 0.962187 & 0.997723 & 1.354483 \\ 
0021 &    2.2945217 &   59.1497811 & --- & bet Cas & DSCTC & 0.272195 & 0.284138 & 0.377964 & 0.622496 \\ 
0025 &    2.3526731 &  $-$45.7474253 & --- & --- & --- & 0.015074 & 0.019376 & 0.021859 & 0.051003 \\ 
0027 &    2.5801942 &   46.0722722 & --- & --- & --- & 0.014914 & 0.016286 & 0.027913 & 0.046962 \\ 
0033 &    2.8160733 &  $-$15.4679794 & --- & --- & --- & 0.027421 & 0.028911 & 0.031828 & 0.046545 \\ 
0039 &    3.3089633 &   15.1835936 & --- & gam Peg & BCEP & 0.956366 & 0.839646 & 0.742393 & 0.785749 \\ 
0045 &    3.6506856 &   20.2067003 & --- & NSV    99 & undef & 0.001080 & 0.001705 & 0.002336 & 0.007366 \\ 
0048 &    3.6600689 &  $-$18.9328653 & --- & AE Cet & LB: & 0.001156 & 0.001833 & 0.002564 & 0.008143 \\ 
\hline
\end{tabular}
\begin{tabular}{cccccccccc} \hline
\multicolumn{1}{c}{(1)} & \multicolumn{1}{c}{(11)} & \multicolumn{1}{c}{(12)} & \multicolumn{1}{c}{(13)} & \multicolumn{1}{c}{(14)} & \multicolumn{1}{c}{(15)} & \multicolumn{1}{c}{(16)} & \multicolumn{1}{c}{(17)} & \multicolumn{1}{c}{(18)} & \multicolumn{1}{c}{(19)} \\ 
\multicolumn{1}{c}{HR} & \multicolumn{1}{c}{$f_{45}^{\star}$ ($\times 10^{-9}$)} & \multicolumn{1}{c}{$f_{52}^{\star}$ ($\times 10^{-9}$)} & \multicolumn{1}{c}{$f_{58}^{\star}$ ($\times 10^{-9}$)} & \multicolumn{1}{c}{$f_{63}^{\star}$ ($\times 10^{-9}$)} & \multicolumn{1}{c}{$f_{72}^{\star}$ ($\times 10^{-9}$)} & \multicolumn{1}{c}{$f_{80}^{\star}$ ($\times 10^{-9}$)} & \multicolumn{1}{c}{$f_{86}^{\star}$ ($\times 10^{-9}$)} & \multicolumn{1}{c}{$f_{99}^{\star}$ ($\times 10^{-9}$)} & \multicolumn{1}{c}{$f_{110}^{\star}$ ($\times 10^{-9}$)} \\ \hline
0003 & 0.045110 & 0.048056 & 0.055965 & 0.055715 & 0.047546 & 0.042668 & 0.039957 & 0.034183 & 0.026974 \\ 
0005 & 0.015360 & 0.014287 & 0.014343 & 0.013207 & 0.010510 & 0.008935 & 0.007667 & 0.006164 & 0.004537 \\ 
0015 & 0.949220 & 0.642224 & 0.456478 & 0.335422 & 0.225124 & 0.161805 & 0.127347 & 0.090751 & 0.061170 \\ 
0021 & 0.589519 & 0.485483 & 0.416094 & 0.351972 & 0.266133 & 0.212027 & 0.175507 & 0.135903 & 0.098852 \\ 
0025 & 0.086732 & 0.095887 & 0.110133 & 0.104018 & 0.087565 & 0.078870 & 0.073035 & 0.061796 & 0.052358 \\ 
0027 & 0.044648 & 0.038470 & 0.034585 & 0.029620 & 0.023448 & 0.019148 & 0.016498 & 0.013264 & 0.010151 \\ 
0033 & 0.047392 & 0.042263 & 0.038594 & 0.033799 & 0.027282 & 0.022794 & 0.019054 & 0.014683 & 0.011011 \\ 
0039 & 0.494316 & 0.326709 & 0.220809 & 0.159055 & 0.101730 & 0.069933 & 0.053288 & 0.035159 & 0.023059 \\ 
0045 & 0.026858 & 0.035779 & 0.051488 & 0.062204 & 0.079683 & 0.091648 & 0.090080 & 0.092416 & 0.082166 \\ 
0048 & 0.032928 & 0.047326 & 0.070281 & 0.085556 & 0.108490 & 0.123857 & 0.123218 & 0.124770 & 0.105232 \\ 
\hline
\end{tabular}
\begin{tabular}{cr@{$\;\pm\;$}rr@{$\;\pm\;$}rr@{$\;\pm\;$}rc} \hline
\multicolumn{1}{c}{(1)} & \multicolumn{2}{c}{(20)} & \multicolumn{2}{c}{(21)} & \multicolumn{2}{c}{(22)} & \multicolumn{1}{c}{(23)} \\ 
\multicolumn{1}{c}{HR} & \multicolumn{2}{c}{$T_{\rm eff}$ (K)} & \multicolumn{2}{c}{$\log g$} & \multicolumn{2}{c}{[M/H]} & \multicolumn{1}{c}{$c_{\rm sc}$ ($\times 10^{-17}$)} \\ \hline
0003 &   4732 &   21 &    3.088 &    0.320 & 0.000 &    0.072 & 1.8003 \\ 
0005 &   5762 &   50 &    4.809 &    0.291 &    0.200 &    0.095 & 0.1705 \\ 
0015 &  12827 &  101 &    4.484 &    0.116 &    0.500 &    0.000 & 0.5109 \\ 
0021 &   6917 &   81 &    3.229 &    0.099 &   $-$0.085 &    0.153 & 2.3213 \\ 
0025 &   4824 &   29 &    2.522 &    0.312 & 0.000 &    0.043 & 3.0634 \\ 
0027 &   6181 &   31 &    1.299 &    0.107 &   $-$0.801 &    0.158 & 0.3045 \\ 
0033 &   6217 &   76 &    3.912 &    0.211 &   $-$0.386 &    0.126 & 0.3484 \\ 
0039 &  21414 &  393 &    4.192 &    0.330 &   $-$2.000 &    0.000 & 0.1248 \\ 
0045 &   3748 &   16 &    1.202 &    0.070 &    0.429 &    0.080 & 11.7681 \\ 
0048 &   3750 &    3 &    0.770 &    0.117 &    0.500 &    0.010 & 15.6778 \\ 
\hline
\end{tabular}
\begin{tabular}{cccc@{$\;\pm$}cc@{$\;\pm$}cc@{$\;\pm\;$}cc@{$\;\pm\;$}cc@{$\;\pm\;$}c} \hline
\multicolumn{1}{c}{(1)} & \multicolumn{1}{c}{(24)} & \multicolumn{1}{c}{(25)} & \multicolumn{2}{c}{(26)} & \multicolumn{2}{c}{(27)} & \multicolumn{2}{c}{(28)} & \multicolumn{2}{c}{(29)} & \multicolumn{2}{c}{(30)} \\ 
\multicolumn{1}{c}{HR} & \multicolumn{1}{c}{Johnson $B_{\rm Simbad}$} & \multicolumn{1}{c}{Johnson $V_{\rm Simbad}$} & \multicolumn{2}{c}{Johnson $B_{\rm CK04}$} & \multicolumn{2}{c}{Johnson $V_{\rm CK04}$} & \multicolumn{2}{c}{standard $B$} & \multicolumn{2}{c}{standard $G$} & \multicolumn{2}{c}{standard $R$} \\ \hline
0003 & 5.65 & 4.61 & 5.649 & 0.011 & 4.583 & 0.010 & 5.128 & 0.010 & 4.680 & 0.010 & 4.357 & 0.008 \\ 
0005 &  --- &  --- & 6.691 & 0.013 & 5.985 & 0.009 & 6.307 & 0.012 & 6.044 & 0.009 & 5.862 & 0.009 \\ 
0015 & 1.95 & 2.06 & 1.914 & 0.010 & 2.021 & 0.011 & 1.834 & 0.010 & 1.983 & 0.011 & 2.144 & 0.012 \\ 
0021 & 2.61 & 2.27 & 2.580 & 0.015 & 2.248 & 0.010 & 2.347 & 0.014 & 2.268 & 0.011 & 2.234 & 0.008 \\ 
0025 & 4.89 & 3.87 & 4.916 & 0.011 & 3.863 & 0.008 & 4.399 & 0.010 & 3.958 & 0.009 & 3.653 & 0.008 \\ 
0027 & 5.44 & 5.04 & 5.386 & 0.012 & 4.978 & 0.009 & 5.133 & 0.013 & 5.009 & 0.010 & 4.932 & 0.008 \\ 
0033 & 5.38 & 4.89 & 5.358 & 0.012 & 4.865 & 0.008 & 5.075 & 0.012 & 4.906 & 0.009 & 4.797 & 0.007 \\ 
0039 & 2.61 & 2.84 & 2.568 & 0.010 & 2.770 & 0.013 & 2.529 & 0.011 & 2.719 & 0.013 & 2.917 & 0.013 \\ 
0045 & 6.38 & 4.80 & 6.342 & 0.015 & 4.714 & 0.010 & 5.668 & 0.014 & 4.879 & 0.010 & 4.326 & 0.014 \\ 
0048 & 6.12 & 4.46 & 6.101 & 0.015 & 4.383 & 0.009 & 5.389 & 0.009 & 4.554 & 0.009 & 3.981 & 0.012 \\ 
\hline
\end{tabular}
\end{table*}

\subsection{The fitting procedure}
\label{sec:fitting_procedure}

\begin{figure*}
\includegraphics[width=\columnwidth]{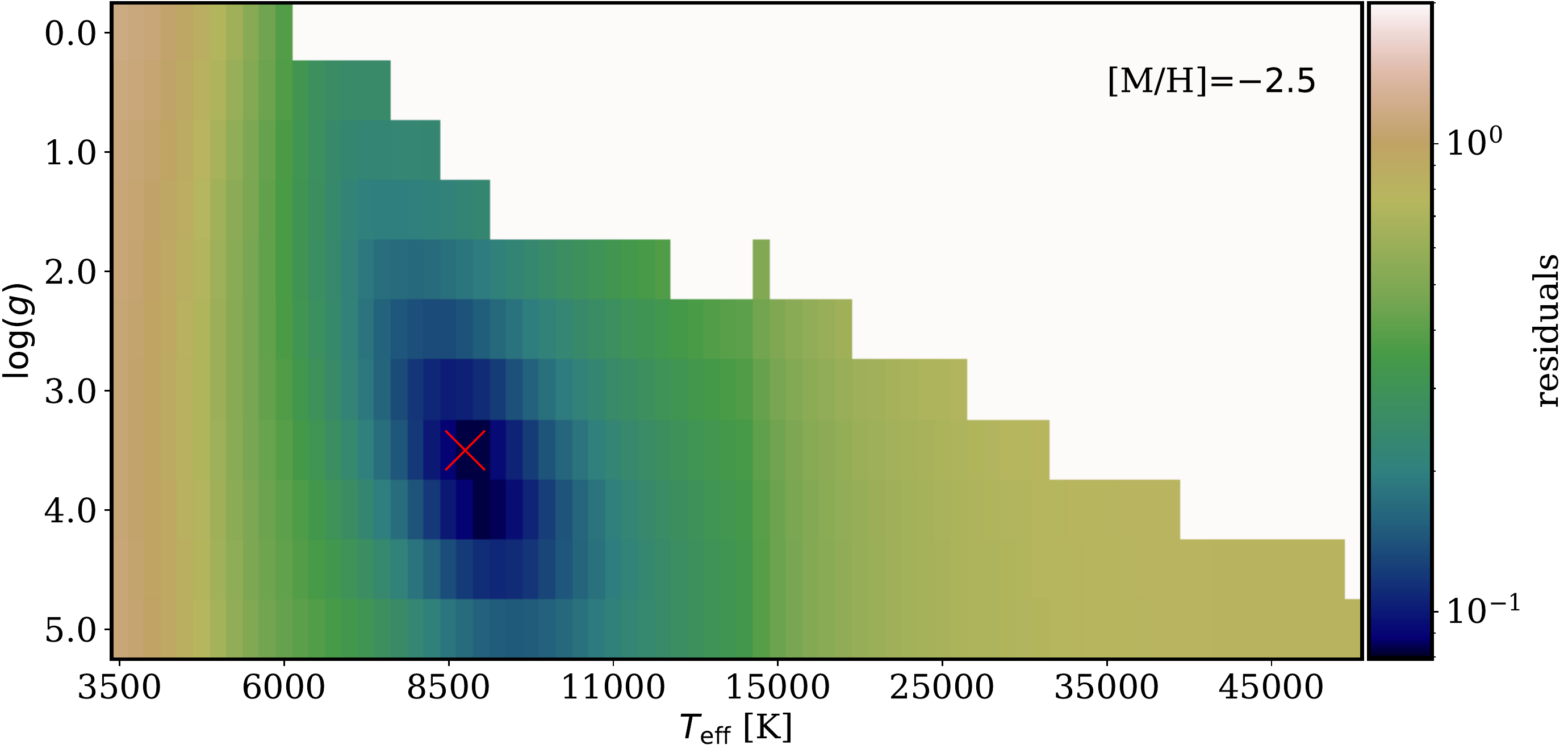}
\hfill
\includegraphics[width=\columnwidth]{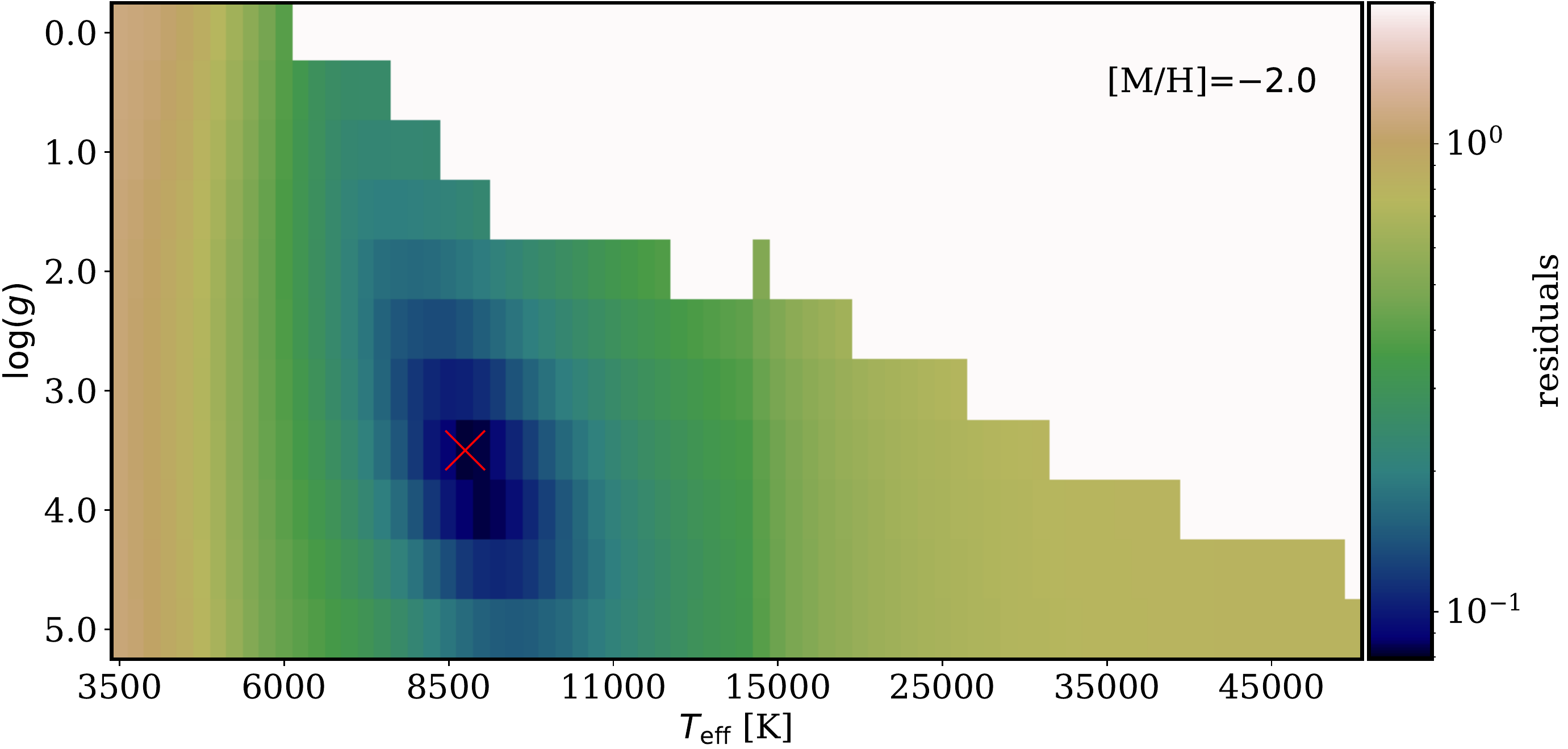}
\vskip 2mm
\includegraphics[width=\columnwidth]{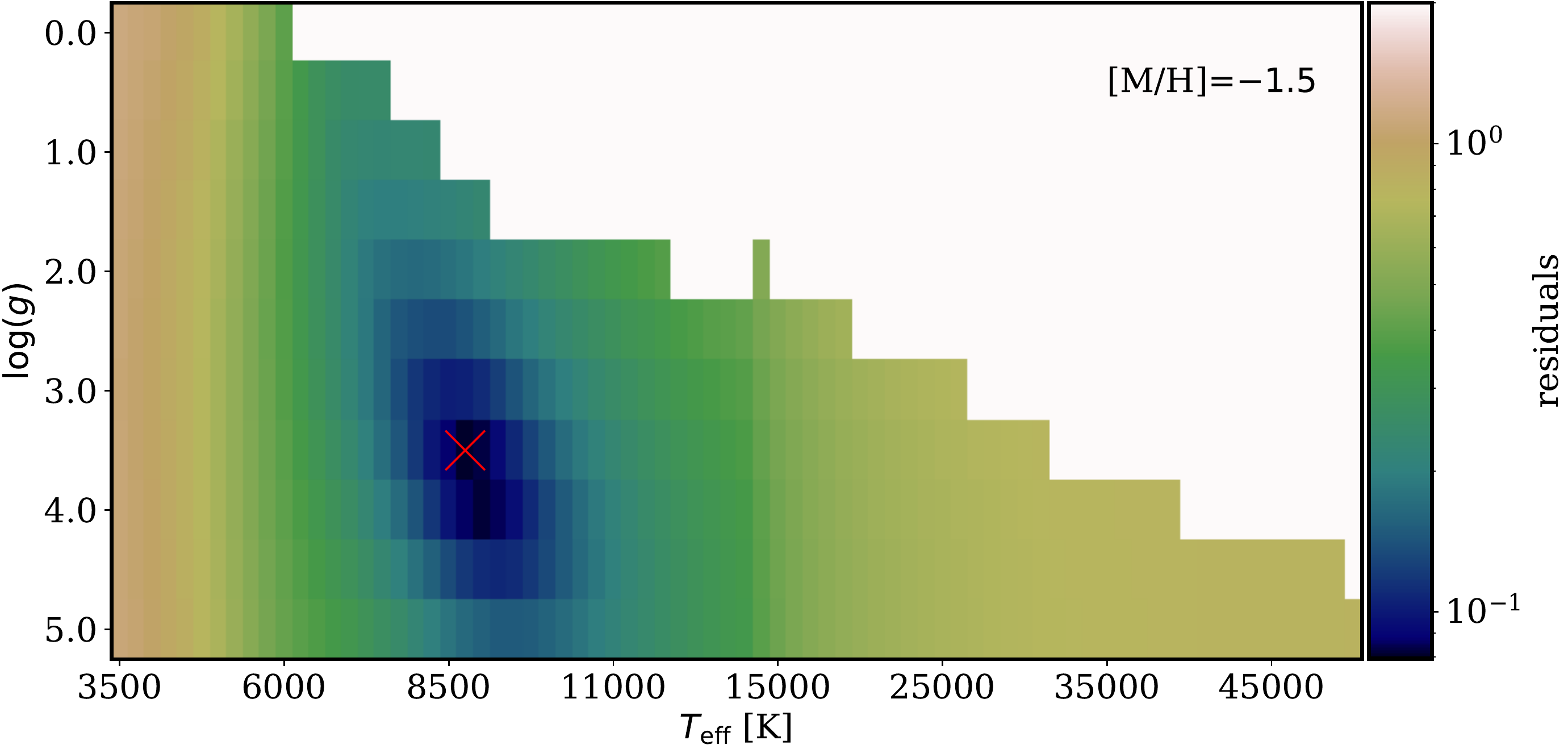}
\hfill
\includegraphics[width=\columnwidth]{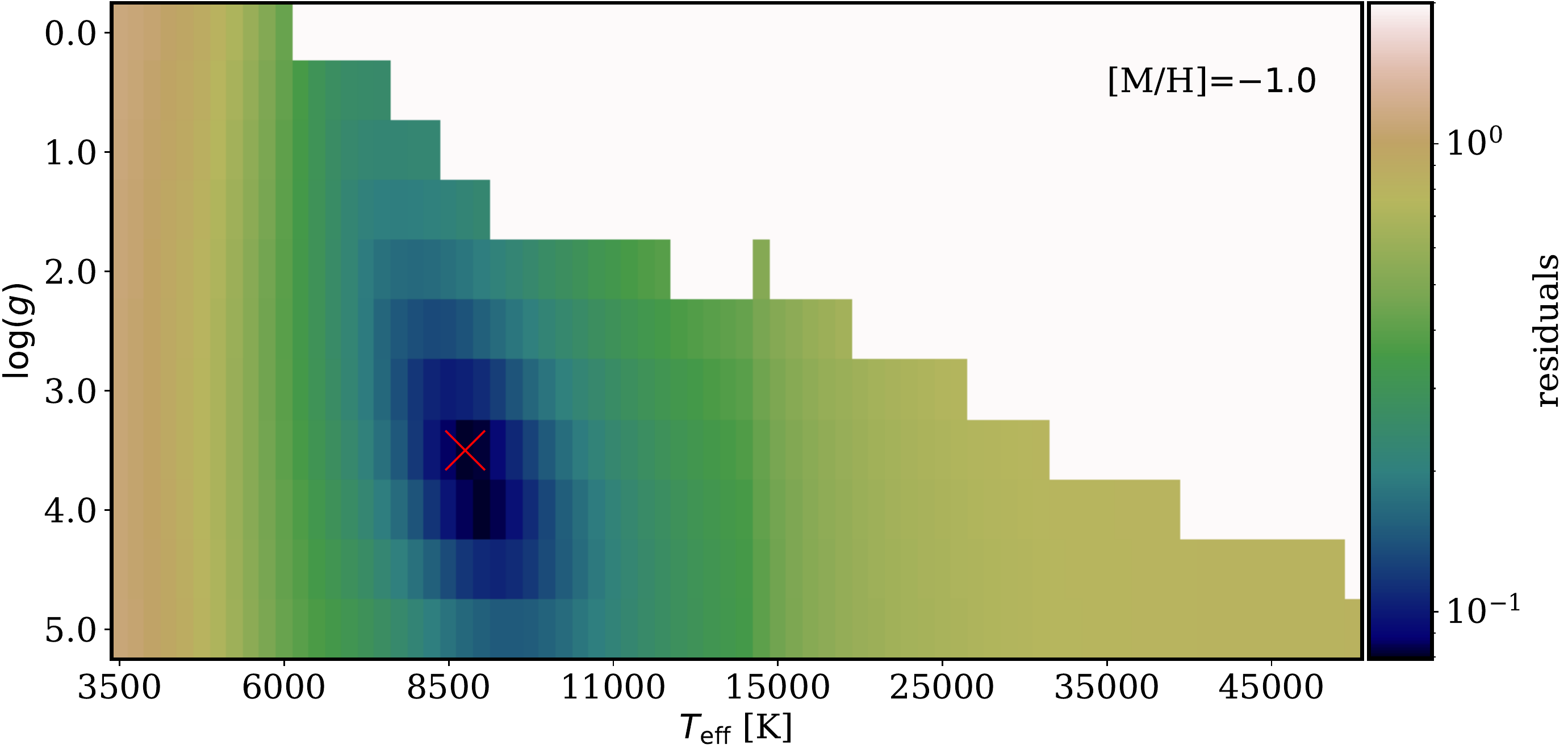}
\vskip 2mm
\includegraphics[width=\columnwidth]{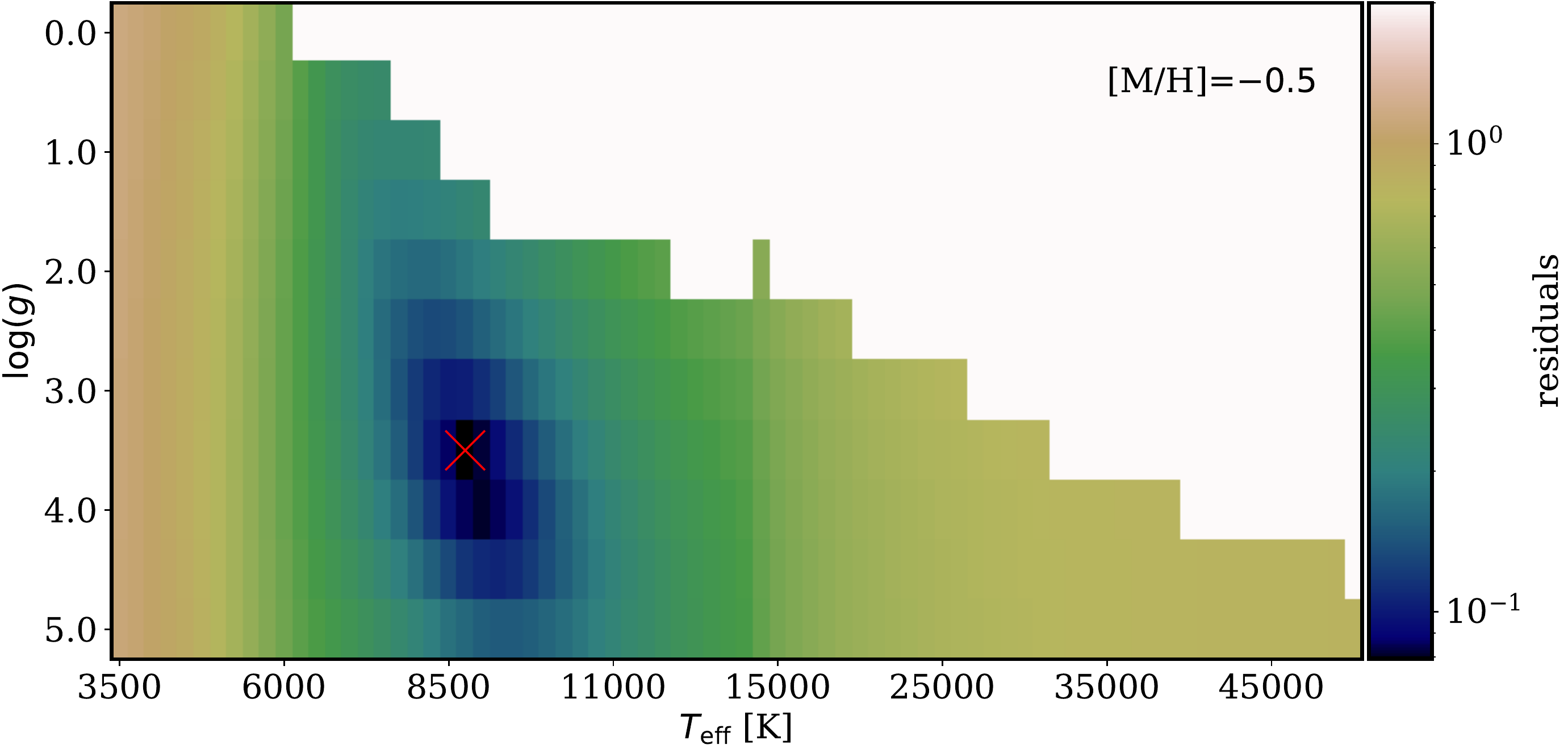}
\hfill
\includegraphics[width=\columnwidth]{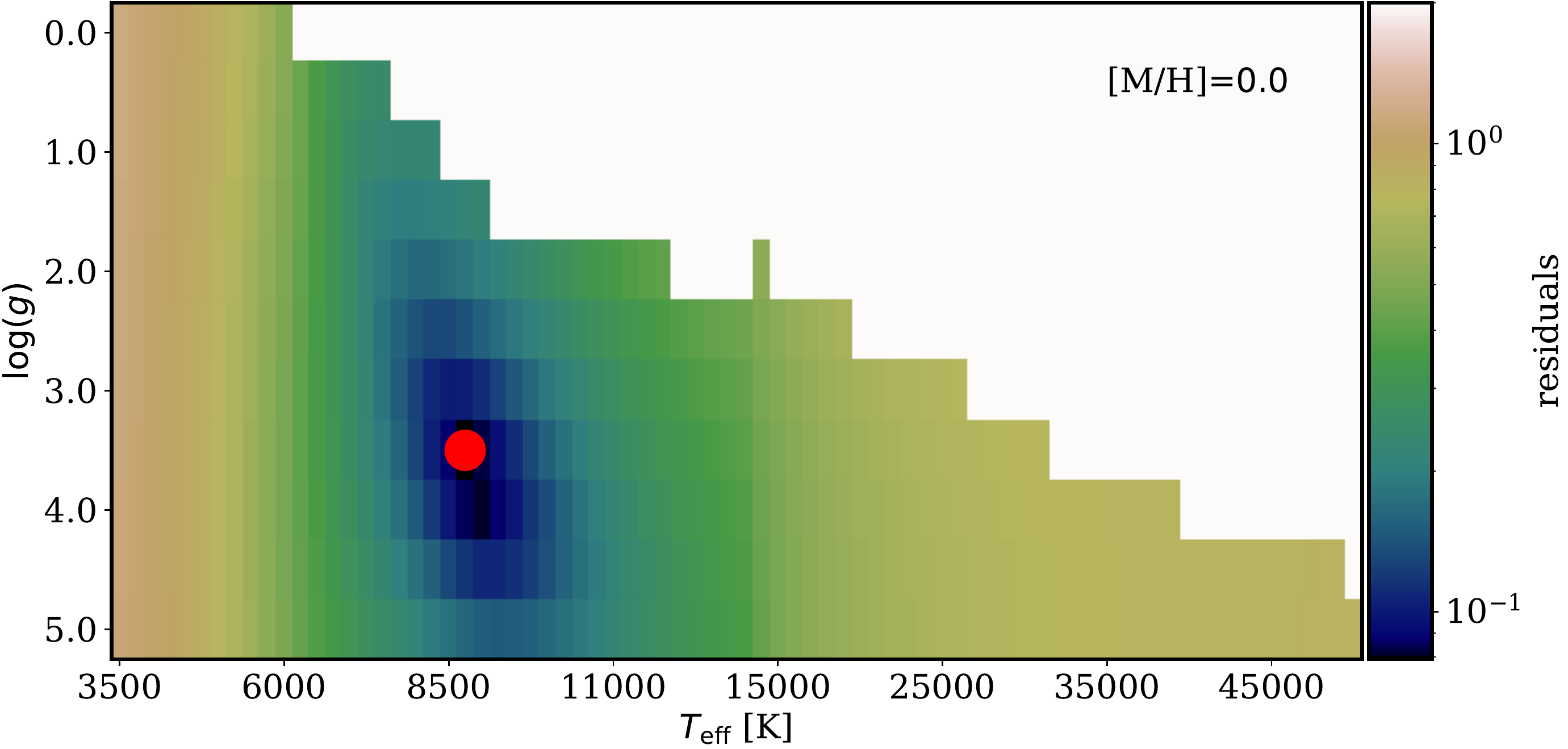}
\vskip 2mm
\includegraphics[width=\columnwidth]{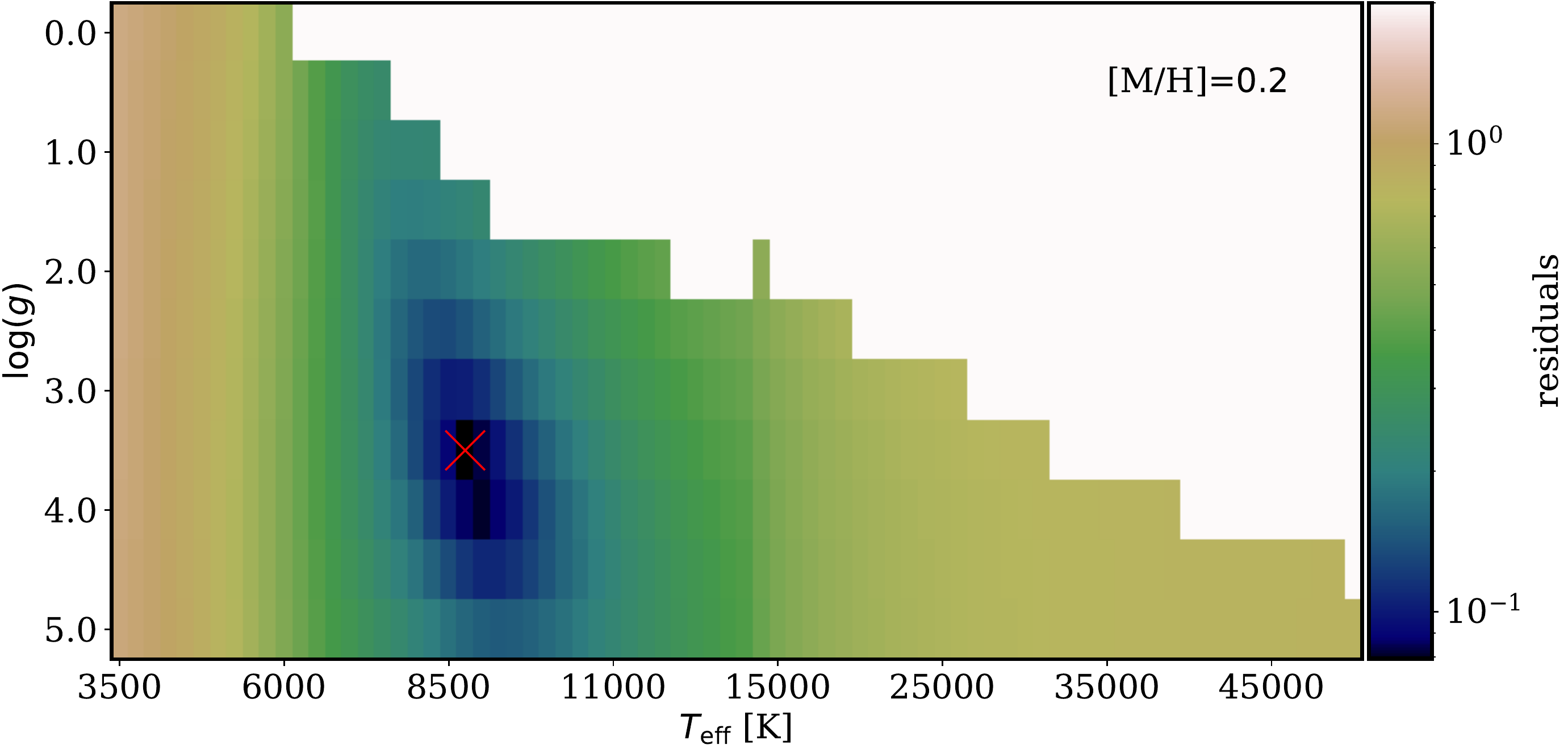}
\hfill
\includegraphics[width=\columnwidth]{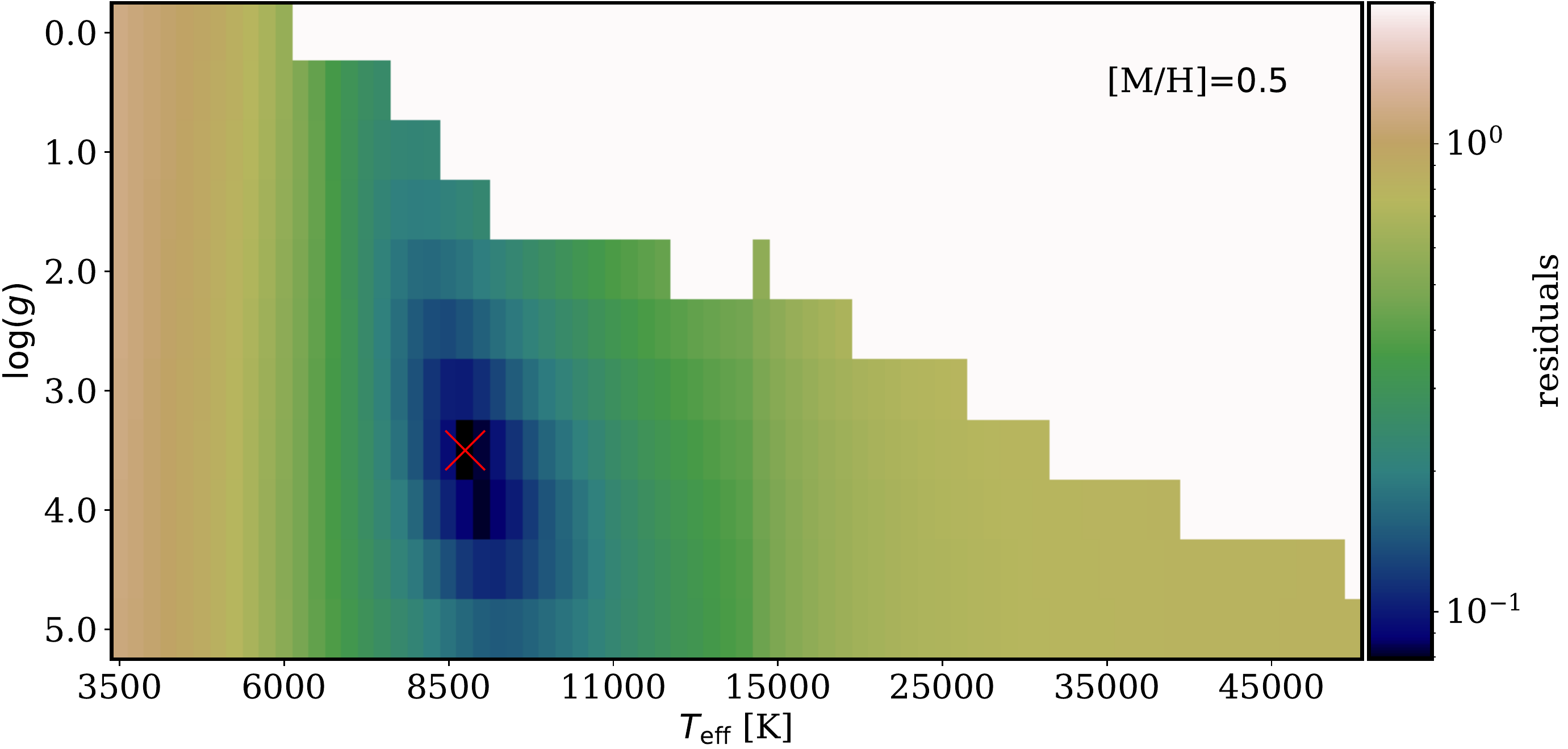}
\caption{Example of computation of initial atmospheric stellar parameters in
the fit of the C13 photometric data of the star HR7254 to the CK04 models.
Each plot shows the residuals obtained after determining the best scaling
factor when fitting the modeled spectra with different $T_{\rm eff}$ and $\log
g$ values at a fixed metallicity ([M/H], given in the upper right corner). The
residuals were computed as $\sqrt{\chi^2/N_{\rm bands}}$, with $\chi^2$ given
by Eq.~(\ref{eq:objective_function}).  The white areas indicate the stellar
parameter space not available in the models. The red cross in each panel marks
a local minimum, i.e., the $T_{\rm eff}$ and $\log g$ values that provide the
best fit at the considered [M/H]. A large red filled circle has been
overplotted in the panel where the global minimum is reached.}
\label{fig:fitting_step1}
\end{figure*}

\begin{figure*}
\includegraphics[height=0.50\columnwidth]{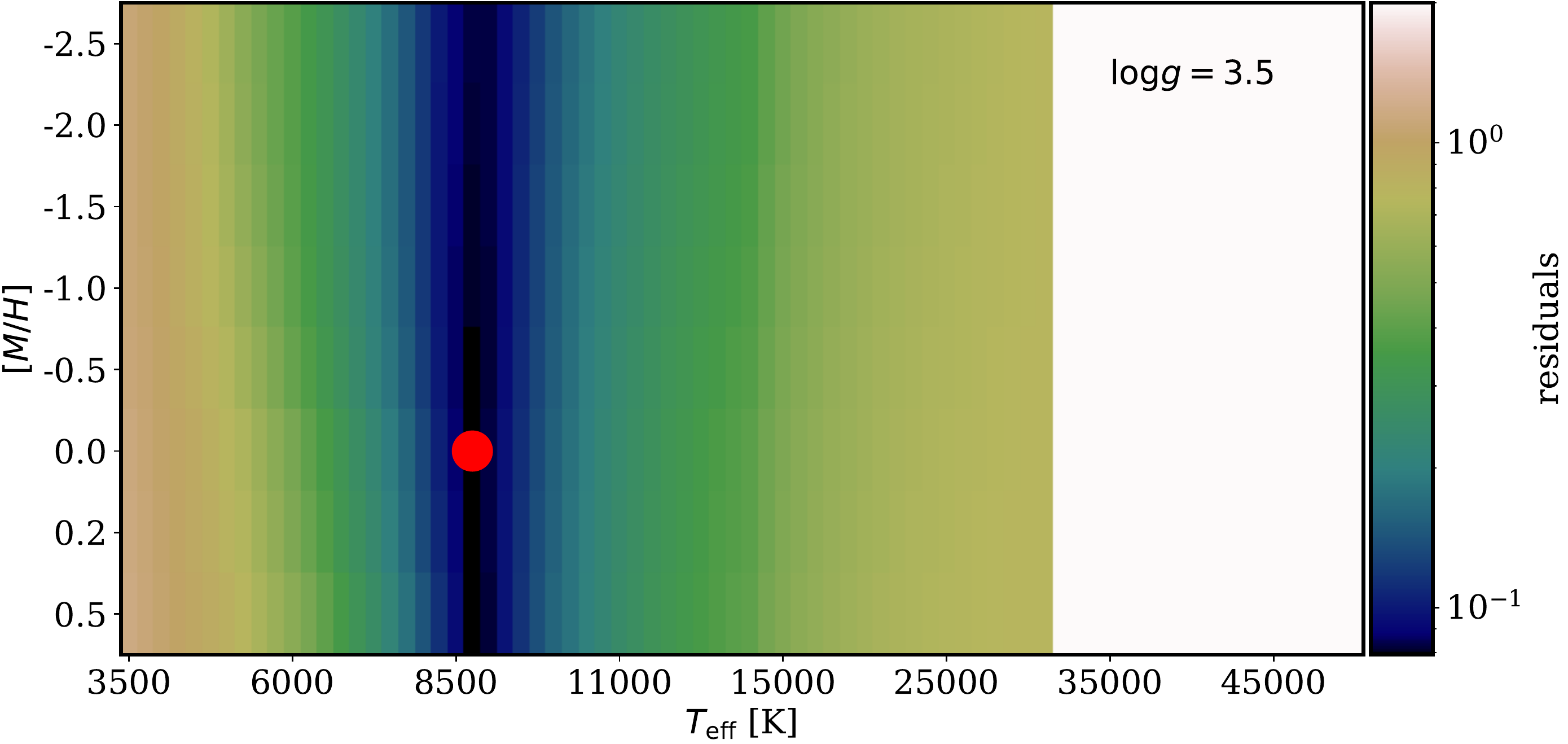}
\hspace{5mm}
\includegraphics[height=0.50\columnwidth]{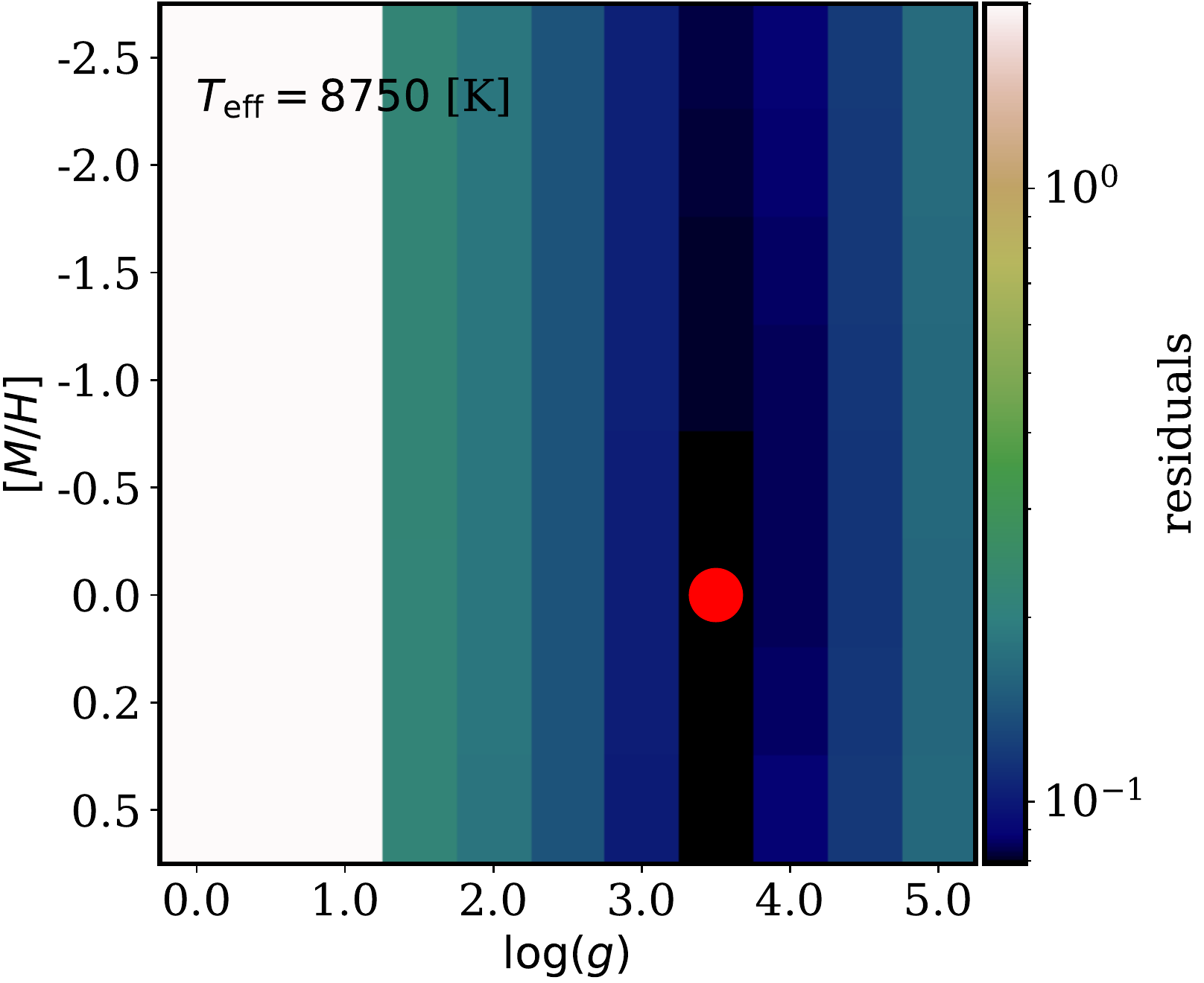}
\caption{Residuals of the same fit displayed in Fig.~\ref{fig:fitting_step1},
corresponding to the star HR7254. These two panels show two perpendicular views
in the 3D parameter space around the global minimum found at 
$T_{\rm eff}=8750$~K, ${\rm [M/H]}=0.0$ and $\log g=3.5$ in the first step of
the fitting procedure discussed in Sect.~\ref{sec:fitting_procedure}.}
\label{fig:fitting_step1_additional}
\end{figure*}

The first step before attempting to estimate the synthetic RGB photometry of
the bright star sample was the fit of the spectral energy distribution of each
star to the stellar model spectrum that best matched the available 13C photometric
data. For that purpose, we have chosen the Stellar Atmosphere Models by
\citet[hereafter CK04]{2003IAUS..210P.A20C}, as provided by the STScI web
page\footnote{\url{https://www.stsci.edu/hst/instrumentation/reference-data-for-calibration-and-tools/astronomical-catalogs/castelli-and-kurucz-atlas}}.
An important advantage of these models is that they can be easily interpolated
using the Python package
stsynphot \citep[][available
online\footnote{\url{https://stsynphot.readthedocs.io/en/latest/index.html}}]{2018stsynphot}
for any arbitrary $T_{\rm eff}$, [M/H] and $\log g$ selection (within the
parameter space covered by the models).
Since the bright star sample is constituted by nearby objects,
it is safe to assume that most of them will not be strongly affected by
interstellar reddening. For that reason, we have not included any correction for
this effect in the subsequent fits. In any case, here we are aiming at
providing accurate photometry of the observed (i.e., uncorrected) RGB fluxes,
and the fact of excluding this type of correction will only translate in a
systematic deviation of the derived stellar parameters.

It is important to highlight that it is not the aim of this work to determine
accurate stellar atmospheric parameters for each star, but to obtain a good fit
to the available 13C photometric data in order to derive reliable spectral
energy distributions that facilitate the proper computation of synthetic RGB
magnitudes. The comparison with observed spectra, shown in
section~\ref{subsec:kiehling_comparison},
indicates that this is actually the case.

The initial 13C stellar magnitudes were converted into absolute flux densities
\mbox{(erg~s$^{-1}$~cm$^{-2}$~\AA$^{-1}$)} using the conversion given by JM75
(the Fortran~IV code provided in their Table~5 was transformed to Python to
facilitate this task).  The selection of the best CK04 model for each star was
accomplished in two steps:
\begin{enumerate}
\item[\bf Step 1:] Initial determination of the atmospheric stellar parameters
using all the available CK04 models at their pre-computed sampling grid in the
parameter space. In particular, the CK04 atlas provides models for abundances
[M/H]=$-2.5$, $-2.0$, $-1.5$, $-1.0$, $-0.5$, $0.0$, $+0.2$ and $+0.5$, with
effective temperatures ranging from 3500 to 50000~K, and $\log g$ (surface
stellar gravity, with \,$g$\, in cm~s$^{-2}$) from 0.0 to 
5.0~dex. A simple chi-square minimization process
was performed, by adjusting the arbitrary constant $c_{\rm sc}$ necessary to
scale each model prediction to the absolute C13 flux densities. For this
purpose, the stellar fluxes for the star $f^{\star}_i$ and the model $f^{\rm
CK04}_i$ in each of the \mbox{$N_{\rm bands}$} photometric bandpasses 
were normalized by their mean values
\begin{align}
f^{\star}_{{\rm nor},i} &= 
\frac{f^{\star}_i}{\langle f^{\star} \rangle}, 
& \;\; {\rm with} & \;\; &
\langle f^{\star} \rangle &= \frac{1}{N_{\rm bands}}
\sum_{i=1}^{N_{\rm bands}} f^{\star}_i, \\
f^{\rm CK04}_{{\rm nor},i} &= 
\frac{f^{\rm CK04}_i}{\langle f^{\rm CK04} \rangle}, 
& \;\; {\rm with} & \; &
\langle f^{\rm CK04} \rangle &= \frac{1}{N_{\rm bands}}
\sum_{i=1}^{N_{\rm bands}} f^{\rm CK04}_i.
\end{align}
The initial stellar fluxes \,$f_i^{\star}$\, (with \,$i=33,\, 35,\ldots, 110$
indicating the 13C photometric band), computed from the published photometric
data, are listed in Table~\ref{tab:bigtable},
columns (5)--(17).

An objective function to be minimized was defined at this point as
\begin{align}
\label{eq:objective_function}
\chi^2(c_{\rm nor}) = \sum_{i=1}^{N_{\rm bands}} \left(
f^{\star}_{{\rm nor},i}-c_{\rm nor} f^{\rm CK04}_{{\rm nor},i}
\right)^2,
\end{align}
where $c_{\rm nor}$ is an intermediate scaling factor between the normalized
stellar and model fluxes. By equating to zero the first derivative of the last
equation, it is immediate to obtain
\begin{align}
c_{\rm nor} = \frac{\sum_{i=1}^{N_{\rm bands}} f^{\star}_{{\rm nor},i}\,
f^{\rm CK04}_{{\rm nor},i}}%
{\sum_{i=1}^{N_{\rm bands}}\left(f^{\rm CK04}_{{\rm nor},i}\right)^2}.
\end{align}

Finally, the sought scaling factor $c_{\rm sc}$
is computed as
\begin{align}
c_{\rm sc} = c_{\rm nor} \frac{\langle f^{\star} \rangle}%
{\langle f^{\rm CK04} \rangle},
\end{align}
which allows to convert the CK04 model fluxes into absolute flux densities as 
\begin{align}
\label{eq:flux_normalization}
f^{\star} = c_{\rm sc} f^{\rm CK04}.
\end{align}

At this stage, the whole parameter space was sampled using all the initially
available combinations of the three stellar parameters. The global minimum of
the objective function in this comprehensive search led to an initial guess for
$T_{\rm eff}$, [M/H], and $\log g$, which were refined in the subsequent step.
A graphical example of this initial computation is shown in
Figs.~\ref{fig:fitting_step1} and~\ref{fig:fitting_step1_additional}. The
comparison of these plots indicates that, not surprisingly, $T_{\rm eff}$ is
the main parameter governing the variation of the $\chi^2$ value, followed by
$\log g$. These two parameters are correlated, as shown by the oblique
orientation of the minimum valley in Fig.~\ref{fig:fitting_step1}. In addition,
the role of [M/H] is quite small, as revealed by the almost undetectable
variation of the residuals with this parameter in
Fig.~\ref{fig:fitting_step1_additional}. This result gives support to the idea
that the derived stellar parameters should not be considered as extremely
accurate.

\item[\bf Step 2:] Refinement of the atmospheric stellar parameters: the
initial $T_{\rm eff}$, [M/H], and $\log g$ values were employed as the starting
point in the parameter space to compute a more refined solution. For that
purpose, we used a numerical minimization process based on the Nelder-Mead
method \citep{1965NelderMead} with the help of the Python package
lmfit \citep[][available
online\footnote{\url{https://lmfit.github.io/lmfit-py/}}]{2014lmfit}. During this process,
the CK04 model predictions were interpolated at arbitrary locations within the
valid stellar atmospheric parameter space, as required by the objective
function to be minimized. The use of a good starting point, obtained in the
previous step, facilitated the convergence of this process.  The final $T_{\rm
eff}$, [M/H] and $\log g$ parameters for each star are given in
Table~\ref{tab:bigtable}, columns (18)--(20).

\subsection{Uncertainties in the fitting procedure}
\label{sec:uncertainties_fitting}

\begin{figure}
\includegraphics[width=\columnwidth]{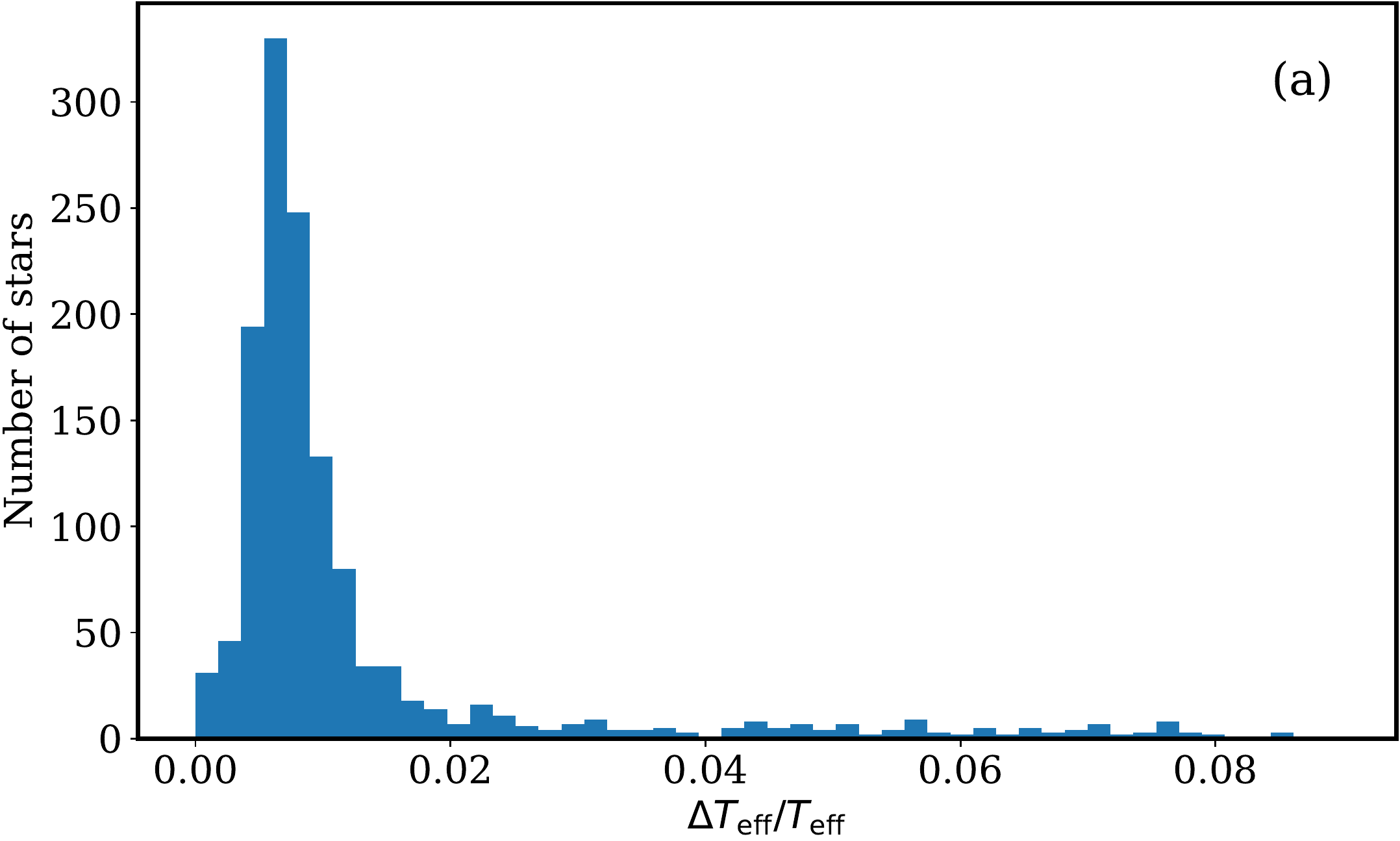}
\vskip 2mm
\includegraphics[width=\columnwidth]{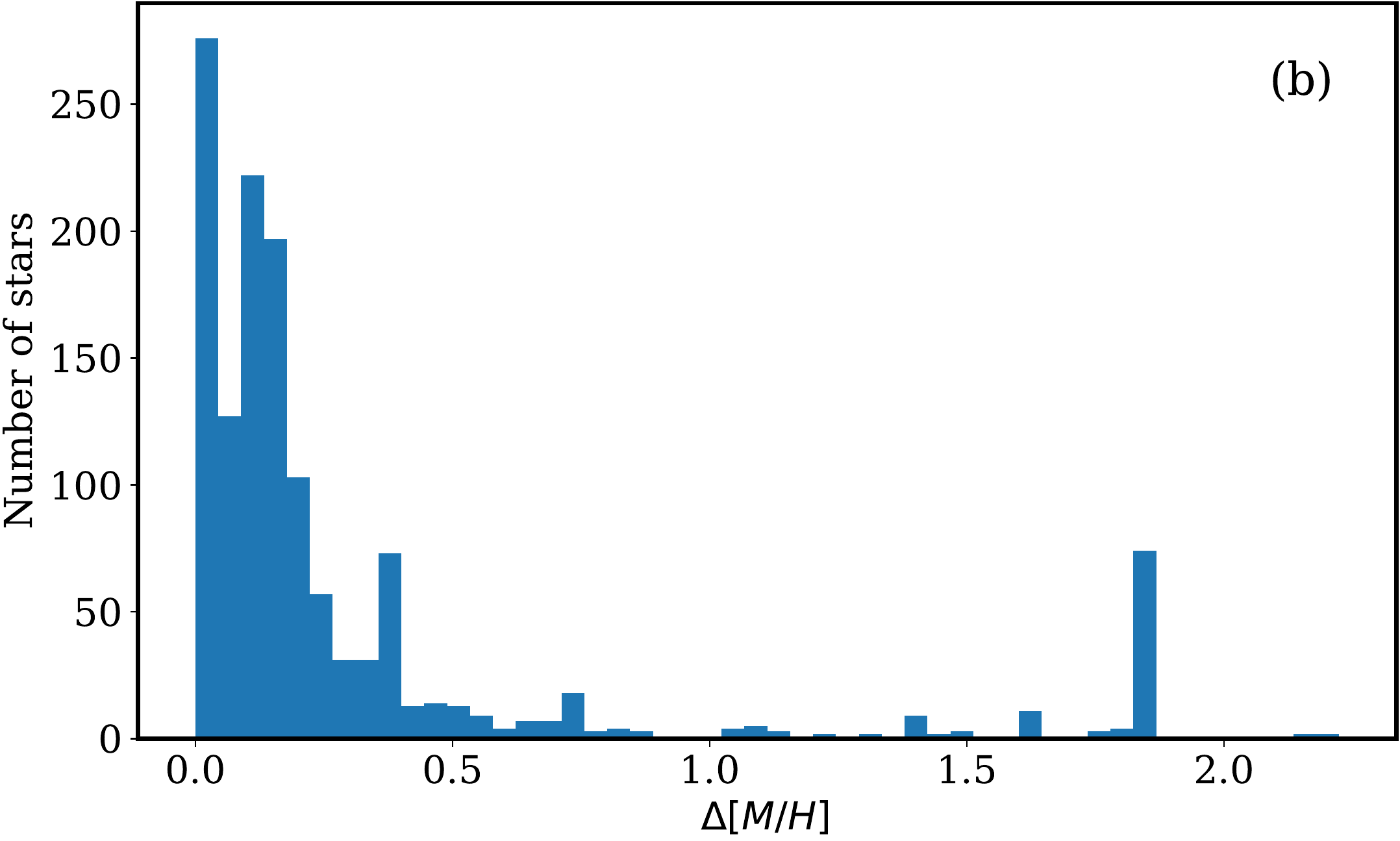}
\vskip 2mm
\includegraphics[width=\columnwidth]{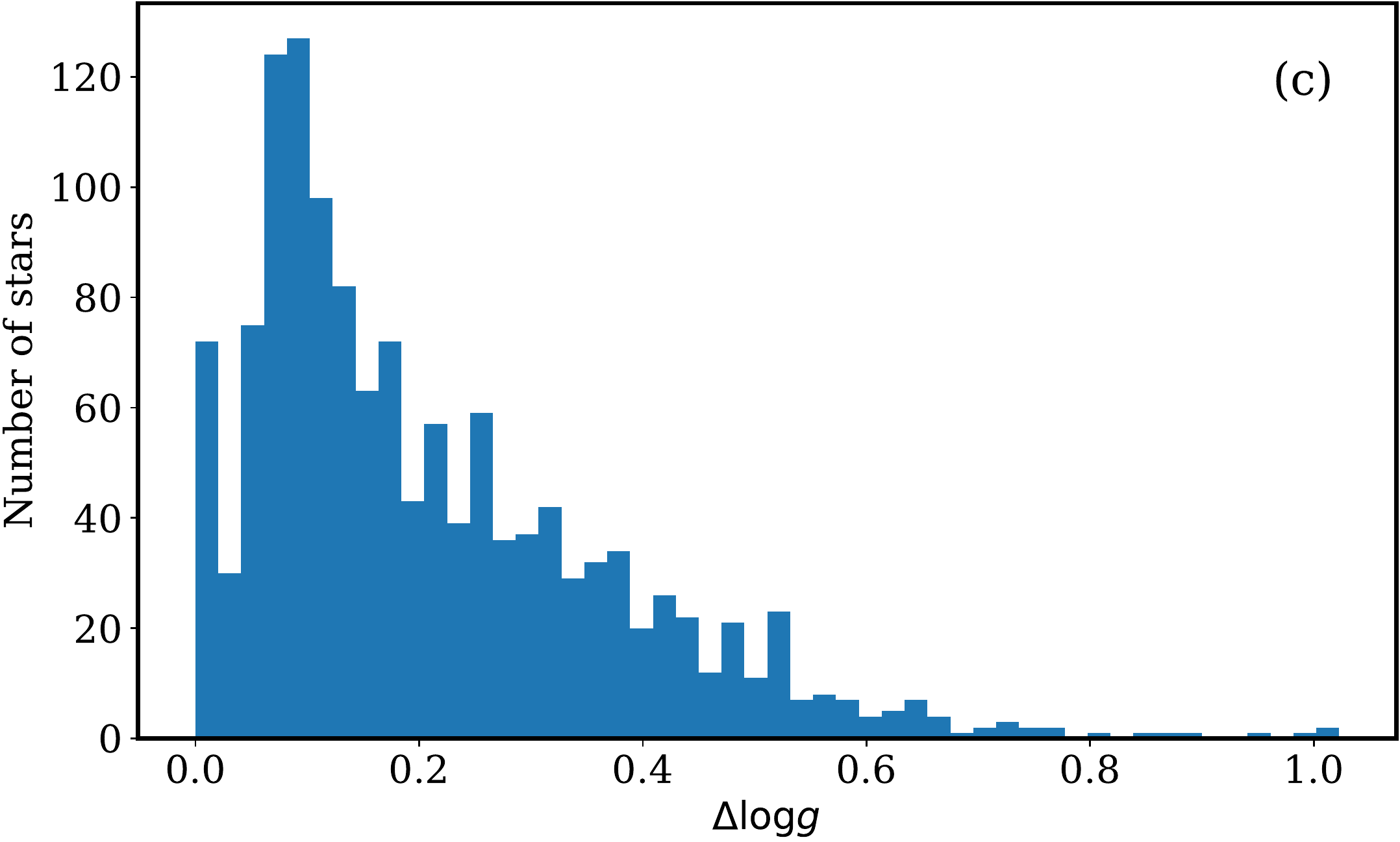}
\caption{Distribution of random uncertainties in the derived stellar
atmospheric parameters in the final sample of fitted CK04 models, estimated from the bootstrapping method described in
Sect.~\ref{sec:uncertainties_fitting}. Panel~(a): relative errors in effective
temperature. Panels~(b) and~(c): absolute errors in metallicity and surface
gravity. It is important to highlight that these uncertainties
are simply lower limits to the expected random errors in the stellar
atmospheric parameters, as explained in the text.}
\label{fig:simul_stderr_starpar}
\end{figure}

Two main sources of uncertainties have been considered: i) systematic errors
due to the inability of the adopted CK04 models to properly
reproduce the spectral energy distribution of the stars, and ii) random errors
in the fitted C13 photometric data. 

Since there is not an easy way to fix model fits exhibiting C13 residuals with
a systematic variation as a function of wavelength,
we decided to get rid of those stars with unreliable fits,
following the criteria described in Sect.~\ref{sec:cleaning_the_sample}. 

On the other hand, the impact of random errors in the fitting process has been
determined by generating bootstrapped absolute flux densities, using for that
purpose the already mentioned probable error of~0.02~mag in the C13 photometric
data, and repeating the whole fitting procedure in each simulated set of C13
photometric data for every star. This allowed obtaining an
initial estimate of the
uncertainties in the fitted stellar parameters, which are given in the
corresponding columns (18)--(20) of Table~\ref{tab:bigtable}, and graphically
displayed in Fig.~\ref{fig:simul_stderr_starpar}. 
It is important to stress that these random uncertainties must
not be understood as the actual uncertainties associated to the derived
atmospheric stellar parameters. As previously mentioned, no attempt to
introduce an interstellar reddening correction has been performed. Furthermore,
we are constrained to the capability of CK04~models to provide an accurate
modelling of actual stellar spectra, without considering, for example, the
impact of the assumed line opacities and chemical abundances, to mention some
of the additional relevant parameters that should be considered to obtain
reliable physical descriptions of the stellar atmospheres. Once the CK04~set of
stellar atmosphere models has been adopted, the quoted random errors are just
a lower limit to the actual $T_{\rm eff}$, [M/H] and $\log g$ uncertainties for
each star. Nevertheless, the quoted values can be compared with the sampling of
the atmospheric parameters at which the CK04~models are provided. In
particular, the effective temperature is sampled with $\Delta T_{\rm
eff}/T_{\rm eff}$ ranging from 0.02 (hot~stars) to 0.07 (cool stars), whereas
gravity and metallicity are sampled at $\Delta(\log g)=0.5$, and $\Delta{\rm
[M/H]}$ ranging from~0.2 to~0.5, respectively. Comparing these numbers with the
distributions displayed in Fig.~\ref{fig:simul_stderr_starpar}, it is clear
that in most cases the derived uncertainties are within the considered sampling
steps, which indicates that, at least from the point of view of the fitting
procedure, each one of the CK04~models is different enough from its neighbouring
models within the 3D parameter space even after bootstrapping the C13
photometric data. This reinforces the usefulness of Step~2 (see
Sect.~\ref{sec:fitting_procedure}) devoted to the refinement of the atmospheric
stellar parameters.
In addition, the bootstrapping method provided a collection of bootstrapped
fitted spectra associated to every single star, that were employed later to
estimate random uncertainties in the synthetic photometry performed on the CK04
model fits.

It is important to highlight that although the numerical minimization procedure
adopted for this work is robust, the computation time spent on each individual
fit is not negligible. In particular, the adopted Nelder-Mead method required
the evaluation of the objective function in many points (typically a few
hundreds) within the 3D parameter space defined by $T_{\rm eff}$, $\log g$ and
[M/H], which in turn translated into the corresponding number of interpolations
of the CK04 models. The median computation time for every fit amounted to a few
minutes. For that reason, we decided to generate a number of bootstrapped
spectra not excessively large, in order to apply this technique to the whole
final sample and not only to a representative subsample of stars. Finally, this
total number of bootstrapped spectra for each individual star was set to 30,
which leads to uncertainties in the uncertainties\footnote{Here we are using
the approximation provided by normally distributed data, for which the
fractional uncertainty on the standard deviation can be approximated by
$(2n-2)^{-1/2}$ \citep[see e.g.][Section 3.7]{squires2001}, with $n$ the number
of samples.} of $\sim 13$\%.


\end{enumerate}

\begin{figure}
\includegraphics[width=\columnwidth]{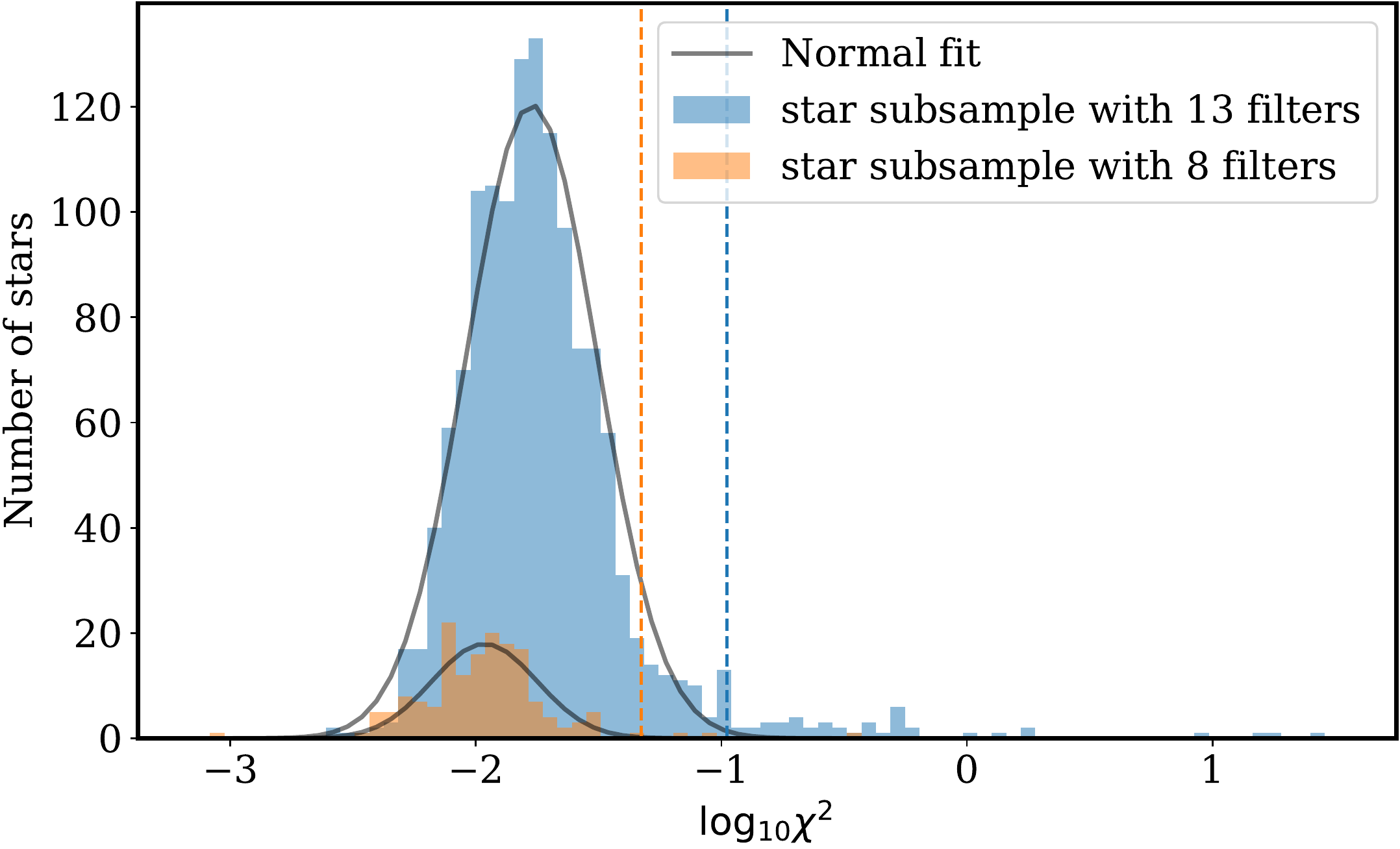}
\caption{Histogram of minimum values of the objective function obtained during
the fitting procedure of CK04 models. The initial sample (1522 stars) were
segregated according to the number of available photometric bandpasses (13
or~8). The black lines indicate the robust fit to
Normal distributions. The vertical dashed lines mark the 3$\sigma$ location
above the median value, which corresponds to $\chi^2_{3\sigma}\simeq 0.105$ for
the~1359 stars observed with~13 filters, and $\chi^2_{3\sigma}\simeq 0.047$ for
the~163 stars observed with~8 filters.}
\label{fig:hist_chisqr}
\end{figure}

\subsection{Cleaning the sample}
\label{sec:cleaning_the_sample}

\begin{figure}
\includegraphics[width=\columnwidth]{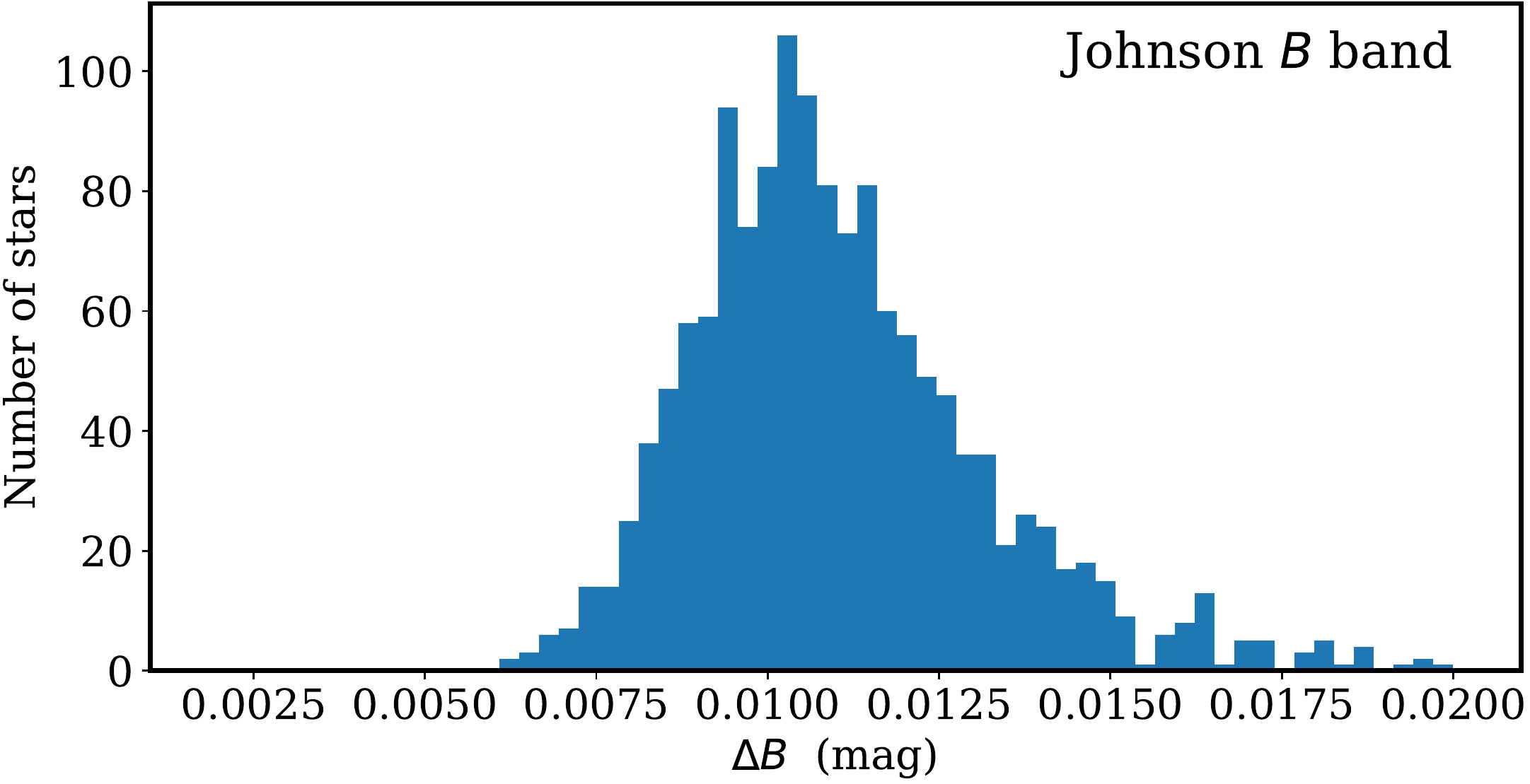}
\vskip 2mm
\includegraphics[width=\columnwidth]{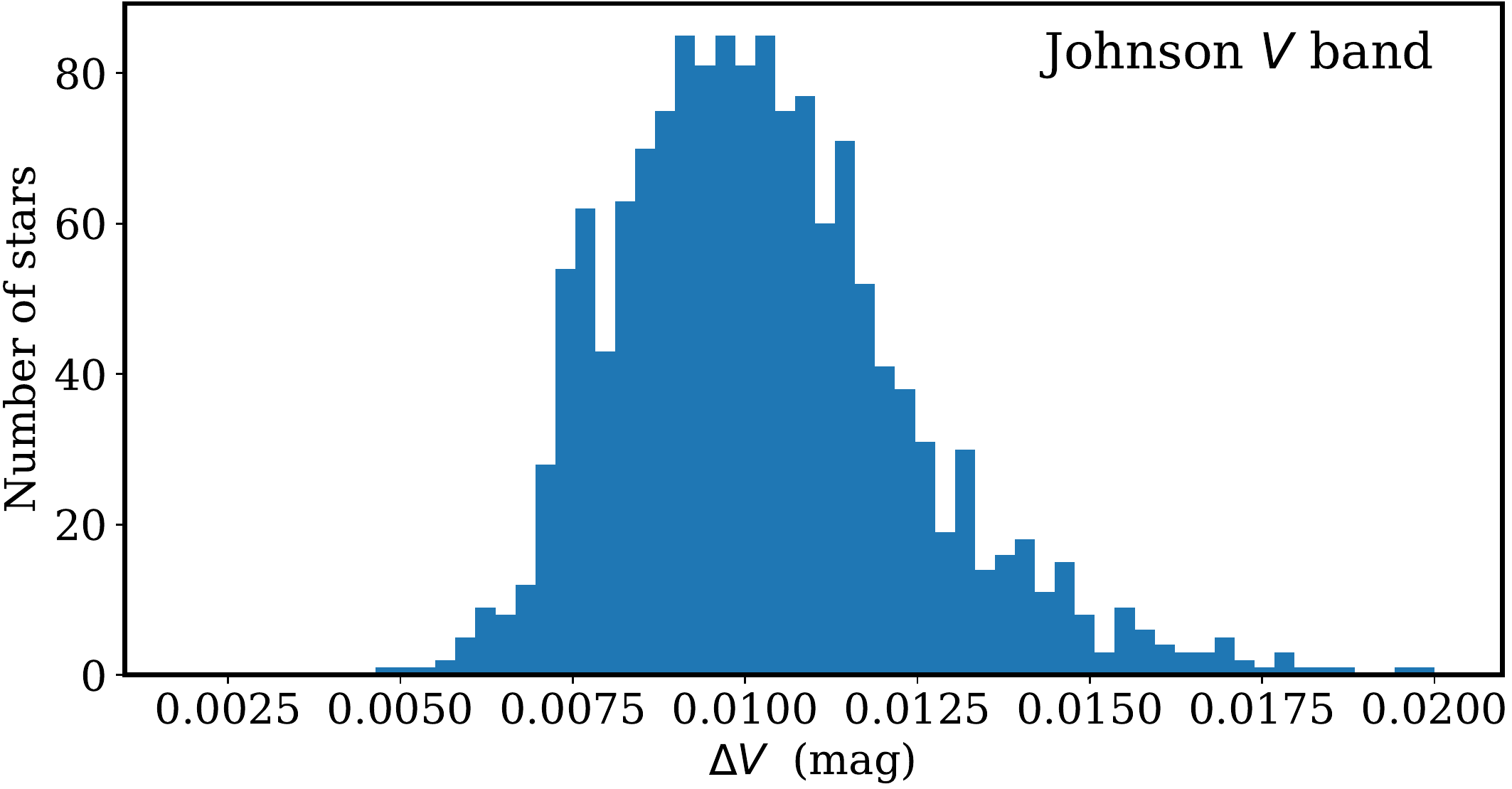}
\caption{Distribution of random uncertainties in the synthetic Johnson~$B$ 
and~$V$ measurements performed on the final sample of fitted CK04 models, 
estimated from the
bootstrapping method described in Sec.~\ref{sec:uncertainties_fitting}. The
median values are \mbox{$\Delta B=0.011$~mag} (top panel) and \mbox{$\Delta
V=0.010$~mag} (bottom panel).}
\label{fig:sigma_BV_bootstrapping}
\end{figure}

\begin{figure}
\includegraphics[width=\columnwidth]{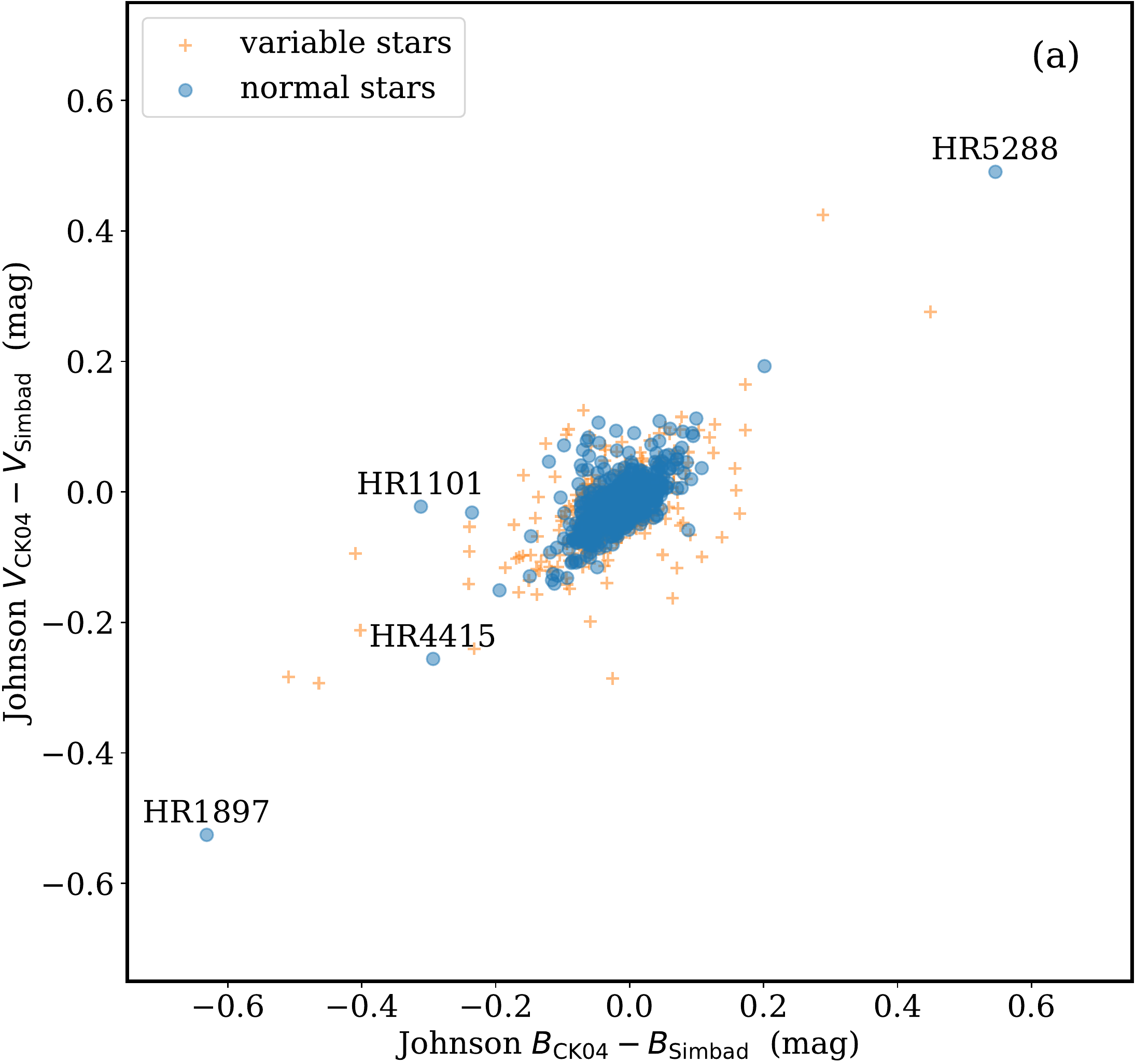}
\vskip 2mm
\includegraphics[width=\columnwidth]{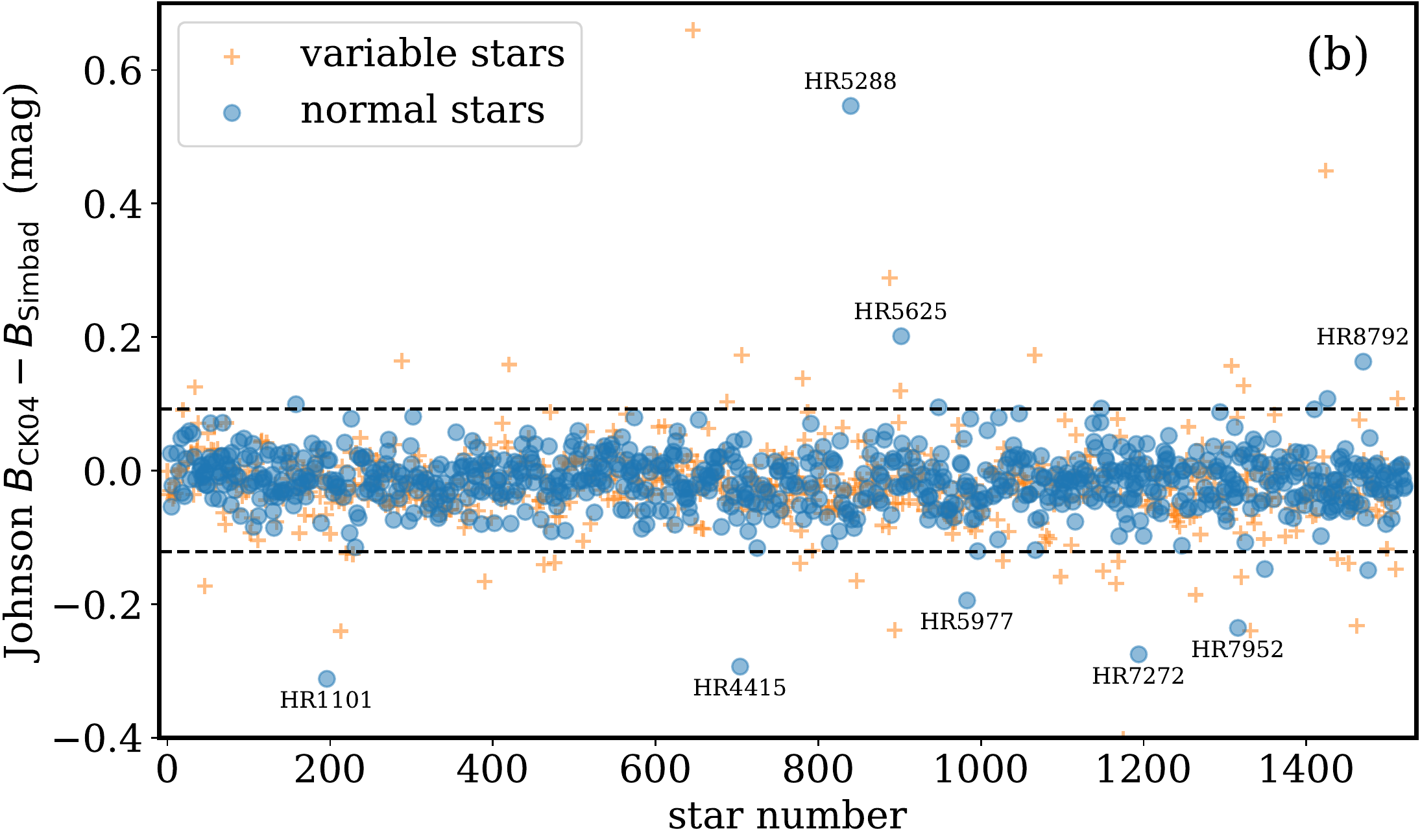}
\vskip 2mm
\includegraphics[width=\columnwidth]{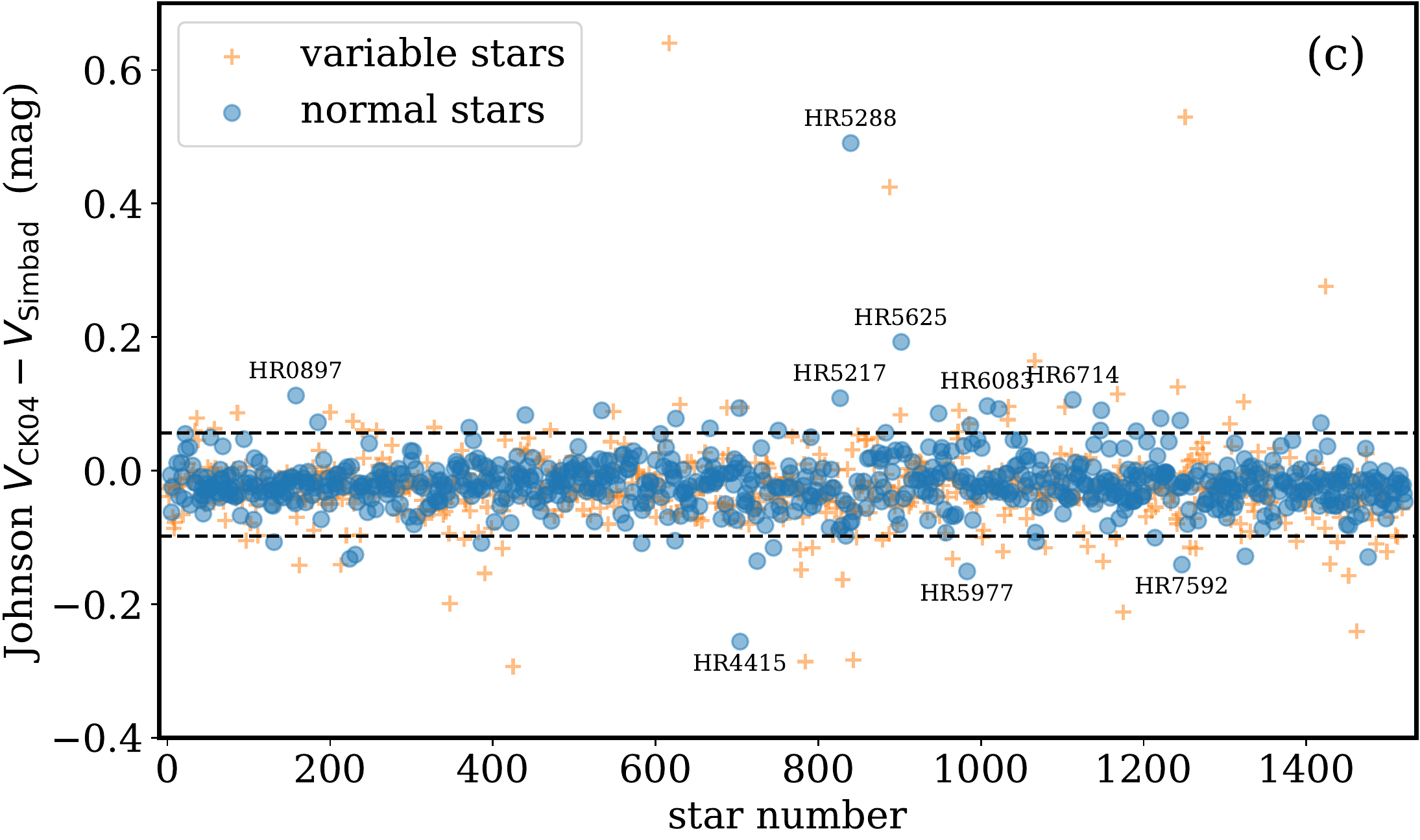}
\caption{Differences between the synthetic Johnson~$B$ and~$V$ magnitudes
computed using the CK04 model fits and the tabulated~$B$ and~$V$ data available
in Simbad. Known variable stars (according to Simbad) are plotted with
orange crosses, whereas normal stars (i.e., non-variable) are
represented with blue filled circles. Panel~(a) shows that, not
surprisingly, the errors
in~$B$ and~$V$ are correlated. The individual analysis of the differences
within each photometric band are shown in panels~(b) and~(c). The dashed lines
in the last two panels correspond to the robust $\pm 3\sigma$ region around the
median. A few stars, exhibiting large deviations, are labelled
in each panel.}
\label{fig:bv_simbad_comparison}
\end{figure}

\begin{table*}
\centering
\caption{Double stars from the JM75 sample for which the individual fluxes of
the two stars have been coadded to derive the resulting integrated~$B$ and~$V$
magnitudes (last two columns). Star\#1 is the initial star identification
provided in JM75. Star\#2 is the companion star (in some cases the companion
star is brighter than Star\#1). For some stars the magnitude in one of the two
filters is missing. The quoted~$B$ and~$V$ magnitudes for the individual stars,
as well as the spectral type,
correspond to the data retrieved from the Simbad database.
Stars marked with an asterisk in the first column were finally
removed from the final sample due to large discrepancies between the Simbad 
data and the predicted $B$ and $V$ magnitudes, as explained in 
Sect.~\ref{sec:cleaning_the_sample}. The fourth column in
Table~\ref{tab:bigtable} provides the name of the companion star for
the double star systems incorporated in the final stellar sample.}
\label{tab:double_stars}
\begin{tabular}{l@{\;\;}r@{\;\;}rlcl@{\;}r@{\;\;}rlcr@{\;\;}r}
\hline
\multicolumn{5}{c}{Star\#1} & \multicolumn{5}{c}{Star\#2} & 
\multicolumn{2}{c}{Star\#1+\#2} \\
Name & $B_1$ & $V_1$ & SpT$_1$ & \mbox{\phantom{X}} &
Name & $B_2$ & $V_2$ & SpT$_2$ & \mbox{\phantom{X}} & $B_{1+2}$ & $V_{1+2}$ \\
\hline
HR0545 & 4.558 & 4.589 & A0V      & &  HR0546 & 4.490 & 4.520 & A2IV        & &  3.771 & 3.801 \\
\makebox[0mm][r]{$^{\rm *}$}%
HR0897 & 3.330 & 3.180 & A3IV--V  & &  HR0898 & 4.200 & 4.110 & A1V         & &  2.928 & 2.796 \\
HR1211 & 6.190 & 6.090 & A1V      & &  HR1212 & 5.590 & 4.700 & G6.5III     & &  5.096 & 4.434 \\
\makebox[0mm][r]{$^{\rm *}$}%
HR1897 & 6.300 & 6.390 & O9.5IV   & &  $\theta^2$ Ori B& 6.290 & 6.380 & B2--B5 & &  5.542 & 5.632 \\
\makebox[0mm][r]{$^{\rm *}$}%
HR1948 & 1.790 & 1.880 & O9.2Ib   & &  HR1949 & 3.550 & 3.730 & O9.5II--III & &  1.594 & 1.698 \\
HR2298 & 4.583 & 4.398 & A8V      & &  HR2299 & 6.990 & 6.600 & F5V         & &  4.471 & 4.264 \\
HR2735 & 6.046 & 5.623 & F0/3     & &  HR2736 & 4.774 & 3.746 & K0III       & &  4.481 & 3.569 \\
HR2890 & ---   & 3.000 & A0--A2   & &  HR2891 & ---   & 1.900 & A1.5IV      & &  ---   & 1.564 \\
HR3890 & 3.250 & 2.990 & A9       & &  HR3891 & 6.080 & 5.990 & B7III       & &  3.173 & 2.924 \\
\makebox[0mm][r]{$^{\rm *}$}%
HR3925 & 5.680 & 5.810 & B5V    & &  HD85980B & 8.260 & 8.230 & ---         & &  5.584 & 5.699 \\
\makebox[0mm][r]{$^{\rm *}$}%
HR4057 & 3.130 & ---   & K1III    & &  HR4058 & 3.400 & ---   & G7III       & &  2.504 & ---   \\
HR4374 & 5.410 & 4.770 & G2V      & &  HR4375 & 4.790 & 4.250 & F8.5V       & &  4.304 & 3.727 \\
\makebox[0mm][r]{$^{\rm *}$}%
HR4730 & 1.100 & 1.280 & B0.5IV   & &  HR4731 & 1.410 & 1.580 & B1V         & &  0.491 & 0.667 \\
HR4825 & 3.800 & 3.440 & F1--F2V  & &  HR4826 & 3.850 & 3.490 & F0--F2V     & &  3.072 & 2.712 \\
HR5054 & 2.271 & 2.220 & A1.5V    & &  HR5055 & 4.050 & 3.880 & A1--A7IV--V & &  2.078 & 2.007 \\
HR5328 & 7.080 & 6.690 & F2V      & &  HR5329 & 4.740 & 4.510 & A7IV        & &  4.621 & 4.373 \\
HR5459 & 0.720 & 0.010 & G2V      & &  HR5460 & 2.210 & 1.330 & K1V         & &  0.475 & $-$0.272 \\
HR5475 & 4.792 & 4.893 & B9III    & &  HR5476 & 5.979 & 5.761 & A6V         & &  4.478 & 4.490 \\
HR5477 & 4.590 & 4.510 & ---      & &  HR5478 & 4.560 & 4.510 & ---         & &  3.822 & 3.757 \\
HR5505 & 4.853 & 4.801 & A0V      & &  HR5506 & 3.610 & 2.450 & K0II--III   & &  3.310 & 2.332 \\
HR5788 & 5.380 & 5.130 & F0IV     & &  HR5789 & 4.390 & 4.140 & F0IV        & &  4.023 & 3.773 \\
\makebox[0mm][r]{$^{\rm *}$}%
HR5977 & 5.350 & 4.870 & F5V      & &  HR5978 & 5.640 & 5.160 & F4(V)       & &  4.733 & 4.253 \\
\makebox[0mm][r]{$^{\rm *}$}%
HR5984 & 2.550 & 2.620 & B1V      & &  HR5985 & 4.870 & 4.890 & B2V         & &  2.429 & 2.493 \\
\makebox[0mm][r]{$^{\rm *}$}%
HR6406 & 4.670 & 3.330 & M5Ib--II & &  HR6407 & 6.030 & 5.322 & G5III+F2V   & &  4.397 & 3.169 \\
HR6484 & 5.395 & 5.398 & A0V      & &  HR6485 & 4.476 & 5.510 & B9.5III     & &  4.088 & 4.113 \\
HR6896 & 6.320 & 4.860 & K1/2III  & &  21 Sgr B & 7.680 & 7.390 & ---       & &  6.047 & 4.759 \\
HR7293 & 7.400 & ---   & G3V      & &  HR7294 & 7.220 & ---   & G2V         & &  6.554 & ---   \\
HR7921 & 6.672 & 5.638 & G8IIb    & &  49 Cyg B & 8.150 & 8.090 & B9.9      & &  6.424 & 5.530 \\
HR7947 & 5.530 & 4.960 & F8V      & &  HR7948 & 5.260 & 4.250 & K1IV        & &  4.634 & 3.795 \\
\makebox[0mm][r]{$^{\rm *}$}%
HR8148 & 7.500 & 6.680 & G7V      & &  HD202940B & 10.980 & 9.960 & K5      & &  7.457 & 6.609 \\
HR8309 & 5.140 & 4.700 & F7V      & &  HR8310 & 6.670 & 6.120 & F3V         & &  4.903 & 4.440 \\
HR8545 & 6.830 & 6.220 & G2V      & &  53 Aqr B & 6.960 & 6.320 & G3V       & &  6.140 & 5.516 \\
HR8558 & 4.890 & 4.490 & F2IV/V   & &  HR8559 & 4.790 & 4.340 & F2IV/V      & &  4.086 & 3.660 \\
HR9074 & 6.910 & 6.400 & F8       & &  BD+32\,4747B & 7.210 & 6.660 & G1    & &  6.297 & 5.770 \\
\hline
\end{tabular}
\end{table*}

\begin{table}
\centering
\caption{References for the $B$ and~$V$ photometric data extracted from the
Simbad database and employed for the comparison with the synthetic magnitudes
computed in the fitted CK04 models. (1) ADS bibcode, (2) number of stars with
$B$ photometry, (3) number of stars with $V$ photometry, and (4) reference.}
\label{tab:bibcode_bv}
\begin{tabular}{l@{}r@{\;\; }r@{\;\;}l}
\hline
Bibcode & $B$ & $V$ & Reference \\
\hline
\texttt{1965LowOB...6..167A } &    1 &    1  & \citet{1965LowOB...6..167A} \\
\texttt{1966CoLPL...4...99J } &    4 &    4  & \citet{1966CoLPL...4...99J} \\
\texttt{1967ArA.....4..375L } &    2 &    2  & \citet{1967ArA.....4..375L} \\
\texttt{1968ArA.....4..425L } &    1 &    1  & \citet{1968ArA.....4..425L} \\
\texttt{1969ArA.....5..149L } &    1 &    1  & \citet{1969ArA.....5..149L} \\
\texttt{1969ArA.....5..161L } &    2 &    2  & \citet{1969ArA.....5..161L} \\
\texttt{1969ArA.....5..231L } &    1 &    1  & \citet{1969ArA.....5..231L} \\
\texttt{1978A\&AS...34....1N} &    3 &    3  & \citet{1978A_AS...34....1N} \\
\texttt{1982A\&AS...47..221R} &    2 &    1  & \citet{1982A_AS...47..221R} \\
\texttt{1985A\&AS...61..331O} &    2 &    2  & \citet{1985A_AS...61..331O} \\
\texttt{1991A\&AS...89..415O} &   10 &    7  & \citet{1991A_AS...89..415O} \\
\texttt{1993A\&AS..100..591O} &   52 &   51  & \citet{1993A_AS..100..591O} \\
\texttt{1997JApA...18..161Y } &    5 &    6  & \citet{1997JApA...18..161Y} \\
\texttt{2000A\&A...355L..27H} &  391 &  390  & \citet{2000A_A...355L..27H} \\
\texttt{2001AJ....122.3466M } &    0 &    1  & \citet{2001AJ....122.3466M} \\
\texttt{2002A\&A...384..180F} &   37 &   39  & \citet{2002A_A...384..180F} \\
\texttt{2002yCat.2237....0D } &  788 &  765  & \citet{2002yCat.2237....0D} \\
\texttt{2006AJ....132..111J } &    6 &    6  & \citet{2006AJ....132..111J} \\
\texttt{2009ApJ...694.1085V } &    0 &   17  & \citet{2009ApJ...694.1085V} \\
\texttt{2011A\&A...531A..92R} &    7 &    7  & \citet{2011A_A...531A..92R} \\
\texttt{2012yCat.1322....0Z } &    4 &    4  & \citet{2012yCat.1322....0Z} \\
\texttt{2013ApJ...764..114H } &    1 &    1  & \citet{2013ApJ...764..114H} \\
\texttt{2014ApJ...794...36H } &    1 &    1  & \citet{2014ApJ...794...36H} \\
\hline
\end{tabular}
\end{table}

After applying the previously described fitting method to the initial sample of
1522 stars, we discovered that some of the CK04 model fits were not reliable.
After sorting the stars by the minimum value of the objective function obtained
at the end of the minimization process, we visually examined all the individual
fits to establish potential biases in the fitting procedure. At this
point the following criteria were sequentially employed to remove stars from
the initial sample:
\begin{enumerate}

\item Stars with large objective function values in the
minimization process: we analysed the histograms of the
resulting $\chi^2$ (Fig.~\ref{fig:hist_chisqr}), segregating the sample in two
groups, depending on the number of photometric bands available for each star
(13 and 8 bands for 1359 and 163 stars, respectively). In both cases, although
the distributions are roughly normal, there is a tail of stars with large
$\chi^2$ values. The $3\sigma$ cut (shown as the dotted vertical lines) is
$\chi^2_{3\sigma} \simeq 0.105$ and~$0.047$, for the subsamples with 13 and~8
bands, respectively.  We decided to remove the 48 stars with $\chi^2$ above
these threshold values.  Among them, 43 stars are known variables according to
Simbad or exhibit emission lines (not considered in our fitting procedure), and
5 show large red or blue fluxes that could not be properly fitted with the CK04
models.

\item Stars with large discrepancies in synthetic Johnson~$B$ and~$V$
magnitudes when compared with the available data in the Simbad
database\footnote{\url{http://simbad.u-strasbg.fr/simbad/}}. These magnitudes
were computed in the VEGA system, using as reference the flux density (factor
$f_r(\lambda)$ in Eq.~\ref{eq:mag_flux_densities}) of the Vega spectrum
\texttt{alpha\_lyr\_stys\_010.fits}, available at the CALSPEC
database\footnote{\url{https://www.stsci.edu/hst/instrumentation/reference-data-for-calibration-and-tools/astronomical-catalogs/calspec}}
\citep{2014PASP..126..711B}. In addition, the spectral sensitivity curves for
the~$B$ and~$V$ filters from \citet[see their Table~1]{2012PASP..124..140B}
were employed\footnote{With this election of flux density reference spectrum
and spectral sensitivity curves, the integrated number of photons
$N_{\gamma,r}$ (see Eq.~\ref{eq:mag_photons}) are 1286455 and
873896~photons~s$^{-1}$~cm$^{-2}$, for the $B$ and $V$ bands, 
respectivetly. In
addition, the averaged number of photons for the reference spectrum $\langle
n_{\gamma,r} \rangle$ (see Eq.~\ref{eq:averaged_photons}) are~1401.67 and
996.80~photons~s$^{-1}$~cm$^{-2}$~\AA$^{-1}$, for the $B$ 
and $V$ bands.}. We
have chosen these two classical bandpasses because their wavelength coverage
overlaps with that of typical RGB Bayer-like photometric systems. The expected
random errors in these synthetic measurements, estimated from
the bootstrapping strategy described in Sect.~\ref{sec:uncertainties_fitting}
and displayed in Fig.~\ref{fig:sigma_BV_bootstrapping}, are small: $\Delta
B\sim 0.011$~mag and $\Delta V\sim 0.010$~mag. Since some stars in the original
JM75 sample were flagged as double (i.e., more than one star were observed
simultaneously), the corresponding flux coaddition was performed to compute the
expected~$B$ and~$V$ magnitudes in those cases (see
Table~\ref{tab:double_stars}; the name of the companion stars
are also provided in the fourth column of Table~\ref{tab:bigtable}).
Interestingly, the spectra of these double stars did not lead to large $\chi^2$
values in Fig.~\ref{fig:hist_chisqr}, mainly because the light of the combined
spectrum was dominated by the brightest star in the system or because in several
cases the difference in spectral type was not large. The comparison between
the synthetic and the tabulated Simbad magnitudes is shown in
Fig.~\ref{fig:bv_simbad_comparison}.  It is important to highlight that the
Simbad measurements come from a relatively high number of different sources
(see Table~\ref{tab:bibcode_bv}).  Thus, it is expected that the compiled data
are heterogeneous in photometric quality and not completely free from
systematic offsets. This problem, together with additional
sources of systematic errors, such as the inability of the adopted CK04 model
fitting procedure to reproduce the actual star spectra in all cases, the
presence of unaccounted stellar variability, and the inherent photometric
uncertainties within the C13 photometric data themselves, make the correlation
exhibited by the magnitude differences in
Fig.~\ref{fig:bv_simbad_comparison}(a) not unexpected. In any case, an
independent analysis of $B_{\rm CK04}\!-\!B_{\rm Simbad}$ and $V_{\rm
CK04}\!-\!V_{\rm Simbad}$ has been performed, as displayed in
Figs.~\ref{fig:bv_simbad_comparison}(b) and~(c). We decided to follow a
statistical approach to remove from the star sample those objects with large
deviations in either $B$ or~$V$. For that purpose, we first computed the median
values ($-0.014$~mag and $-0.021$~mag for $B_{\rm CK04}\!-\!B_{\rm Simbad}$ and
$V_{\rm CK04}\!-\!V_{\rm Simbad}$, respectively) and rejected those stars
outside the robust $\pm3\sigma$ interval around the median value ($\pm
0.107$~mag and $\pm0.077$~mag). A total of 128~stars were removed from the
sample, being $\sim2/3$ of this rejected subsample (82~stars) constituted by
known variables.

\end{enumerate}

\begin{figure}
\includegraphics[width=\columnwidth]{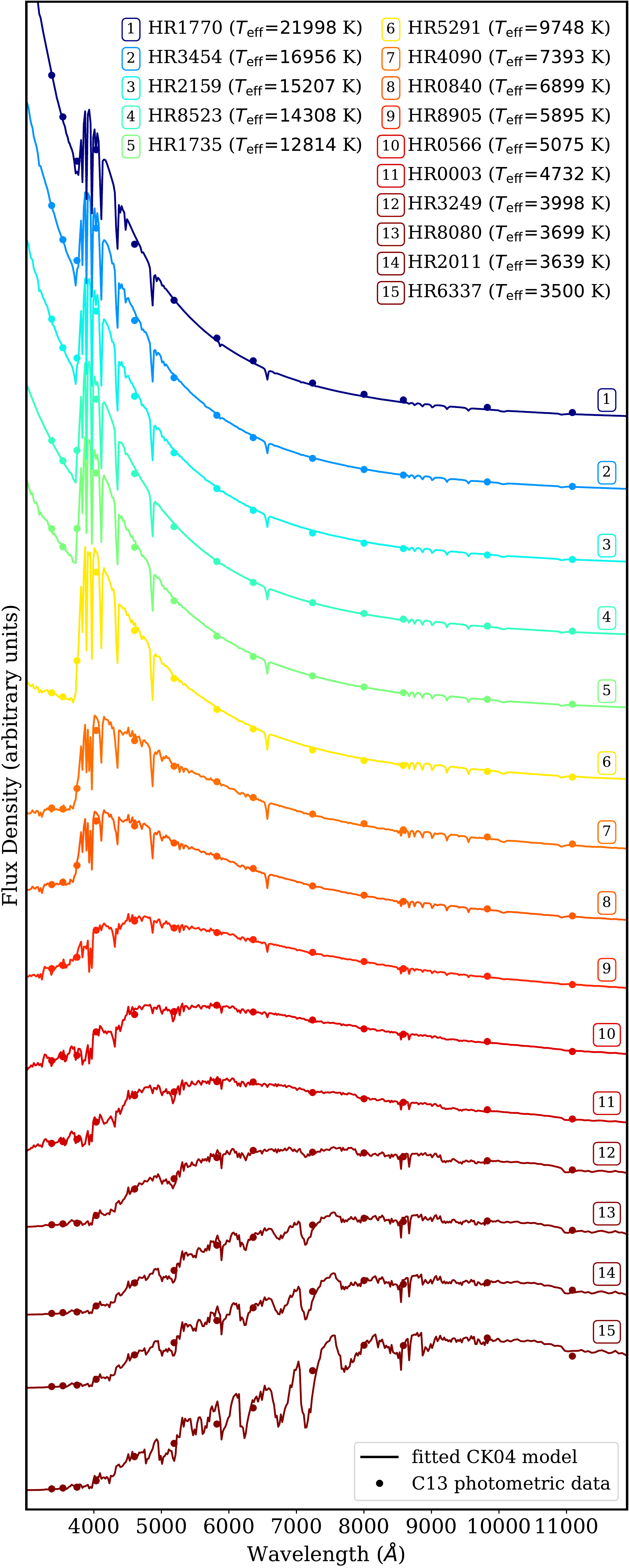}
\caption{Examples of spectrum fits to stars with different effective
temperature. The filled circles correspond to the C13 photometric data (flux
densities in arbitrary units) from JM75. The continuous lines are the fitted
CK04 models. The spectra have been numbered by decreasing
fitted $T_{\rm eff}$ (from Table~\ref{tab:bigtable}), and their 
identifications are given in the upper key.}
\label{fig:sp_sequence_teff}
\end{figure}

After this cleaning process, the final sample of fitted CK04 models is formed
by 1346~stars, that constitute the UCM (Universidad Complutense
de Madrid) library of spectrophotometric standards.  Although an important
fraction of them ($\sim 44$\%, 594 stars) are still classified as known
variables, we have decided to keep these objects in the final list since no
significant differences with the Simbad tabulated measurements
have been found. Some of these variables are eclipsing binaries, and most of
the time they constitute suitable reference stars. In any case,
columns~(5) and~(6) in Table~\ref{tab:bigtable} provide an
additional name and the corresponding classification \citep[according
to][]{2017ARep...61...80S}\footnote{See also a detailed
description of the variability types in
\url{http://cdsarc.u-strasbg.fr/viz-bin/getCatFile_Redirect/?-plus=-\%2b\&B/gcvs/vartype.txt}}
for the variable stars, and potential users of the derived RGB magnitudes
should be well aware of this.

A selection of spectrum fits for stars exhibiting a wide range in effective
temperature is shown in Fig.~\ref{fig:sp_sequence_teff}. 

\subsection{Comparison with Kiehling 1987}
\label{subsec:kiehling_comparison}

\begin{figure*}
\includegraphics[width=\columnwidth]{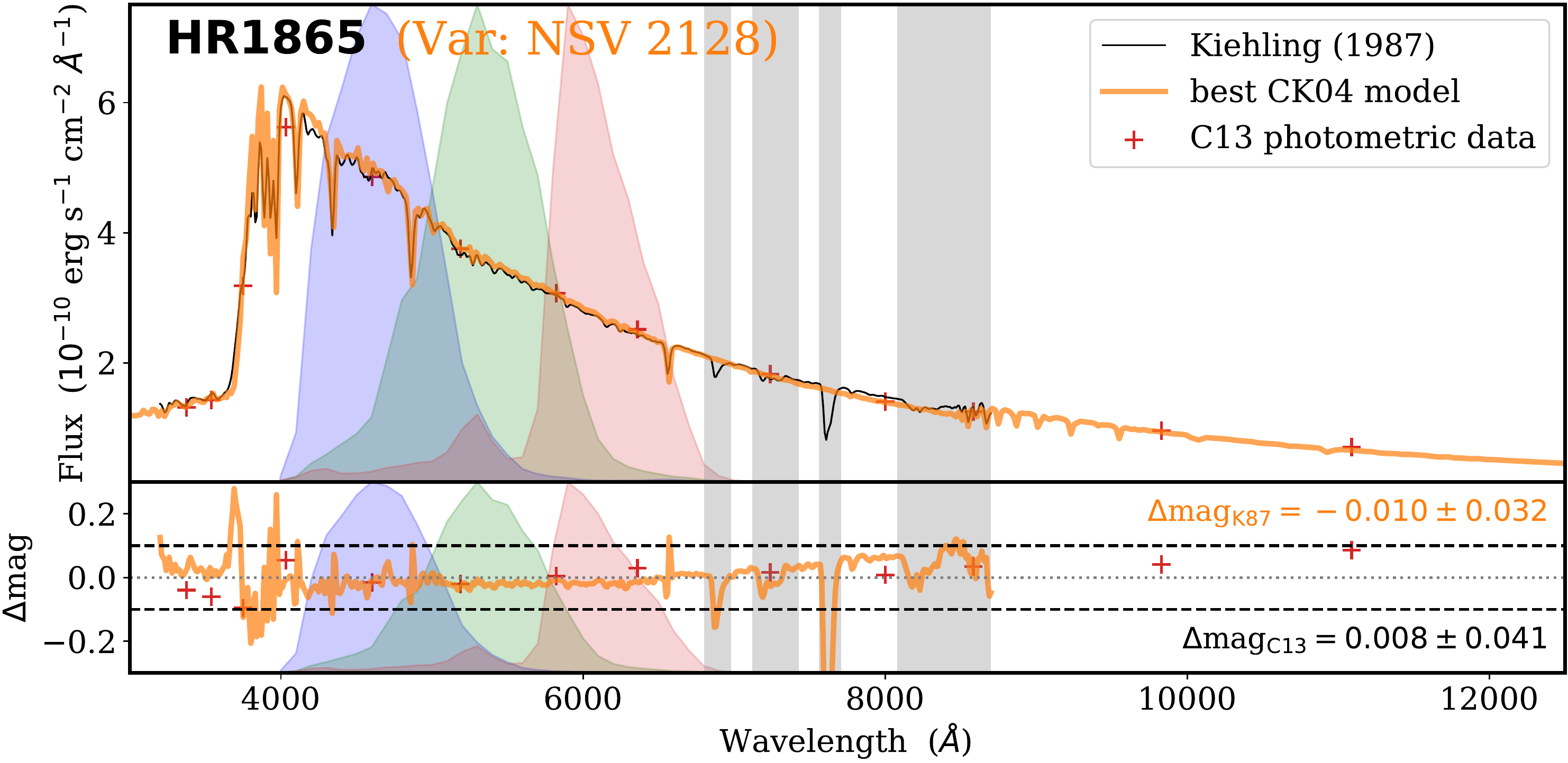}
\hfill
\includegraphics[width=\columnwidth]{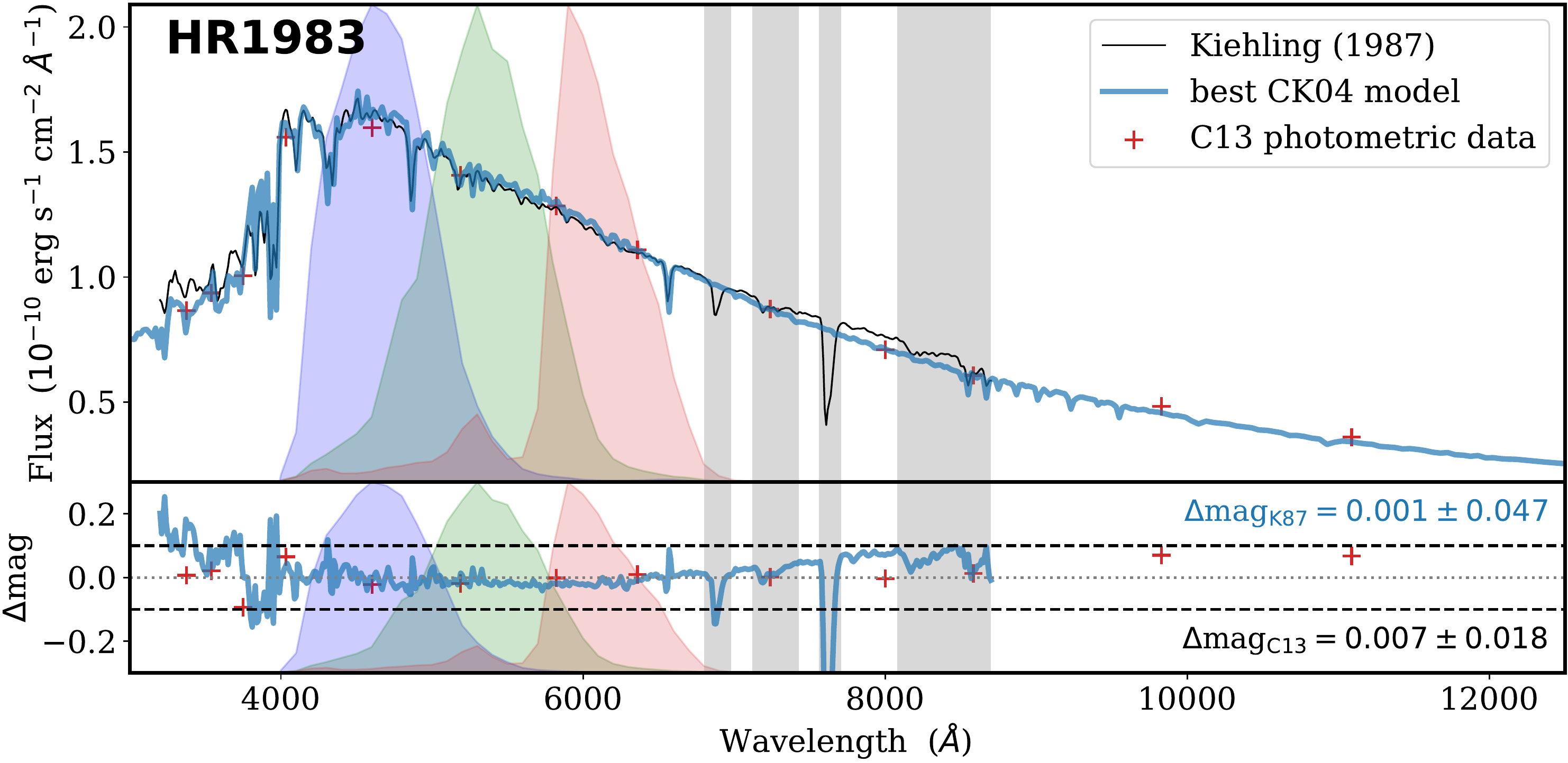}
\vskip 1mm
\includegraphics[width=\columnwidth]{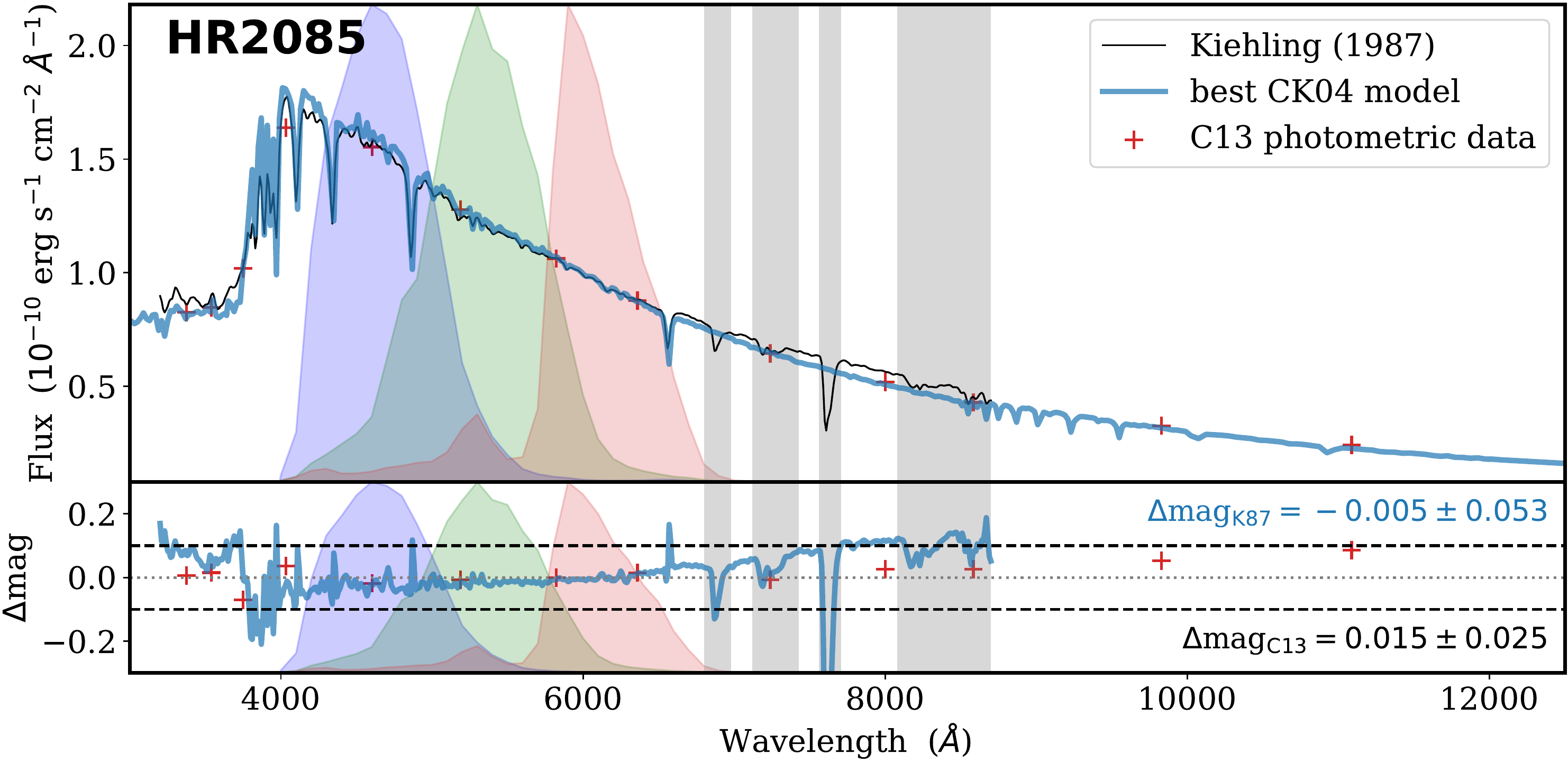}
\hfill
\includegraphics[width=\columnwidth]{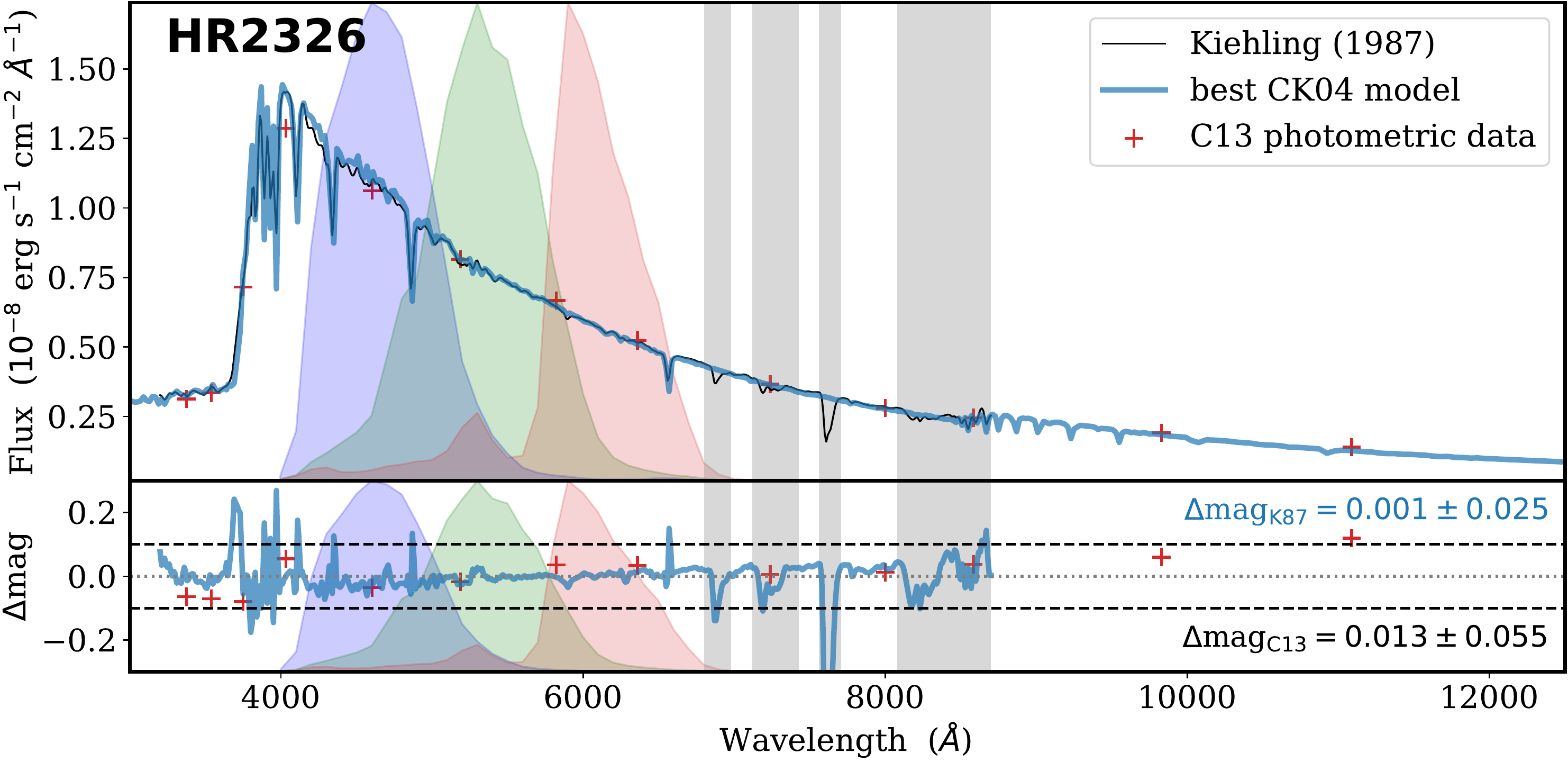}
\caption{Comparison between the best CK04 fitted models and the
spectrophotometric data from K87, for 4 stars (out of 39) in common with our
final sample (all the plots corresponding to the 39 stars in common with K87
are provided in Appendix~A). CK04 models are
represented with thick blue lines for non-variable stars and thick orange lines
for variable stars, while the K87 spectra are plotted with thin black lines.
For each star, the upper panel displays the flux densities. The lower panel
shows the corresponding residuals, in magnitudes, obtained when dividing the
fluxes from CK04 models by the K87 measurements (continuous blue/orange line)
or by the JM75 photometric data (red crosses); the dotted line
marks the $\Delta$mag=0.0 level, whereas the dashed lines encompass the
$\pm0.1$~mag interval. For the variable stars one additional identification
name is given in parenthesis. The C13 photometric data from
JM75 are plotted with red crosses. The standard RGB spectral sensitivity
curves, as defined in Section~\ref{sec:rgb_definition}, are shown with blue,
green, and red shaded areas, whereas the vertical grey bands correspond to the
wavelength intervals where conspicuous telluric absorptions appear in the K87
spectra. See Sec.~\ref{subsec:kiehling_comparison} for additional details.}
\label{fig:comparison_kieh87_ck04}
\end{figure*}

To assess the quality of the fitting procedure just described, we have
compared the compiled C13 photometric data and the fitted CK04 models with
observed
flux calibrated spectra. For that purpose, we have chosen the data from
\citet[hereafter K87]{1987A&AS...69..465K}, who published high-quality spectral
energy distributions for 60~bright stars in the \mbox{[3200--8600] \AA}
wavelength interval, with typical internal flux errors of 0.02~mag
(above~4000~\AA) and~0.05~mag (below~4000~\AA), when comparing observations
from different nights. Fortunately, there were 39 stars in common between our
final star sample and that from K87. Although 25~of them are known variable
stars, according to Simbad, there is a very good agreement between K87 with
both the C13 color photometric data and the corresponding CK04 fitted models.

The graphical comparison is shown in Fig.~\ref{fig:comparison_kieh87_ck04}
(extended by Fig.~A1). The wavelength regions
of strong O$_2$ and H$_2$O telluric absorptions (wavelength intervals
\mbox{[6850--6950]~\AA}, \mbox{[7150--7350]~\AA}, \mbox{[7550--7650]~\AA}, and
\mbox{[8150--8350]~\AA}) were not corrected in the~K87 data, and have been
marked with a grey background in the displayed plots. The C13 measurements are
plotted with filled red circles. The best CK04 model fits to these data are
shown with thick blue (for non-variable stars) or orange (for variable stars)
lines. The K87 spectra are overplotted with thin black lines. It is very
important to highlight that the K87 spectra displayed in this figure are not
fits to the C13 data, but calibrated spectral energy distributions,
independently calibrated from the JM75 data. The resolution of these spectra
was slightly reduced using a Gaussian kernel of 600~km/s in order to match the
spectral resolution exhibited by the CK04 models\footnote{The kernel width was
determined using the \texttt{movel} utility of the \reduceme\ package
\citep{1999PhDT........12C}; \url{https://reduceme.readthedocs.io/en/latest/}}.
The 39 stars are sorted by HR name, and for each one two panels are
shown: the top plot represents the flux density
\mbox{(erg~s$^{-1}$~cm$^{-2}$~\AA$^{-1}$)}; the lower panel shows the 
corresponding residuals, in magnitudes, obtained when dividing
the fluxes from CK04 models by the K87 measurements (continuous blue/orange
line) or by the JM75 photometric data (filled red circles).
The dotted horizontal line in the residuals panel sets the
$\Delta{\rm mag}=0$ level, while the two dashed lines encompass the $\pm0.1$~mag
interval. For each star, the residuals panel also provides two residuals
summaries: $\Delta{\rm mag}_{\rm K87}$, the median and robust standard
deviation when dividing the best CK04 model by the corresponding K87 spectrum
(avoiding the marked telluric absorption regions), and $\Delta{\rm mag}_{\rm
C13}$, the median and robust standard deviation when computing the ratio
between the synthetic C13 fluxes (computed with the best CK04 model and the
transmission curves displayed in Fig.~\ref{fig:response_curves}) and the C13
measurements compiled in this work. In this sense, $\Delta{\rm mag}_{\rm K87}$
provides an indication of how well the K87 spectra agree with the JM75
photometric measurements, while $\Delta{\rm mag}_{\rm C13}$ summarizes the
quality of the CK04 model fit to the JM75 data. The scatter in both parameters
is in most cases (for non-variable stars) below $\pm0.05$~mag.



\section{Synthetic RGB magnitudes}
\label{sec:synthetic_rgb_magnitudes}

\subsection{Definition of a camera-independent RGB standard system}
\label{sec:rgb_definition}

\begin{figure}
\includegraphics[width=\columnwidth]{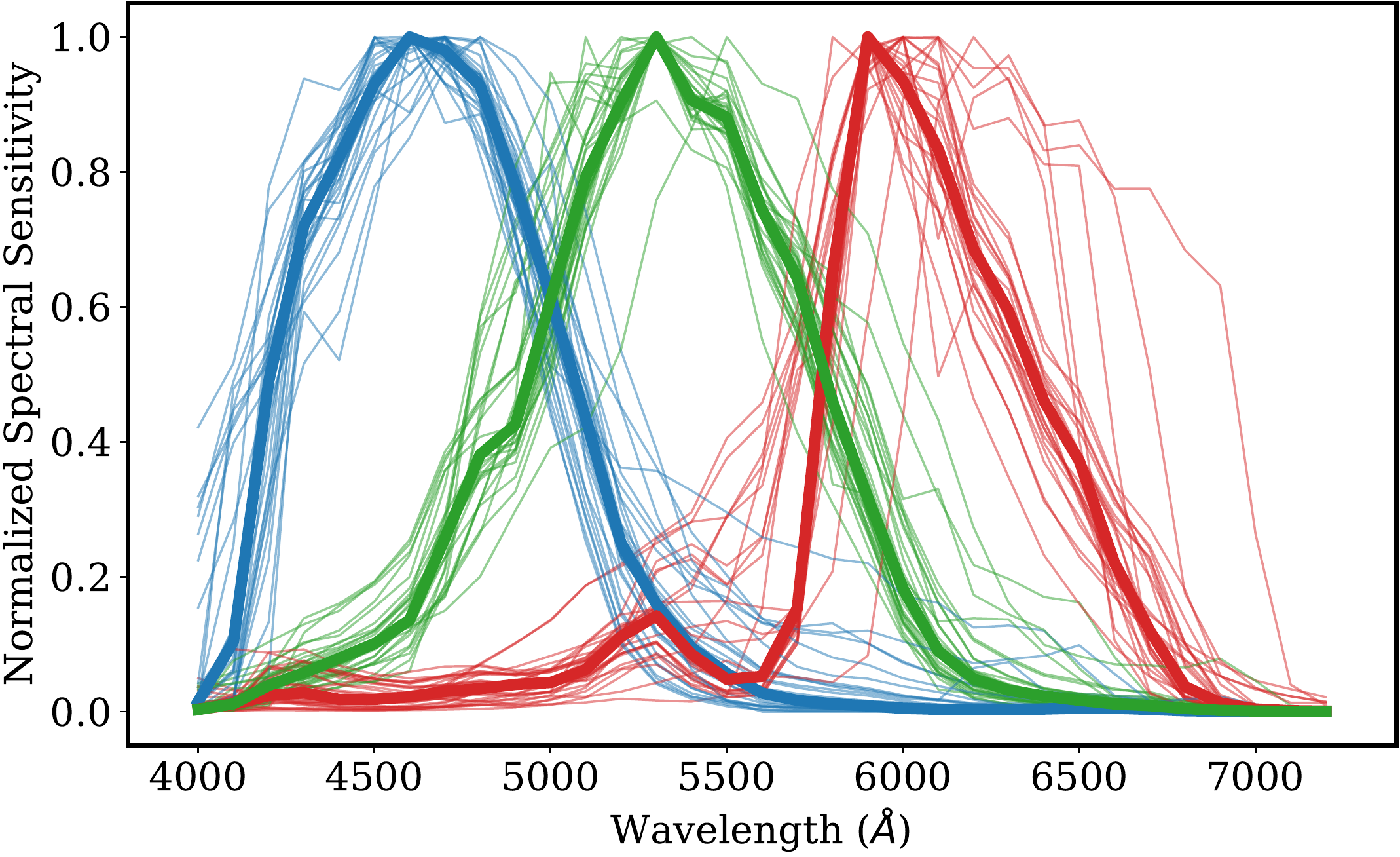}
\caption{Comparison of RGB spectral sensitivity curves: the thin lines
correspond to the 28 cameras measured by
\citet{6475015}: Canon 1DMarkIII, Canon 20D, Canon 300D, Canon 40D, Canon 500D,
Canon 50D, Canon 5DMarkII, Canon 600D, Canon 60D, Hasselblad H2, Nikon D3X,
Nikon D200, Nikon D3, Nikon D300s, Nikon D40, Nikon D50, Nikon D5100, Nikon
D700, Nikon D80, Nikon D90, Nokia N900, Olympus E-PL2, Pentax K-5, Pentax Q,
Point Grey Grasshopper 50S5C, Point Grey Grasshopper2 14S5C, Phase One, and
SONY NEX-5N. Then median values at each sampled wavelength are
plotted as the thick lines, and the corresponding values are listed in 
Table~\ref{tab:median_RGB_table}.}
\label{fig:rgb_standard}
\end{figure}

\begin{table}
\centering
\caption{Renormalized median spectral sensitivity curves computed from the
empirical sensitivity curves corresponding to the 28
cameras measured by \citet{6475015} and displayed in Fig.~\ref{fig:rgb_standard}.
Column description: (1)~wavelength (in \AA); and~(2)--(4) spectral sensitivity at each
considered wavelength, for the median B, G and R bandpasses, respectively. In order
to constraint the curves at the borders of the initial 4000--7200~\AA\ interval
sampled by the \citet{6475015} measurements, two additional rows have been introduced
in the table, at 3990 and 7210~\AA, forcing the curves to drop rapidly to zero
at these wavelengths. The information
in this table is electronically available at~\url{http://guaix.ucm.es/rgbphot}.}
\label{tab:median_RGB_table}

\begin{tabular}{cccc} \hline
(1) & (2) & (3) & (4) \\
$\lambda$ (\AA) & standard $B$ & standard $G$ & standard $R$ \\ \hline

3990 & 0.0000000 & 0.0000000 & 0.0000000 \\
4000 & 0.0150428 & 0.0030840 & 0.0034970 \\
4100 & 0.1039736 & 0.0113570 & 0.0103892 \\
4200 & 0.4892935 & 0.0388725 & 0.0238270 \\
4300 & 0.7202255 & 0.0575395 & 0.0277668 \\
4400 & 0.8216436 & 0.0791910 & 0.0180056 \\
4500 & 0.9308637 & 0.1006650 & 0.0180360 \\
4600 & 1.0000000 & 0.1360232 & 0.0218683 \\
4700 & 0.9802917 & 0.2571178 & 0.0299132 \\
4800 & 0.9275882 & 0.3809050 & 0.0339620 \\
4900 & 0.7807393 & 0.4251800 & 0.0401877 \\
5000 & 0.6143757 & 0.6113000 & 0.0430846 \\
5100 & 0.4338580 & 0.7933000 & 0.0625152 \\
5200 & 0.2491595 & 0.9033850 & 0.1111756 \\
5300 & 0.1594246 & 1.0000000 & 0.1419566 \\
5400 & 0.0947855 & 0.9064100 & 0.0849897 \\
5500 & 0.0567221 & 0.8807233 & 0.0478130 \\
5600 & 0.0273214 & 0.7437300 & 0.0521587 \\
5700 & 0.0166126 & 0.6428150 & 0.1533735 \\
5800 & 0.0114100 & 0.4597650 & 0.6503433 \\
5900 & 0.0084778 & 0.3175050 & 1.0000000 \\
6000 & 0.0048916 & 0.1819950 & 0.9353758 \\
6100 & 0.0034230 & 0.0897230 & 0.8337379 \\
6200 & 0.0029658 & 0.0485390 & 0.6858826 \\
6300 & 0.0032853 & 0.0316045 & 0.5929939 \\
6400 & 0.0037959 & 0.0229870 & 0.4600072 \\
6500 & 0.0051010 & 0.0167670 & 0.3717754 \\
6600 & 0.0050765 & 0.0112830 & 0.2205769 \\
6700 & 0.0032660 & 0.0088190 & 0.1200198 \\
6800 & 0.0011294 & 0.0046771 & 0.0371512 \\
6900 & 0.0005127 & 0.0016871 & 0.0126072 \\
7000 & 0.0002948 & 0.0007490 & 0.0037169 \\
7100 & 0.0001017 & 0.0003077 & 0.0012105 \\
7200 & 0.0000616 & 0.0001488 & 0.0005449 \\
7210 & 0.0000000 & 0.0000000 & 0.0000000 \\
\hline
\end{tabular}
\end{table}

Since the spectral sensitivity curves of Bayer-like color filter systems vary
between different camera models, we have decided to define a particular set of
median sensitivity curves that can be adopted as a camera-independent RGB
standard system. For that purpose, we initially compared the 28 spectral
sensitivity curves measured by \citet{6475015}, which are displayed in
Fig.~\ref{fig:rgb_standard} (thin lines), and then we computed the median value
at each sampled wavelength (from 4000~\AA\ to 7200~\AA, with
$\Delta\lambda=100$~\AA). The resulting median curves, that are represented
with thick lines in the same figure and are tabulated in
Table~\ref{tab:median_RGB_table}, can be adopted to define a new standard
RGB system.

We have checked that the spectral sensitivity curves gathered
by \citet{6475015} already included response curves that are quite similar to
those found in more recent camera models \citep[see e.g. Fig.~1
in][]{2019sanchezdemiguel_etal}, and we have decided not to include these
additional curves in order to avoid biasing the median responses to a particular
camera manufacturer.

\begin{table*}
\centering
\caption{Characteristic parameters of the standard photometric system employed
to determine the synthetic RGB magnitudes. The quoted numbers correspond to the
spectral sensitivity curves listed in Table~\ref{tab:median_RGB_table}.  Column
description (see more details in Sect.~\ref{sec:rgb_definition}): (1)~bandpass
name; (2)~wavelength at the peak of the spectral sensitivity curve; (3)~average
wavelength; (4)~pivot wavelength (5)~bandpass equivalent width; (6)~bandpass
r.m.s.~width; (7)~integrated number of photons, as defined in
Eq.~\ref{eq:mag_photons}, for the reference flux density of the AB system;
(8)~averaged number of photons, given by Eq.~\ref{eq:averaged_photons}, for the
reference flux density of the AB system; (9)~magnitude difference when using
the reference flux density of the ST~system
(Eq.~\ref{eq:f_ref_ST}) instead of the one for the AB~system
(Eq.~\ref{eq:f_ref_AB}); and (10)~magnitude difference when using the Vega
spectrum \texttt{alpha\_lyr\_stys\_010.fits} as the reference
flux density instead of the reference flux density of the AB
system (Eq.~\ref{eq:f_ref_AB}).}
\label{tab:standard_rgb_parameters}
\begin{tabular}{cc@{\,\,\,\,\,\,}c@{\,\,\,\,\,\,}c@{\,\,\,\,}c@{\,\,\,\,}c@{\,\,\,\,}c@{\,\,\,\,}c@{\,\,\,\,\,}c@{\,\,\,\,\,}c} \hline
(1) & (2) & (3) & (4) & (5) & (6) & (7) & (8) & (9) & (10) \\
RGB & wpeak & avgwave & pivot & equivwidth & rmswidth &
$N_{\gamma,r}$ & $\langle n_{\gamma,r} \rangle$ &
$m_{\rm AB}\!-\! m_{\rm ST}$ & 
$m_{\rm AB}\! -\! m_{\rm Vega}$ \\[3pt]
bandpass & (\AA) & (\AA) & (\AA) & (\AA) & (\AA) &
(photons~s$^{-1}$~cm$^{-2}$) &
(photons~s$^{-1}$~cm$^{-2}$~\AA$^{-1}$) &
(mag) & (mag) \\ \hline
standard $B$ & 4600.00 & 4691.29 & 4680.11 & 846.89 & 331.52 &
993989 & 1173.69 & \phantom{$-$}0.341 & $-$0.124 \\
standard $G$ & 5300.00 & 5323.53 & 5308.56 & 916.67 & 395.38 &
948922 & 1035.18 & \phantom{$-$}0.067 & $-$0.024 \\
standard $R$ & 5900.00 & 6006.96 & 5989.64 & 684.86 & 426.98 &
628378 & \phantom{1}917.52 & $-$0.195 & \phantom{$-$}0.103 \\
\hline
\end{tabular}
\end{table*}

Some basic bandpass properties that can be easily derived, once the
spectral sensitivity curves are defined, are provided in
Table~\ref{tab:standard_rgb_parameters}: in particular, the
wavelength at the peak of the spectral sensititivy curve (\texttt{wpeak}
property in synphot), the average and pivot wavelengths, as defined in
\citet[\texttt{avgwave} and \texttt{pivot} properties in
synphot]{1986HiA.....7..833K} computed as
\begin{align}
{\rm avgwave} = 
\frac{\int_{\lambda_i}^{\lambda_f} \lambda\;T(\lambda)\;{\rm d}\lambda}%
{\int_{\lambda_i}^{\lambda_f} T(\lambda)\;{\rm d}\lambda}
\end{align}
and
\begin{align}
{\rm pivot} =
\frac{\int_{\lambda_i}^{\lambda_f} \lambda\;T(\lambda)\;{\rm d}\lambda}%
{\int_{\lambda_i}^{\lambda_f} \frac{T(\lambda)}{\lambda}\; {\rm d}\lambda},
\end{align}
the bandpass equivalent width (\texttt{equivwidth} in synphot), which is the
normalization factor in Eq.~\ref{eq:averaged_photons}, and the r.m.s.\ of the
bandpass following \citet[\texttt{rmswidth} in synphot]{1986HiA.....7..833K}
\begin{align}
{\rm rmswidth} = \sqrt{
\frac{\int_{\lambda_i}^{\lambda_f} (\lambda - {\rm avgwave})^2
\,T(\lambda)\;{\rm d}\lambda}%
{\int_{\lambda_i}^{\lambda_f} T(\lambda)\;{\rm d}\lambda}
}.
\end{align}

The synthetic RGB magnitudes computed in this work, and described in the next
section, were computed in the AB~system, in which the flux density reference
spectrum is defined to exhibit a constant flux density per unit frequency
\citep{1983ApJ...266..713O}, given by 
\begin{align}
f_r(\nu)=10^{-48.60/2.5} \simeq
3.6308\times 10^{-20}~{\rm erg}~{\rm s}^{-1}~{\rm cm}^{-2}~{\rm
Hz}^{-1}.
\end{align}
This expression can easily be converted into a flux density
per unit wavelength as 
\begin{align}
\label{eq:f_ref_AB}
f_r(\lambda)=\frac{c}{\lambda^2}\,f_r(\nu) 
\simeq \frac{0.10885}{\lambda^2}\;
\text{erg}~\text{s}^{-1}~\text{cm}^{-2}~{\text{\AA}}^{-1},
\end{align}
with $\lambda$ expressed in \AA. Inserting the last expression in
Eqs.~\ref{eq:mag_photons} and~\ref{eq:averaged_photons}, we can compute
additional relevant parameters for each RGB spectral sensitivity curve defined
in this section, such as $N_{\gamma,r}$ and $\langle n_{\gamma,r} \rangle$, 
which are also given in Table~\ref{tab:standard_rgb_parameters}.

\begin{figure}
\includegraphics[width=\columnwidth]{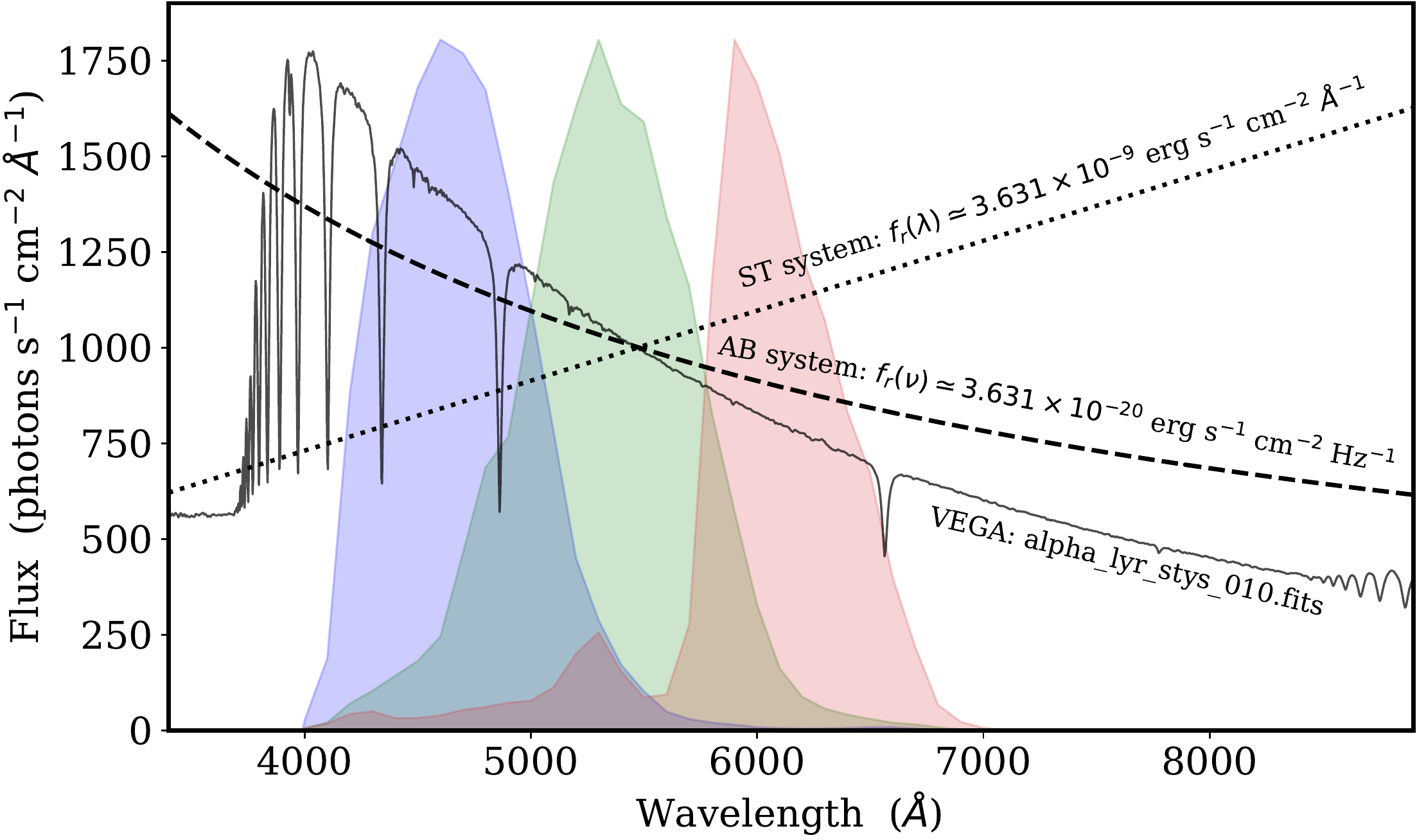}
\caption{Comparison between the flux density (in
photons~s$^{-1}$~cm$^{-2}$~\AA$^{-1}$) of the reference
spectrum used to define the AB magnitudes (dashed black line;
Eq.~\ref{eq:f_ref_AB}), the ST magnitudes (dotted black line;
Eq.~\ref{eq:f_ref_ST}), and Vega magnitudes (full black line). The standard RGB
spectral sensitivity curves, defined in Table~\ref{tab:median_RGB_table}, are
shown with blue, green and red shaded areas. The units employed in this figure
(known in synphot as \texttt{PHOTLAM}) allow a direct visual comparison with
the $\langle n_{\gamma,r}\rangle$ values listed in
Table~\ref{tab:standard_rgb_parameters}.}
\label{fig:vega_ab_st_comparison}
\end{figure}

For completeness, and although in this paper we strongly advocate the use of
AB~magnitudes, we also indicate, in the last two columns of
Table~\ref{tab:standard_rgb_parameters}, the expected offsets when measuring 
ST~magnitudes, in which the reference flux densitity per unit wavelength is 
constant \citep{1986HiA.....7..833K} and equal to
\begin{align}
\label{eq:f_ref_ST}
f_r(\lambda) = 10^{-21.10/2.5} \simeq 3.6308\times 10^{-9}
~\text{erg}~\text{s}^{-1}~\text{cm}^{-2}~{\text{\AA}}^{-1},
\end{align}
or when employing Vega magnitudes, using for that purpose
the flux density provided by the spectrum
\texttt{alpha\_lyr\_stys\_010.fits} as reference. As expected, the differences
between the three different systems are smaller in the $G$~band, where the
corresponding reference flux densities intersect (See
Fig.~\ref{fig:vega_ab_st_comparison}).

\subsection{Synthetic RGB magnitudes for the bright star sample}
\label{sec:rgb_magnitudes_bright_stars}

\begin{figure}
\includegraphics[width=\columnwidth]{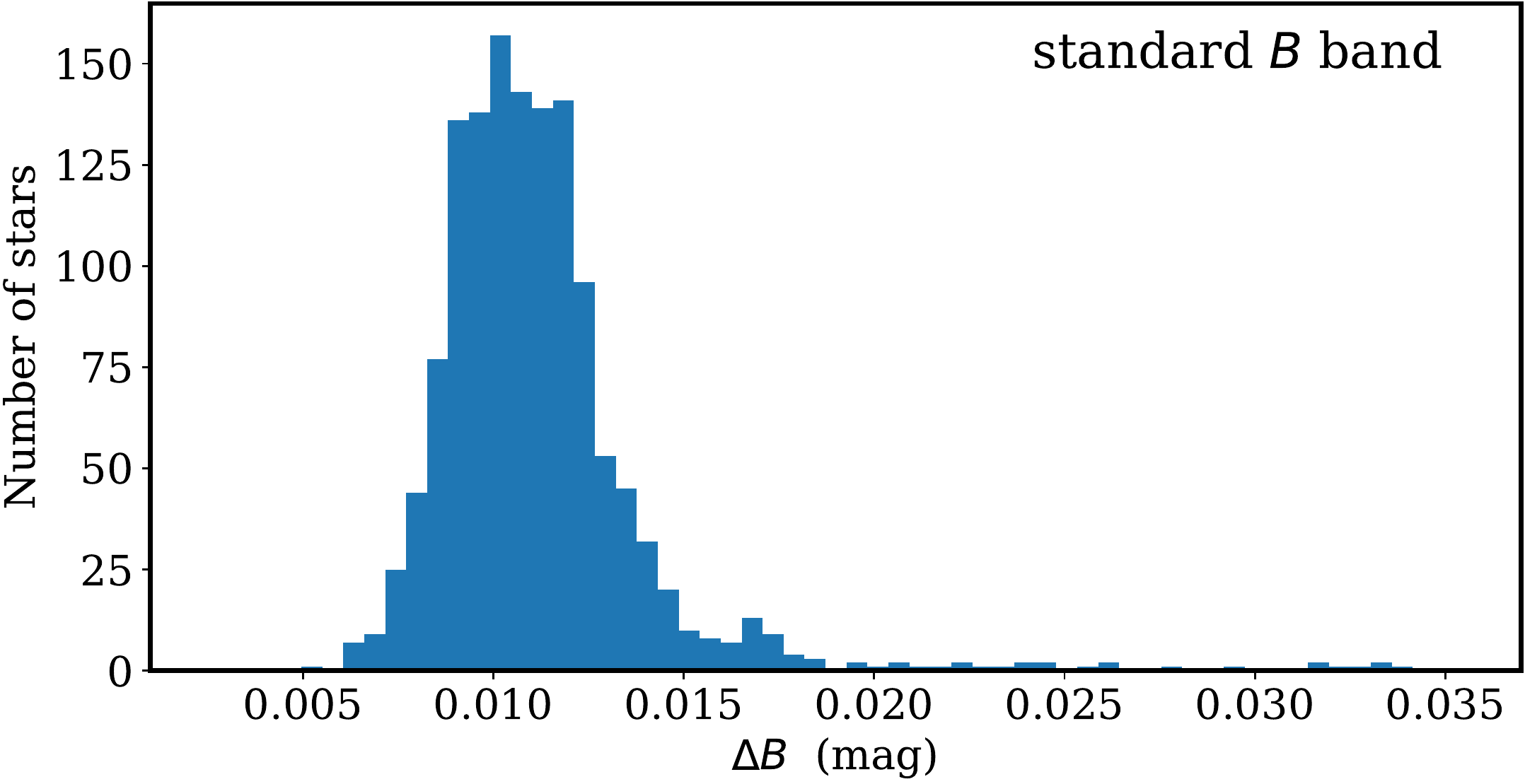}
\vskip 2mm
\includegraphics[width=\columnwidth]{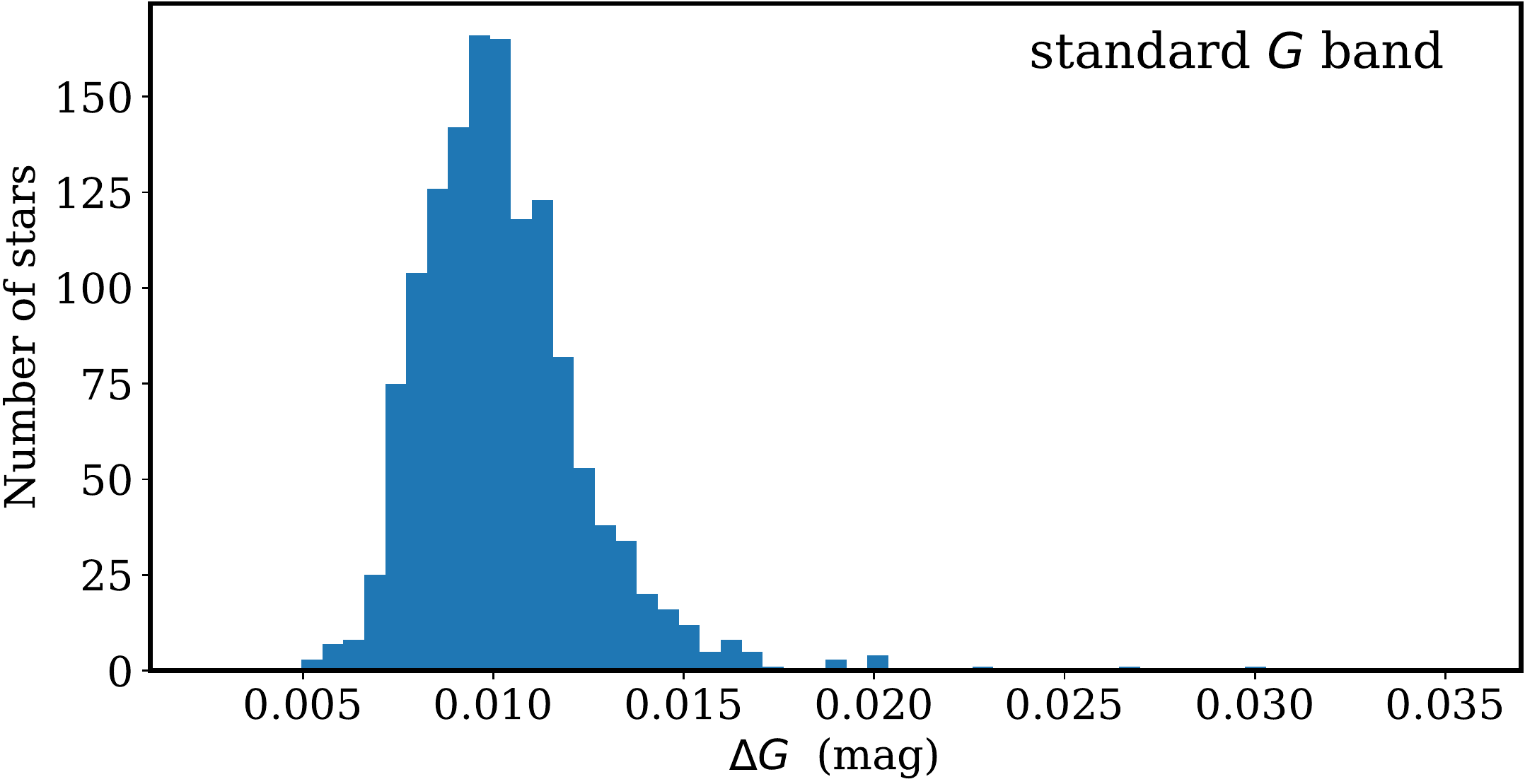}
\vskip 2mm
\includegraphics[width=\columnwidth]{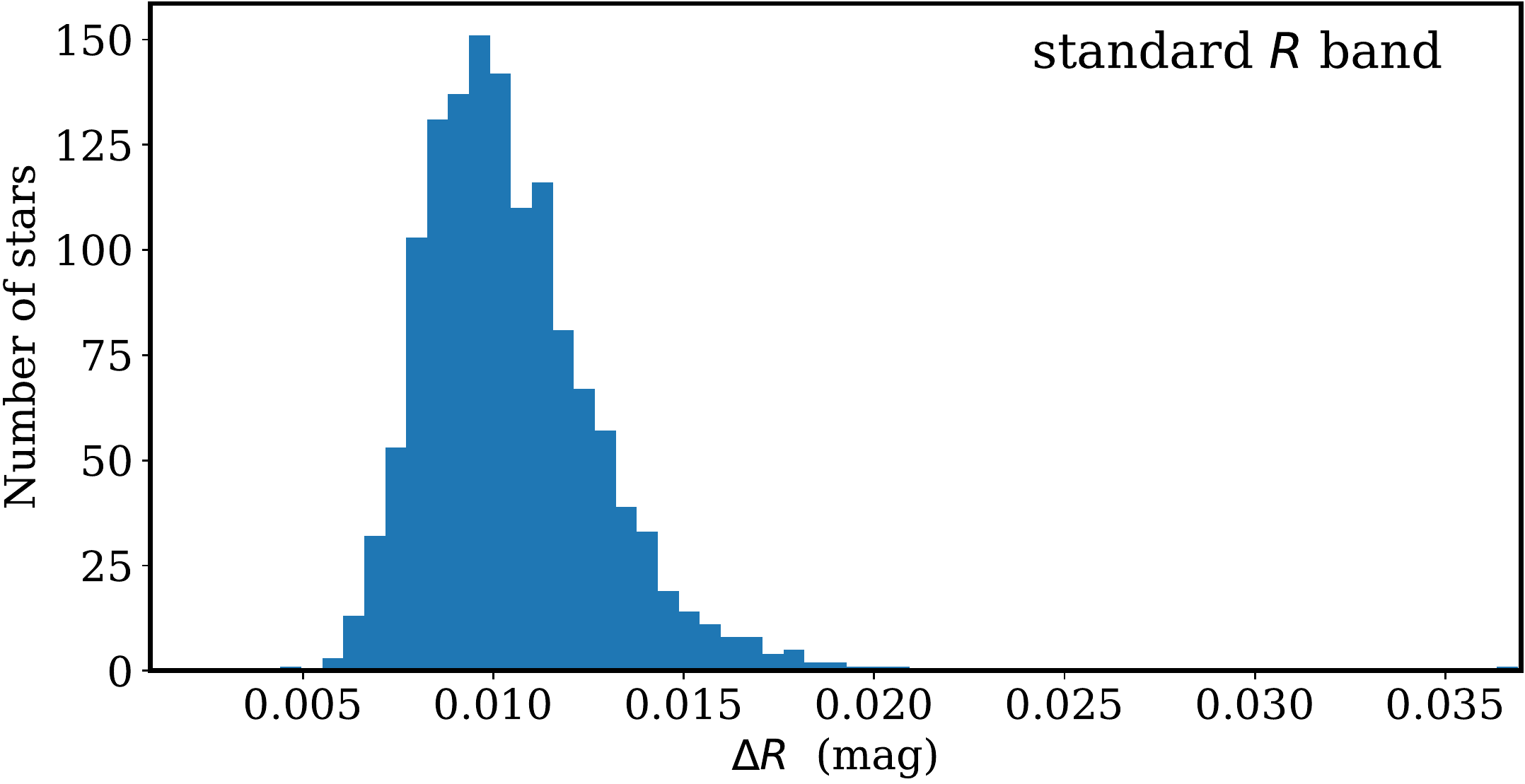}
\caption{Distribution of random uncertainties in the synthetic standard RGB
measurements performed on the final sample of fitted CK04 models, estimated
from the bootstrapped spectra generated as explained in 
Sect.~\ref{sec:uncertainties_fitting}. The median values are \mbox{$\Delta
B=0.011$ mag} (top panel), \mbox{$\Delta G = 0.010$ mag} (middle panel)
\mbox{and $\Delta R=0.010$ mag} (bottom panel).}
\label{fig:simul_stderr_standard_rgb}
\end{figure}

\begin{figure}
\includegraphics[width=\columnwidth]{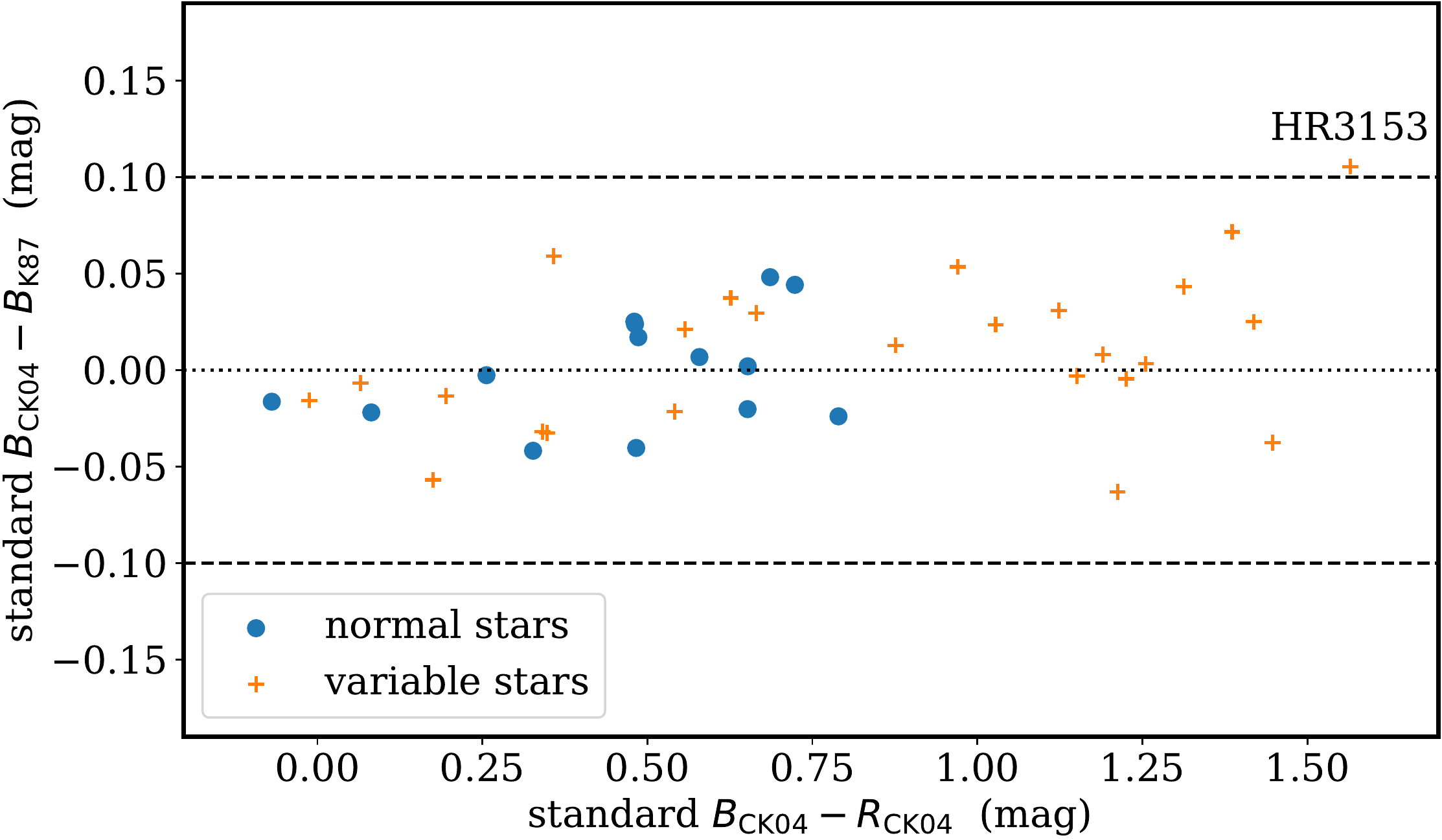}
\vskip 2mm
\includegraphics[width=\columnwidth]{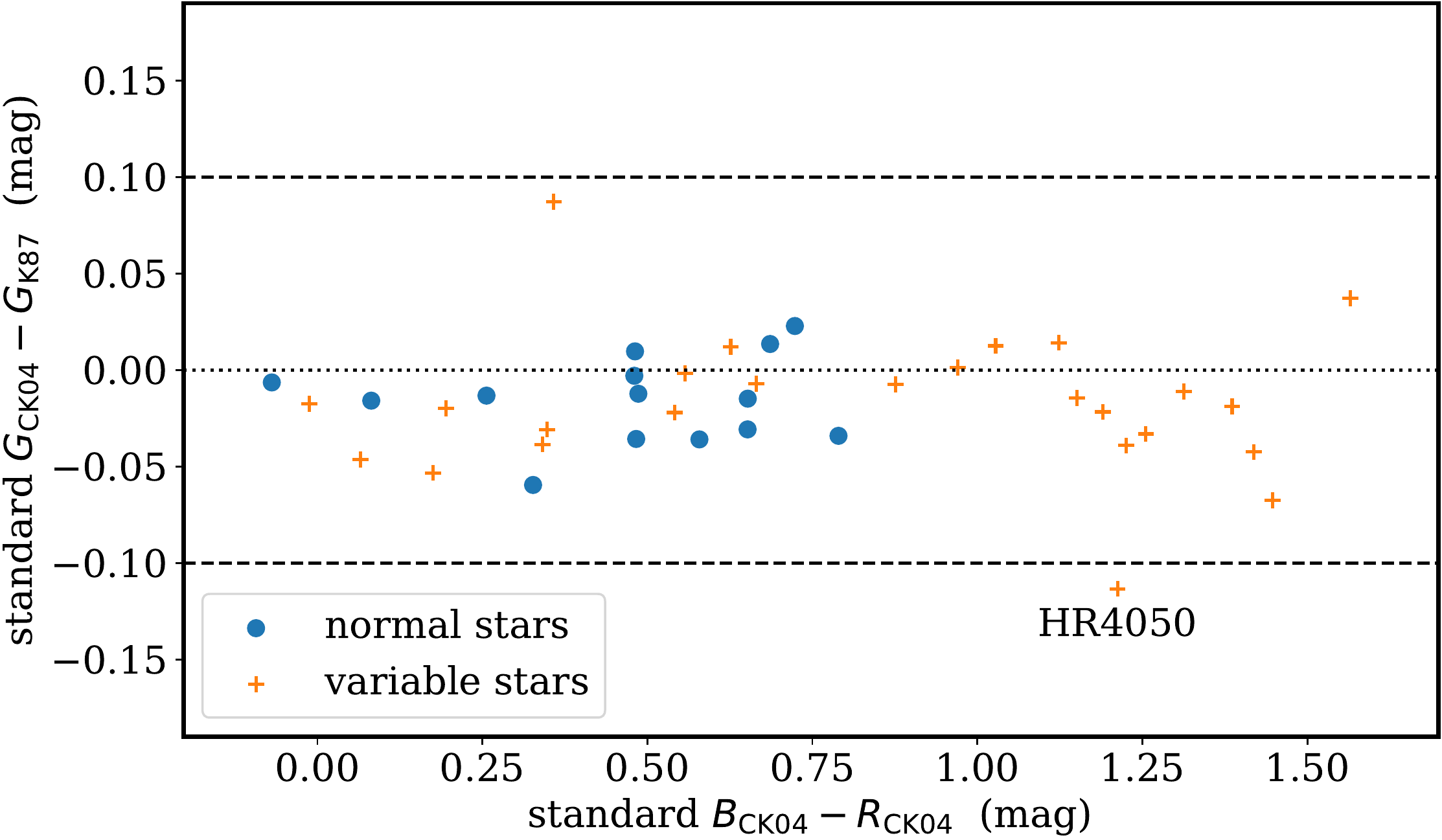}
\vskip 2mm
\includegraphics[width=\columnwidth]{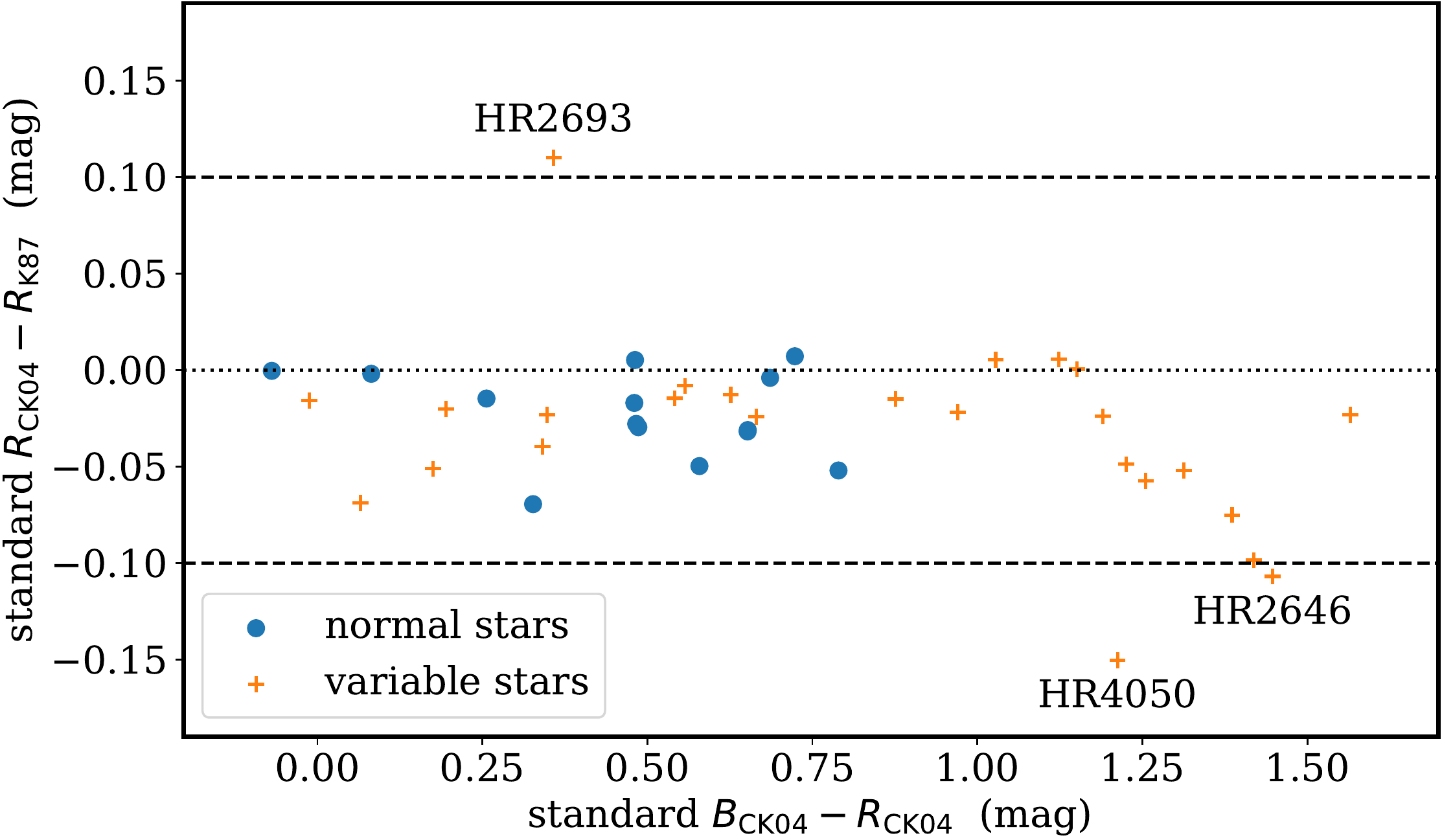}
\caption{Differences between the synthetic RGB magnitudes computed using the
best CK04 model fit and those determined from the spectra available in the K87
sample. Known variable stars are plotted with orange crosses, whereas normal
stars (i.e., non-variable) are represented with blue filled circles. The
horizontal dotted line in each plot marks the \mbox{$\Delta{\rm mag}$=0.0}
level, and the two horizontal dashed lines encompass the $\pm0.1$~mag region.
All the non-variable stars are clearly within the latter interval (the variable
stars beyond the $\pm0.1$ region have been labelled). The median and standard
deviation, using the non-variable stars, are \mbox{$0.000\pm0.028$ mag} for
the~$B$ bandpass (top panel), \mbox{$-0.014\pm0.022$ mag} for the~$G$ bandpass
(middle panel), and \mbox{$-0.023\pm0.022$ mag} for the~$R$ bandpass (bottom
panel).}
\label{fig:rgb_ck04_k87}
\end{figure}

\begin{figure}
\includegraphics[width=\columnwidth]{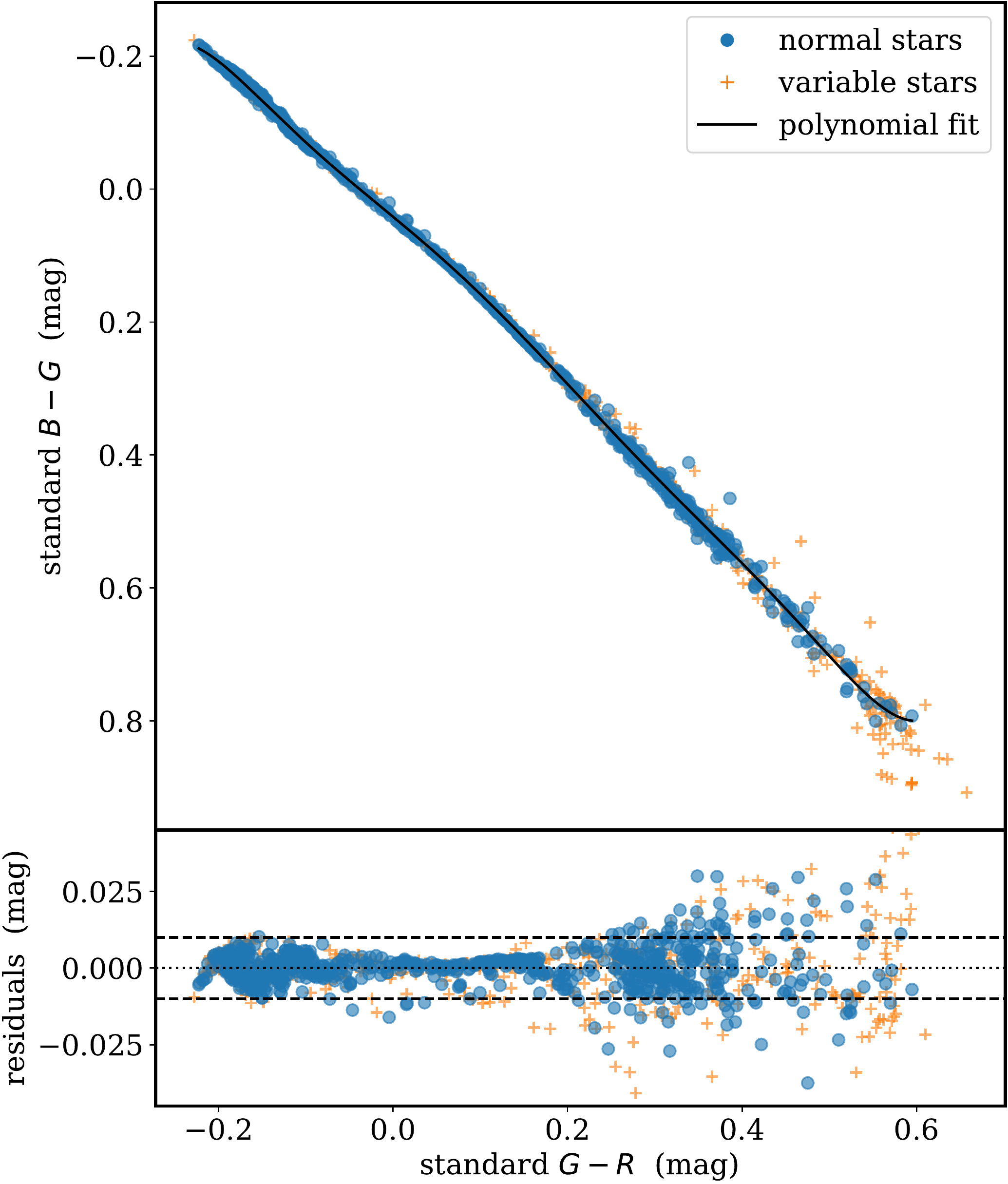}
\caption{Color--color diagram, using the RGB magnitudes measured in the final
sample of 1346 stars, using the standard photometric system defined in
Sect.~\ref{sec:rgb_magnitudes_bright_stars}. The black line shows the 7-degree
polynomial fit given in Eq.~\ref{eq:color_color_fit}. The residuals around the
fit are displayed in the lower panel, where the horizontal dashed lines
encompass the $\pm 0.01$~mag interval. For stars with \mbox{$G-R \leq
0.2$~mag}, the standard deviation of the residuals is 0.004~mag for both normal
(i.e.\ non-variable) and variable stars, whereas for stars with redder colors
the standard deviations are 0.012~mag and~0.032~mag for the normal and variable
stars, respectively.}
\label{fig:color_color_RGB}
\end{figure}

\begin{table*}
\centering
\caption{Comparison between the synthetic RGB magnitudes measured in the best
CK04 fit models and those determined from the K87 flux
calibrated spectra.  The 39 stars are those already shown in
Fig.~\ref{fig:comparison_kieh87_ck04}. Column description: 
(1) HR number;
(2) additional name when the star is a known variable according to Simbad; 
(3)--(5) synthetic RGB magnitudes (AB system), measured in the best CK04 fit, using the
bandpass definitions given in Table~\ref{tab:median_RGB_table}, 
with uncertainties estimated from bootstrapping;
(6)--(8) synthetic RGB magnitudes (AB system), measured in the flux calibrated K87
spectrum.}
\label{tab:ck04_k87}
\begin{tabular}{ccr@{$\;\;\pm\;$}lr@{$\;\;\pm\;$}lr@{$\;\;\pm\;$}lccc} \hline
\multicolumn{1}{c}{(1)} & \multicolumn{1}{c}{(2)} & \multicolumn{2}{c}{(3)} & \multicolumn{2}{c}{(4)} & \multicolumn{2}{c}{(5)} & \multicolumn{1}{c}{(6)} & \multicolumn{1}{c}{(7)} & \multicolumn{1}{c}{(8)} \\ 
\multicolumn{1}{c}{HR} & \multicolumn{1}{c}{Var.Name} & \multicolumn{2}{c}{standard $B$ (CK04)} & \multicolumn{2}{c}{standard $G$ (CK04)} & \multicolumn{2}{c}{standard $R$ (CK04)} & \multicolumn{1}{c}{standard $B$ (K87)} & \multicolumn{1}{c}{standard $G$ (K87)} & \multicolumn{1}{c}{standard $R$ (K87)} \\ \hline
1325 & --- & 4.794 & 0.010 & 4.462 & 0.009 & 4.215 & 0.009 & 4.787 & 4.498 & 4.265 \\ 
1457 & alf Tau & 1.758 & 0.018 & 1.001 & 0.011 & 0.445 & 0.012 & 1.715 & 1.012 & 0.497 \\ 
1654 & eps Lep & 3.969 & 0.012 & 3.296 & 0.012 & 2.819 & 0.011 & 3.972 & 3.310 & 2.818 \\ 
1829 & NSV  2008 & 3.198 & 0.013 & 2.866 & 0.011 & 2.641 & 0.010 & 3.177 & 2.867 & 2.649 \\ 
1865 & NSV  2128 & 2.566 & 0.012 & 2.559 & 0.012 & 2.578 & 0.013 & 2.581 & 2.576 & 2.594 \\ 
1983 & --- & 3.758 & 0.010 & 3.600 & 0.008 & 3.502 & 0.008 & 3.761 & 3.613 & 3.516 \\ 
2085 & --- & 3.804 & 0.010 & 3.740 & 0.009 & 3.722 & 0.010 & 3.826 & 3.756 & 3.724 \\ 
2326 & --- & $-$0.797 & 0.011 & $-$0.774 & 0.010 & $-$0.728 & 0.010 & \makebox[0pt][r]{$-$}0.781 & \makebox[0pt][r]{$-$}0.768 & \makebox[0pt][r]{$-$}0.727 \\ 
2646 & sig CMa & 4.464 & 0.017 & 3.619 & 0.013 & 3.017 & 0.011 & 4.502 & 3.687 & 3.124 \\ 
2693 & NSV  3424 & 2.078 & 0.011 & 1.861 & 0.009 & 1.720 & 0.008 & 2.019 & 1.774 & 1.610 \\ 
2943 & alf CMi A & 0.481 & 0.009 & 0.358 & 0.007 & 0.287 & 0.006 & 0.495 & 0.378 & 0.307 \\ 
3153 & V460 Car & 6.256 & 0.022 & 5.348 & 0.013 & 4.691 & 0.015 & 6.150 & 5.311 & 4.714 \\ 
3185 & rho Pup & 2.884 & 0.010 & 2.771 & 0.008 & 2.709 & 0.008 & 2.941 & 2.824 & 2.760 \\ 
3188 & --- & 4.837 & 0.010 & 4.427 & 0.009 & 4.151 & 0.009 & 4.789 & 4.414 & 4.156 \\ 
3249 & NSV  3973 & 4.338 & 0.012 & 3.643 & 0.013 & 3.148 & 0.012 & 4.331 & 3.665 & 3.172 \\ 
3438 & --- & 4.389 & 0.013 & 4.000 & 0.011 & 3.737 & 0.010 & 4.409 & 4.031 & 3.769 \\ 
3547 & --- & 3.643 & 0.011 & 3.216 & 0.010 & 2.920 & 0.009 & 3.599 & 3.193 & 2.913 \\ 
3748 & alf Hya & 2.761 & 0.015 & 2.103 & 0.010 & 1.637 & 0.011 & 2.730 & 2.089 & 1.632 \\ 
3873 & NSV  4613 & 3.368 & 0.013 & 3.045 & 0.011 & 2.827 & 0.010 & 3.390 & 3.067 & 2.842 \\ 
4050 & V337 Car & 4.226 & 0.012 & 3.511 & 0.012 & 3.014 & 0.011 & 4.289 & 3.624 & 3.164 \\ 
4114 & NSV  4869 & 3.798 & 0.013 & 3.748 & 0.010 & 3.733 & 0.009 & 3.805 & 3.795 & 3.802 \\ 
4362 & FN Leo & 5.565 & 0.017 & 4.731 & 0.009 & 4.147 & 0.010 & 5.540 & 4.774 & 4.245 \\ 
4517 & nu. Vir & 4.873 & 0.032 & 4.143 & 0.019 & 3.617 & 0.018 & 4.869 & 4.176 & 3.675 \\ 
4540 & --- & 3.776 & 0.011 & 3.579 & 0.009 & 3.450 & 0.009 & 3.818 & 3.638 & 3.519 \\ 
4630 & eps Crv & 3.719 & 0.008 & 3.117 & 0.008 & 2.691 & 0.011 & 3.695 & 3.104 & 2.686 \\ 
4763 & gam Cru & 2.591 & 0.013 & 1.763 & 0.014 & 1.205 & 0.019 & 2.520 & 1.782 & 1.281 \\ 
4786 & bet Crv & 3.082 & 0.009 & 2.710 & 0.008 & 2.456 & 0.008 & 3.045 & 2.698 & 2.469 \\ 
4932 & NSV  6064 & 3.314 & 0.012 & 2.921 & 0.011 & 2.649 & 0.011 & 3.285 & 2.928 & 2.673 \\ 
5019 & --- & 5.061 & 0.011 & 4.774 & 0.009 & 4.574 & 0.008 & 5.044 & 4.787 & 4.604 \\ 
5072 & --- & 5.322 & 0.012 & 5.037 & 0.010 & 4.841 & 0.009 & 5.299 & 5.027 & 4.836 \\ 
5235 & NSV 19993 & 2.887 & 0.012 & 2.682 & 0.010 & 2.546 & 0.010 & 2.919 & 2.720 & 2.585 \\ 
5340 & alf Boo & 0.573 & 0.010 & 0.008 & 0.008 & $-$0.397 & 0.010 & 0.520 & 0.007 & \makebox[0pt][r]{$-$}0.375 \\ 
5459 & --- & 0.039 & 0.014 & $-$0.244 & 0.012 & $-$0.441 & 0.011 & 0.014 & \makebox[0pt][r]{$-$}0.241 & \makebox[0pt][r]{$-$}0.424 \\ 
5854 & NSV 20391 & 3.232 & 0.009 & 2.718 & 0.008 & 2.356 & 0.010 & 3.219 & 2.726 & 2.371 \\ 
5868 & lam Ser & 4.640 & 0.014 & 4.431 & 0.011 & 4.292 & 0.010 & 4.673 & 4.462 & 4.316 \\ 
6030 & --- & 4.372 & 0.008 & 3.901 & 0.008 & 3.583 & 0.008 & 4.397 & 3.935 & 3.635 \\ 
6102 & --- & 4.305 & 0.012 & 3.925 & 0.008 & 3.653 & 0.008 & 4.303 & 3.940 & 3.684 \\ 
6159 & NSV  7812 & 5.644 & 0.012 & 4.933 & 0.011 & 4.418 & 0.010 & 5.648 & 4.972 & 4.467 \\ 
6623 & --- & 3.752 & 0.009 & 3.467 & 0.008 & 3.269 & 0.008 & 3.792 & 3.502 & 3.297 \\ 
\hline
\end{tabular}
\end{table*}

Using the median sensitivity curves defined in the previous subsection, we 
measured synthetic RGB magnitudes in the AB system over the final sample of
1346 CK04 models fitted to the C13 photometric data. The results are given in
columns (26)--(28) of Table~\ref{tab:bigtable}. The uncertainties in each case
were derived from the standard deviations of the different values computed when
using the bootstrapped spectra generated during the fitting procedure
(Sect.~\ref{sec:uncertainties_fitting}). Histograms with these uncertainties
are displayed in Fig.~\ref{fig:simul_stderr_standard_rgb}. Not surprisingly,
the median uncertainties are similar to those previously found for the Johnson
$B$ and $V$ filters, in particular \mbox{$\Delta B=0.011$ mag}, \mbox{$\Delta G
= 0.010$ mag}, \mbox{and $\Delta R=0.010$ mag}.

Another interesting exercise is the comparison of the
synthetic RGB magnitudes with those computed using the K87 sample. The
synthetic RGB magnitudes for the 39 stars in common are listed in
Table~\ref{tab:ck04_k87}.  The differences in each bandpass, as a function of
the \mbox{$B-R$} color, are plotted in Fig.~\ref{fig:rgb_ck04_k87}. The median
and standard deviation of the differences, using only the 14~non-variable stars
in this subsample, are \mbox{$0.000\pm0.028$ mag}, \mbox{$-0.014\pm0.022$ mag},
and \mbox{$-0.023\pm0.022$ mag}, for the~$B$, $G$, and $R$ bandpasses
respectively. The dispersion is up to 3~times larger than the typical
uncertainties previously estimated from the bootstrapping analysis, which
should be at least in part attributable to the internal flux errors in the K87
spectra (already mentioned in Sect.~\ref{subsec:kiehling_comparison}), and we
cannot discard the presence of a small systematic offset between the~$G$
and~$R$ measurements when comparing the CK04 and K87 spectra. In addition, most
variable stars in Fig.~\ref{fig:rgb_ck04_k87} follow the same trend exhibited
by the non-variable stars, with a larger dispersion towards redder colors. This
behaviour is consistent with the results shown in
Figs.~\ref{fig:comparison_kieh87_ck04} and~A1, where several
variable stars (plotted with orange lines) already exhibited non-negligible
differences when comparing the best CK04 fitted model with the K87 spectrum.

A color--color diagram is shown in Fig.~\ref{fig:color_color_RGB}, where a
tight correlation between the standard $B-G$ and $G-R$ colors is manifest,
that can be well reproduced by the 7-degree polynomial given by
\begin{align}
\label{eq:color_color_fit}
\nonumber
(B-G) = & \mbox{}\;\;0.04162407 + 1.08718859\,(G-R) + \\ \nonumber
        & +0.31438309\,(G-R)^2 + 6.07961811\,(G-R)^3 +\\ \nonumber
        & -10.8882237\,(G-R)^4 - 65.9762145\,(G-R)^5 +\\
        & +216.580798\,(G-R)^6 - 174.464510\,(G-R)^7,
\end{align}
which is valid for \,$-0.22 < G\!-\!R < 0.59$.
The residuals around this fit, computed using only the normal (i.e.\
non-variable) stars,
exhibit a scatter well constrained within $\pm 0.01$~mag for stars with\;
\mbox{$G-R \leq 0.2$~mag}. Note, however, that the scatter
increases towards redder colors, where the \,$\log g$\, range covered by common
stars is considerably large.

\subsection{Comparison between different RGB systems}

\begin{figure}
\includegraphics[width=\columnwidth]{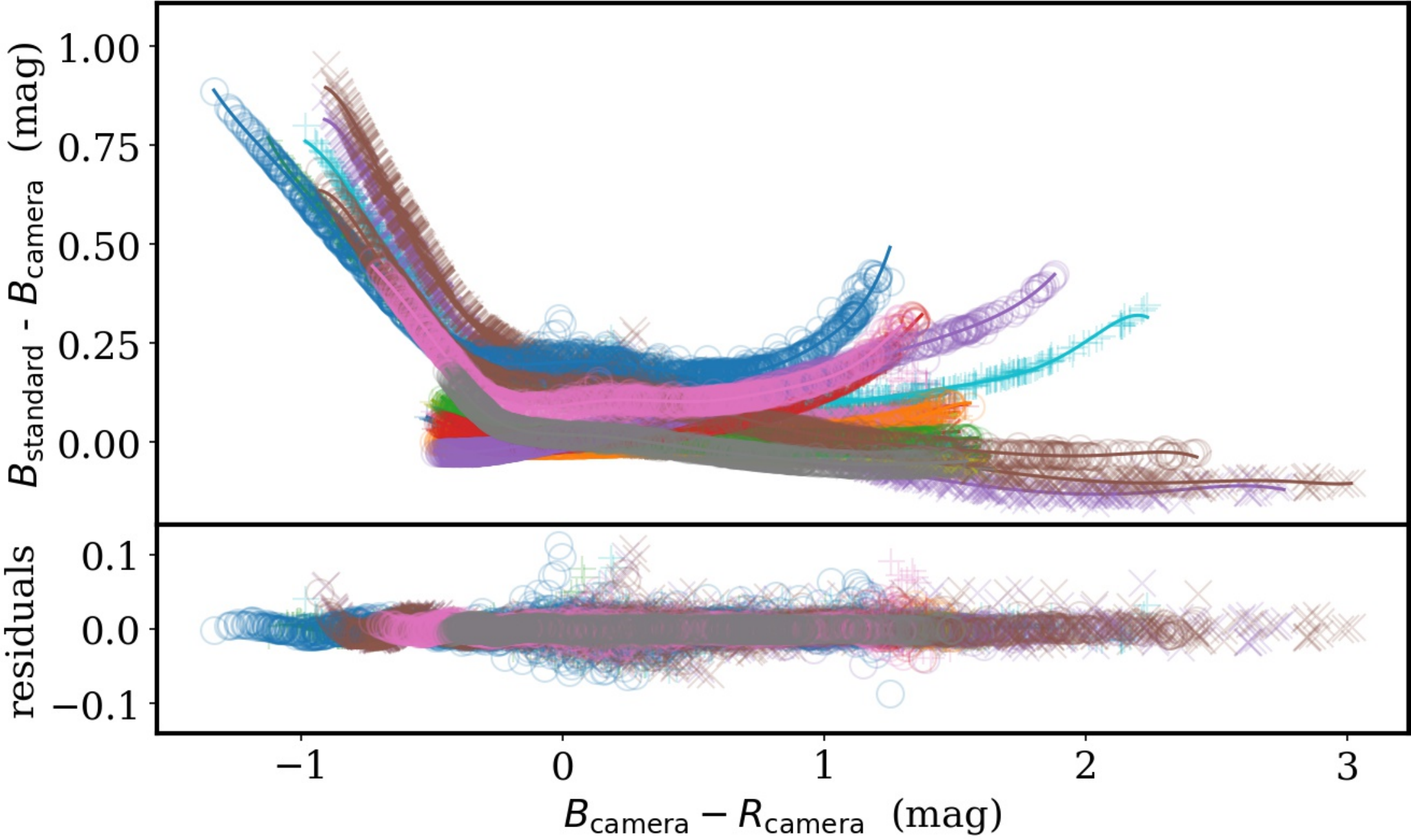}
\vskip 2mm
\includegraphics[width=\columnwidth]{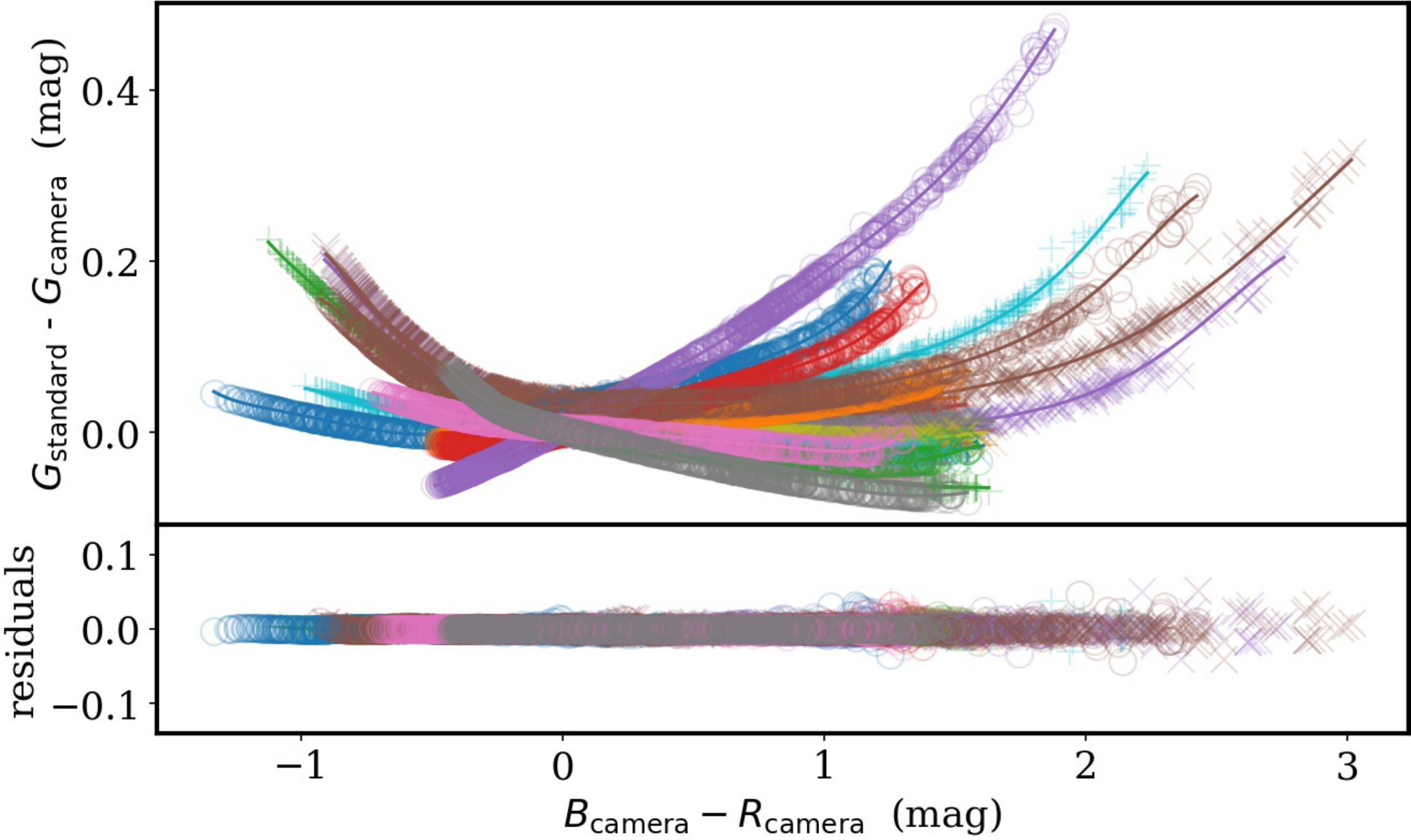}
\vskip 2mm
\includegraphics[width=\columnwidth]{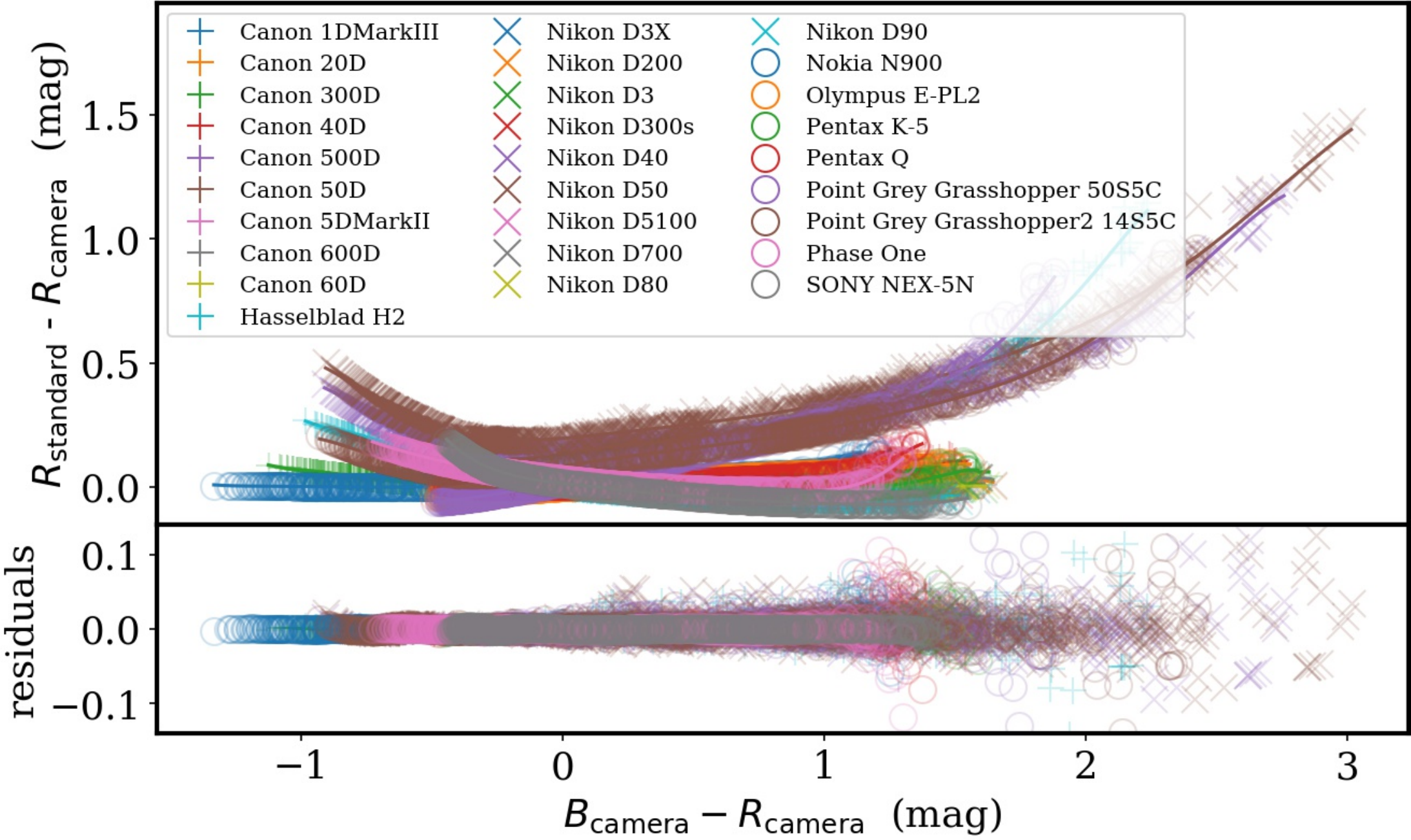}
\caption{Differences between the RGB magnitudes computed using the median
sensitivity curves of Table~\ref{tab:median_RGB_table} (subindex 'standard')
and the ones determined with each one of the 28~individual
sets of RGB sensitivity curves displayed in Fig.~\ref{fig:rgb_standard}
(subindex 'camera'), as a function of the $B-R$~color of the considered camera.
The three RGB filters are represented separately ($B$, $G$, and $R$ from top to
bottom). The differences have been computed for the 1346~stars composing the
final star sample listed in Table~\ref{tab:bigtable}, and for the 28~cameras
listed in the key that appears in the inset of the bottom panel. A 9-degree
polynomial has been fitted to each filter and camera (continuous lines;
polynomial coefficients given in Appendix~B), while
the fit residuals are displayed at the bottom panel of each filter plot.}
\label{fig:gujinwei_comparison}
\end{figure}

In order to test the potential usability of the standard RGB system defined
with the median sensitivity curves, it is important to check how easy will be
the transformation from RGB magnitudes measured with a typical camera, to the
mentioned standard system. For that purpose, we computed the synthetic RGB
magnitudes over the whole final sample of 1346 stars, using the 28~sets
of RGB filters compiled by \citet{6475015} and displayed in
Fig.~\ref{fig:rgb_standard}.

In Fig.~\ref{fig:gujinwei_comparison} we represent the difference between the
RGB magnitudes computed using the median spectral sensitivity curves provided
in Table~\ref{tab:median_RGB_table} and the ones measured with each of the
28~individual camera sets. In most cases, the differences are of the order of
a few tenths of a magnitude, although larger corrections are necessary for the
bluest stars in the $B$ band and for the reddest stars in the $R$ band.
Without entering into
unnecessary details, this behaviour is not surprising considering the
variations of the different RGB sensitivity curves in
Fig.~\ref{fig:rgb_standard}. It is important to emphasize here that, more
important than how large the differences between a particular RGB filter set
and the standard RGB system are, the relevant result is that these differences
can be reasonably well modelled by smooth polynomials. For illustration, the
displayed data for each filter and camera have been fitted using a
9-degree polynomial, overplotted as continuous line with the
same color as the one chosen for each RGB set, with the fit residuals displayed
at the bottom plot of all the panels. The resulting polynomial coefficients for
28~considered cameras are listed in
Table~B1, together with the median and
standard deviation of the residuals. In most cases, the residuals scatter is
below~0.01~mag. Based on these results, we conclude that observations performed
with most cameras can be transformed into the standard RGB system by using
smooth polynomial fits.


\section{Conclusions}
\label{sec:conclusions}

The availability of a catalogue of bright standard stars with RGB magnitudes,
measured in a well-defined photometric system, is essential to scientifically
exploit the large amount of data that at present can be gathered with
high-quality and inexpensive digital cameras equipped with Bayer-like color
filters.

For the work presented in this paper we have used
high-quality 13~medium-narrow-band photometric data to fit stellar
atmosphere models, in order to obtain spectral energy distributions for bright
stars. The reliability of these fits has been checked by comparing the
synthetic Johnson~$B$ and~$V$ magnitudes computed on the fitted models with the
corresponding data available through the Simbad database. This comparison has
shown that the $\pm 3\sigma$ scatter in both bands is around or
below~0.10~mag, being part of this dispersion attributable to the heterogenous
source compilation in the Simbad data. The initial star sample, composed by
1522 objects, was cleaned by removing bad fits and stars with discrepant
Johnson~$B$ and~$V$ magnitudes, reducing the final sample to 1346~objects.
As an additional check, we have compared a subset of 39 fitted models with
actual absolute flux calibrated spectra from the literature, which revealed
that the typical standard deviation in the fitted 13~photometric bandpasses
is~0.05~mag, being part of this dispersion attributable to the flux calibrated
spectra. Furthermore, the fitted 13~band fluxes for each star were randomly
modified (assuming typical uncertainties of~0.02~mag) in order to generate
bootstrapped versions of the fitted models. 

The whole set of fitted star models, that constitute the UCM library of
spectrophotometric standards, together with their associated bootstrapped
versions, have been employed to compute synthetic RGB magnitudes and
uncertainties. Prior to this step, a new RGB photometric system, that can be
used as a standard reference, has been established, by defining standard RGB
sensitivity curves as the median of the corresponding curves of 28~commercial
cameras. The typical random uncertainties in the synthetic RGB magnitudes,
computed only from the bootstrapped spectra, are close to $0.01$~mag. These
uncertainties are 3~times larger when we compare the synthetic RGB magnitudes
with the corresponding values in the subsample of 39 stars with flux calibrated
spectra. However, the unaccounted uncertainties in the flux
calibrated data should have a contribution to this budget. In addition, we
cannot discard a small systematic deviation of $-0.01$ and
$-0.03$~mag in the $G$ and~$R$ bandpasses, although these numbers rely on the
analysis of only 14~non-variable stars.

The feasibility of using the new RGB photometric bandpasses defined in this
work as a standard RGB system, has been demonstrated by computing simple
polynomial transformations that model the differences between the RGB
magnitudes derived employing the standard system and the ones obtained using 28
individual sets of RGB sensitivity curves of real cameras, with a typical
scatter around these polynomial fits within 0.01~mag.

Since non-variable stars constitue ideal radiometric references, one immediate
application of this work is the transformation of the sky in an accessible and
free laboratory for the proper calibration of the high volume of already
existing (and future) digital cameras. Obviously, a proper calibration will
require the corresponding observational effort and subsequent data reduction
and analysis, but the repeatability of measurements should facilitate to reach
calibration accuracies comparable with those achievable in radiometric
laboratories. It is important to emphasize that the synthetic
stellar library include a non-negligible number of variable stars (594 out of
1346 objects), but they have been kept in the final sample because they did not
present discrepant~$B$ and~$V$ magnitudes in the Simbad database. In any case,
their use and validity should be subject to a careful analysis.

We hope that the catalogue of 1346 flux calibrated stellar spectra presented 
here, that by itself already constitutes a library of bright
spectrophotometric standards suitable for spectroscopic calibrations, and the
corresponding synthetic RGB magnitudes, can be used as a
reference for future work on several astronomical fields, where the
collaboration of many observers equipped with high-quality digital cameras may
provide data that facilitate the research advancement.  In addition, this could
help to make citizen science a reality in the realm of astronomy, increasing
the public's interest and understanding of science, highlighting the fact that
scientific research matters.


\section*{Acknowledgements}

The authors are grateful for the exceptionally careful reading
by the referee, whose constructive remarks have helped to improve the paper,
making the text more precise and readable.
The authors acknowledge financial support from the Spanish Programa Estatal de
I+D+i Orientada a los Retos de la Sociedad under grant RTI2018-096188-B-I00,
which is partly funded by the European Regional Development Fund (ERDF),
S2018/NMT-4291 (TEC2SPACE-CM), and ACTION, a project funded by the European
Union H2020-SwafS-2018-1-824603. SB acknowledges Xunta de Galicia for financial
support under grant ED431B 2020/29. The participation of ASdM
was (partially) supported by the EMISSI@N project (NERC grant NE/P01156X/1).
This work has been possible thanks to the extensive use of IPython and Jupyter
notebooks \citep{PER-GRA:2007}.  This research made use of
Astropy,\footnote{\url{http://www.astropy.org}} a community-developed core
Python package for Astronomy \citep{astropy:2013, astropy:2018}, Numpy
\citep{harris2020array}, Scipy \citep{2020SciPy-NMeth}, and Matplotlib
\citep{4160265}. This research has made use of the Simbad database and the
VizieR catalogue access tool, CDS, Strasbourg, France \mbox{(DOI:
10.26093/cds/vizier)}. The original description of the VizieR service was
published in \mbox{A\&AS 143, 23}.

\section*{Data Availability}

The work in this paper has made use of the photometric data published by
\citet[][Table~7, available
online\footnote{\url{http://vizier.u-strasbg.fr/viz-bin/VizieR?-source=II/84}}]{1975RMxAA...1..299J},
\citet[][Table~1]{1976RMxAA...1..327S}, \citet[][Table~1]{1997PASP..109..958B},
the Stellar Atmopshere Models of \citet{2003IAUS..210P.A20C}, as provided by
the STScI web
page\footnote{\url{https://www.stsci.edu/hst/instrumentation/reference-data-for-calibration-and-tools/astronomical-catalogs/castelli-and-kurucz-atlas}},
the database of camera spectral sensitivity database of \citet[][available
online\footnote{\url{http://www.gujinwei.org/research/camspec/db.html}}]{6475015},
the Bright Star Catalogue \citep[][available
online\footnote{\url{https://vizier.u-strasbg.fr/viz-bin/VizieR-3?-source=V/50/catalog}}]{1964cbs..book.....H},
and the General Catalogue of Variable Stars \citep[][available
online\footnote{\url{https://vizier.u-strasbg.fr/viz-bin/VizieR?-source=B/gcvs}}]{2017ARep...61...80S}.

The supplementary material described as Appendices~A and~B is available online
only.

All the results of this paper, together with future additional material, will
be available online at \url{http://guaix.ucm.es/rgbphot}.



\bibliographystyle{mnras}
\bibliography{paper_RGB}





\bsp	
\label{lastpage}
\end{document}